\documentclass{elsart}

 \usepackage{epsfig}

\usepackage{amssymb,amsmath}

\def\GeV{{\ \mbox{GeV}}}
\def\be{\begin{equation}}
\def\ee{\end{equation}}
\newcommand{\thelog}{\log}
\newcommand{\imax}{L}

\begin{document}

\begin{frontmatter}



\title{\bf Quantum chromodynamics at high energy\\
and statistical physics}


\author{S. Munier}

\address{Centre de physique th\'eorique, \'Ecole Polytechnique, 
CNRS, Palaiseau, France}

\begin{abstract}
When hadrons scatter at high energies, strong color fields, 
whose dynamics is described
by quantum chromodynamics (QCD), are
generated at the interaction point.
If one represents these fields in terms of partons (quarks and gluons),
the average number densities of the latter
saturate at ultrahigh energies.
At that point, nonlinear effects become predominant
in the dynamical equations.
The hadronic states that one gets in this regime of 
QCD are generically called ``color glass condensates''.

Our understanding of scattering in QCD
has benefited from recent progress in statistical
and mathematical physics. The evolution of hadronic scattering amplitudes 
at fixed impact parameter in the regime 
where nonlinear parton saturation effects 
become sizable was
shown to be similar to the time evolution of a system 
of classical particles undergoing 
reaction-diffusion processes. The dynamics of such a system 
is essentially governed by
equations in the universality class of the stochastic
Fisher-Kolmogorov-Petrovsky-Piscounov equation, 
which is a stochastic nonlinear
partial differential equation. Realizations of that kind of 
equations (that is, ``events'' in a particle physics language)
have the form of noisy traveling waves. 
Universal properties of the latter
can be taken over to scattering amplitudes in QCD.

This review provides an introduction to the basic
methods of statistical physics useful in QCD,
and summarizes the correspondence between
these two fields
and its theoretical and phenomenological implications.
\end{abstract}


\end{frontmatter}



\tableofcontents

\section{\label{sec:introduction}Introduction to high energy scattering in QCD}


The study of quantum chromodynamics in the high-energy regime
has taken a new soar in the last 15 years with the wealth of experimental 
data that have been collected, first at the electron-proton collider DESY-HERA, 
and then at the heavy-ion collider RHIC.
More energy in the collision enables 
the production of objects of higher mass in
the final state, and thus the discovery of new particles.
But higher energies make it also possible to observe more quantum fluctuations
of the incoming objects, that is to say, to study more deeply the structure of the
vacuum.

The well-established microscopic theory which describes 
the interactions of hadronic objects is
quantum chromodynamics (QCD). (For a comprehensive textbook, 
see Ref.~\cite{Muta:1998vi}).
There are not many known analytical approaches to QCD,
except perturbative expansions of observables 
in powers of the strong coupling constant $\alpha_s$ which,
thanks to asymptotic freedom, is justified for carefully
chosen observables in special kinematical regimes.
But fixed-order calculations in QCD are known to usually have a very limited
range of applicability. This is because in the evaluation of Feynman graphs,
the coupling constant always comes with ``infrared'' and ``collinear'' logarithms
that are related to the phase space that is available to the reaction, that is to say,
to kinematics. Resumming part of these logarithms is mandatory. All of them
is too difficult.
The question is to carefully select the dominant ones, and this is not at all easy.


At the HERA collider, electrons or positrons scattered off protons at the 
center-of-mass energy $\sqrt{s}$,
exchanging a photon of virtuality $Q$. Through the scattering, one
could probe partonic fluctuations of the proton 
(made of quarks and gluons) of transverse momenta $k\sim Q$, and
longitudinal momentum fractions $x\sim Q^2/(Q^2+s)$.


For a long time, the dominant paradigm had been that the {\em collinear logarithms}
$\log Q^2$,
that become large when $Q^2$ is large compared to the QCD confinement scale
$\Lambda^2$, were
the most important ones.
As a matter of fact, searches for new particles or for exotic physics require
to scrutinize matter at very small distances, and hence very large $Q^2$ have
to be considered.
Perturbative series of powers of $\alpha_s\log Q^2$ have to be fully resummed.
The equation that performs this resummation is the celebrated 
Dokshitzer-Gribov-Lipatov-Altarelli-Parisi (DGLAP) equation 
\cite{Gribov:1972ri,Dokshitzer:1977sg,Altarelli:1977zs}.


However, once HERA had revealed its ability to get extremely good statistics
in a regime in which
$Q^2$ is moderate (from $1$ to $100\GeV^2$) and $x$ very small (down to $10^{-5}$)
it became clear that {\em infrared logarithms} ($\log 1/x$) could show up and even 
dominate the measured observables.
The resummation of the series of infrared logs is performed by the 
Balitsky-Fadin-Kuraev-Lipatov (BFKL) equation
\cite{Lipatov:1976zz,Kuraev:1977fs,Balitsky:1978ic}.
The series $\sum (\alpha_s\log 1/x)^k$ (with appropriate coefficients)
is the leading order (LO), while the series $\sum \alpha_s(\alpha_s\log 1/x)^k$
is the next-to-leading order (NLO), which has also been computed 
\cite{Ciafaloni:1998gs,Fadin:1998py}.
The BFKL equation is a linear integro-differential equation.


At ultrahigh energy, the bare BFKL equation seems to violate the Froissart
bound, that states that total hadronic cross sections cannot rise faster than
$(\log^2 s)/m_\pi^2$. The latter is a consequence
of the unitarity of the probability of scattering.
 The BFKL equation predicts a power rise with the
energy of the form $s^\varepsilon$, where $\varepsilon$ is positive and
quite large ($0.3$ to $0.5$ according to the effective value of $\alpha_s$ 
that is chosen).
The point at which the BFKL equation breaks down depends on the value of the
typical transverse momentum which characterizes the observable 
(It is the photon virtuality
$Q$ in the case of deep-inelastic scattering).
One may define the energy-dependent {\em saturation scale} $Q_s(x)$ in such a way
that the BFKL equation holds for $Q>Q_s(x)$.
For $Q\sim Q_s(x)$, the probability for scattering to take place
is of order 1, and for $Q<Q_s(x)$, it would be larger than 1 if one trusted
the BFKL equation.
The saturation scale 
is a central observable, which we shall keep discussing in this review:
It signs the point at which the linear (BFKL) formalism has to be corrected
for nonlinear effects. The regime in which nonlinearities manifest themselves
is a regime of strong color fields, sometimes called the {\em color glass
condensate}
(For the etymology of this term,
see e.g. the lectures of Ref.~\cite{McLerran:2001sr};
for a review, see Ref.~\cite{Iancu:2003xm}).

The fact that unitarity is violated
is not only due to the lack of a hadronic scale in the BFKL equation,
which is a perturbative equation; Introducing confinement in the form of
a cutoff would not help this
particular problem.
It simply means that still higher orders are needed.
The NLO corrections to the BFKL kernel indeed correct this
behavior in such a way that the description of the HERA data 
in the small-$x$ regime
is possible by the BFKL equation.
However, these corrections are 
not enough to tame the power-like growth of cross sections as
predicted by the LO BFKL equation.
It seems that a resummation of contributions of arbitrary order would be needed.

New equations were proposed well before the advent of colliders
able to reach this regime.
Gribov-Levin-Ryskin wrote down a model 
for the evolution of the hadronic scattering cross sections
in the early 80's \cite{Gribov:1981ac,Gribov:1984tu}, and Mueller and Qiu
derived a similar equation from QCD a bit later \cite{Mueller:1985wy}.
These equations are integral evolution equations with a nonlinear term,
which basically takes into account parton saturation effects, that is to say,
recombination or rescattering.
The latter cannot be described in a linear framework such as the BFKL formalism.
Subsequently, more involved QCD evolution equations were derived
from different points of view.
In the 90's, McLerran and Venugopalan 
\cite{McLerran:1993ni,McLerran:1993ka,McLerran:1994vd} proposed a first model,
mainly designed to approach heavy-ion collisions.
Subsequently, Balitsky
\cite{Balitsky:1995ub},
Jalilian-Marian, Iancu, McLerran, Weigert, Leonidov and Kovner (B-JIMWLK)
\cite{JalilianMarian:1997jx,JalilianMarian:1997gr,Iancu:2001ad,Iancu:2000hn,Weigert:2000gi}
worked out QCD corrections to this model, and got equations that reduce
to the BFKL equation in the appropriate limit.
Technically, these equations actually have the form of an infinite hierarchy of
coupled integro-differential equations (in Balitsky's formulation 
\cite{Balitsky:1995ub}), of a functional renormalizaton group equation,
or alternatively, of a Langevin equation 
(in Weigert's formulation \cite{Weigert:2000gi}).
A much simpler equation was derived in 1996 by Balitsky \cite{Balitsky:1995ub}
and rederived by Kovchegov in 1999
\cite{Kovchegov:1999yj,Kovchegov:1999ua}
in a very elegant way
within a different formalism. The obtained equation is called the Balitsky-Kovchegov
equation (BK).
The latter derivation was based on Mueller's color dipole model \cite{Mueller:1993rr},
which proves particularly suited to represent QCD in the high energy limit.

The exciting feature of this kinematical regime of
hadronic interaction from a theoretical point of view
is that the color fields are strong, although, at sufficiently high energies,
the QCD coupling is weak, authorizing a perturbative approach,
and thus analytical calculations.
In such strong field regime, nonlinear effects become crucial.
But the conditions of applicability of the different
equations that had been found had never been quite clear.
Anyway, these equations are extremely difficult to solve, which had probably been
the main obstacle to more rapid theoretical developments in the field
until recently.

\begin{figure}
\begin{center}
\epsfig{file=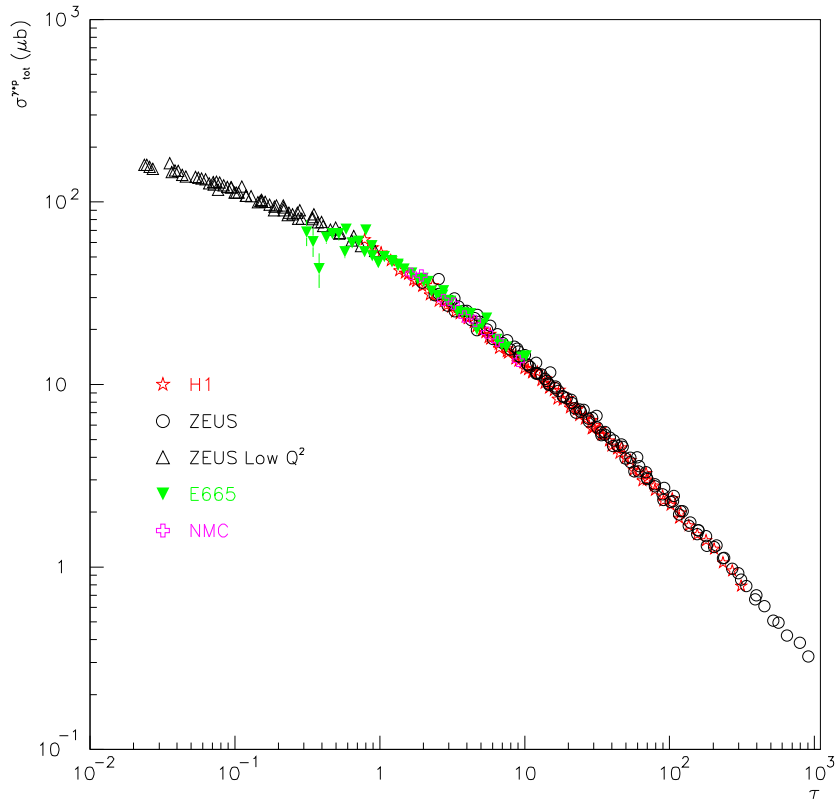,width=0.8\textwidth}
\end{center}
\caption{\label{fig:geometricscaling}
[From Ref.~\cite{Marquet:2006jb}]
Photon-proton total cross section from
the most recent set of deep-inelastic scattering data in the low-$x$ regime
plotted as a function of a single scaling variable 
$\tau=Q^2/Q_s^2(x)$, where $Q$ is the virtuality of the photon and
$Q_s^2(x)\sim\Lambda^2 x^{-0.3}$
is the so-called saturation scale.
Although the cross section is a priori a function of two variables,
all data fall on the same curve. This phenomenon is called
{\em geometric scaling}~\cite{Stasto:2000er}.
}
\end{figure}

Furthermore, for a long time, the phenomenological need for
such a sophisticated formalism was not obvious, since linear
evolution equations such as the DGLAP equation were able to
account for almost all data. But Golec-Biernat and W\"usthoff
showed that unitarization effects may have already
been seen at HERA \cite{GolecBiernat:1998js,GolecBiernat:1999qd}. 
Their model predicted, in particular, that
the virtual photon-proton cross section should
only depend on one single variable $\tau$, made of a combination of
the transverse momentum scale (fixed by the virtuality of the photon $Q$) 
and $x$.
This phenomenon was called ``geometric scaling'' \cite{Stasto:2000er}.
It was found in the HERA data (see Fig.~\ref{fig:geometricscaling}):
This is maybe one of the most
spectacular experimental result from HERA 
in the small-$x$ regime.

This observation has triggered many phenomenological and theoretical works.
Soon after its discovery in the data, 
geometric scaling was shown to be a solution of the Balitsky-Kovchegov (BK)
equation, essentially numerically, with some analytical arguments
(see e.g. \cite{Levin:1999mw,Levin:2000mv,Armesto:2001fa,GolecBiernat:2001if}).
The energy dependence of the saturation scale was eventually precisely computed
by Mueller and Triantafyllopoulos \cite{Mueller:2002zm}.
Later, it was shown that the BK equation is actually in the universality
class of the Fisher-Kolmogorov-Petrovsky-Piscounov (FKPP) equation \cite{Fisher,KPP},
and geometric scaling was found to be implied by the fact that the latter equation
admits {\em traveling wave} solutions \cite{Munier:2003vc}.

A first step beyond the BK equation, in the direction of a full solution
to high energy QCD, was taken by Mueller and Shoshi in 2004 \cite{Mueller:2004sea}.
Actually, they did not solve the B-JIMWLK equations,
but instead, they solved the linear BFKL equation with two 
absorptive boundary conditions, which
they argued to be appropriate to represent the expected nonlinearities.
Geometric scaling {\em violations} were found from their calculation, 
which should show up at any energies.

Subsequently, it was shown that high-energy QCD 
at fixed coupling is actually 
in the universality class of {\em reaction-diffusion processes},
studied in statistical physics,
whose dynamics may be encoded in equations similar to the
{\em stochastic} FKPP equation \cite{Iancu:2004es}.
The Mueller-Shoshi solution was shown to be consistent
with solutions to the latter equation.
So high-energy QCD seems to be in correspondence with
disordered systems studied in statistical physics.
This correspondence has provided a new understanding
of QCD in the high-energy regime, and it has
proven very useful to find more features of high-energy
scattering.
The obtained results go beyond a solution to the B-JIMWLK equation,
which in fact, thanks to the new picture, 
is seen to be incomplete.

{\em Scope}

The goal of this review article is to 
summarize the main ideas behind this conjectured correspondence
between scattering at high-energy in QCD and some
processes studied in statistical physics, as well as to introduce the
QCD reader to the useful technical tools borrowed from
statistical physics.
We also feel that there is a cultural gap to be filled
between statistical physics and particle physics.
Indeed, statistical physicists are used to build simple toy models
which contain the interesting physics,
and whose main properties are likely to be independent of the
details of the model, i.e. {\em universal}.
In QCD, since the theory is well-established,
we are often reluctant to give up some of its features
to work out exact results in a toy model.
One of our aims is to convince the reader that
such a way of thinking is efficient in the case
of high-energy QCD, by exhibiting results for QCD
scattering amplitudes, obtained by looking for the universality
class of the considered process, and 
that are believed to be exact.

Over the last few years, several hundreds of papers have appeared related to this
subject, mainly issued from a very active though restricted community.
Obviously, we cannot give a complete account of this abundant literature.
As a matter of fact, some important recent developments had to be left out,
for which we shall only provide references for the interested reader who
might want to deepen his study in these directions.
Concerning the correspondence itself, we do not
attempt to establish a definite stochastic nonlinear evolution equation
for QCD amplitudes, for to our judgement, this research line
is not mature enough yet: A better understanding of
the very saturation mechanism at work in QCD is definitely needed before one may
come to this issue. 
Furthermore, it is not clear to us that a stochastic formulation
would be a technical progress, since there are not many known methods to
handle complicated stochastic equations.
We feel that the same is true for
the search for effective actions that would include so-called Pomeron loops.
We also do not address the developments based on the boost-invariance symmetry
that scattering amplitudes should have: This would drive us too far off
the main focus of this review.
As for more phenomenological aspects, 
we only discuss the basic features of total cross sections without
attempting to address other observables such as diffraction.
We do also not address the issue of next-to-leading effects such as
the running of the QCD coupling. This discussion, though crucial if one
wants to make predictions for actual colliders,
would probably only be technical in its nature: There is no conceptual difference
between the fixed coupling and the running coupling case.
Here, only basic phenomenological facts brought about
by this new understanding of high-energy QCD are addressed, namely 
geometric scaling and diffusive scaling.


{\em Outline}

The outline goes as follows.
The next section is devoted to describing scattering in QCD
from a $s$-channel point of view, relying essentially on the
parton model or rather on an interpretation useful in the
 high-energy limit, the color dipole model.
Once this picture is introduced, it is not difficult to understand 
the correspondence with reaction-diffusion
processes occuring in one spatial dimension, whose dynamics is captured by
equations in the university class of the Fisher-Kolmogorov-Petrovsky-Piscounov (FKPP)
equation.
We then explain how traveling waves appear in this context. 
In Sec.~\ref{sec:zerodimensional}, we study in greater detail a toy model
for which many technics (field theory, statistical methods) may be
worked out completely. This model however ignores spatial dimensions,
and thus, does not account for traveling waves.
We summarize the state-of-the-art research on equations in the universality class
of the FKPP equation in Sec.~\ref{sec:reviewtraveling}.
Finally, we come back to QCD, discussing the relevance of one-dimensional-like
models in the FKPP class, and showing how noisy traveling waves may show up
in the actual data.


\section{\label{sec:schannel}
Hadronic interactions in a $s$-channel picture
and analogy with reaction-diffusion processes
}

In this section, we shall introduce the physical picture
of high-energy scattering in the parton model.
In particular, the color dipole model \cite{Mueller:1993rr} is described
since it is particularly suited to address high-energy scattering,
especially close to the regime in which nonlinear effects
are expected to play a significant role.
In a second part, we shall argue that high-energy scattering is a peculiar
reaction-diffusion process.

\subsection{\label{sec:partonmodel}Parton model and dipoles}

\subsubsection{General picture}

\begin{figure}
\begin{center}
\begin{tabular}{cc}
\epsfig{file=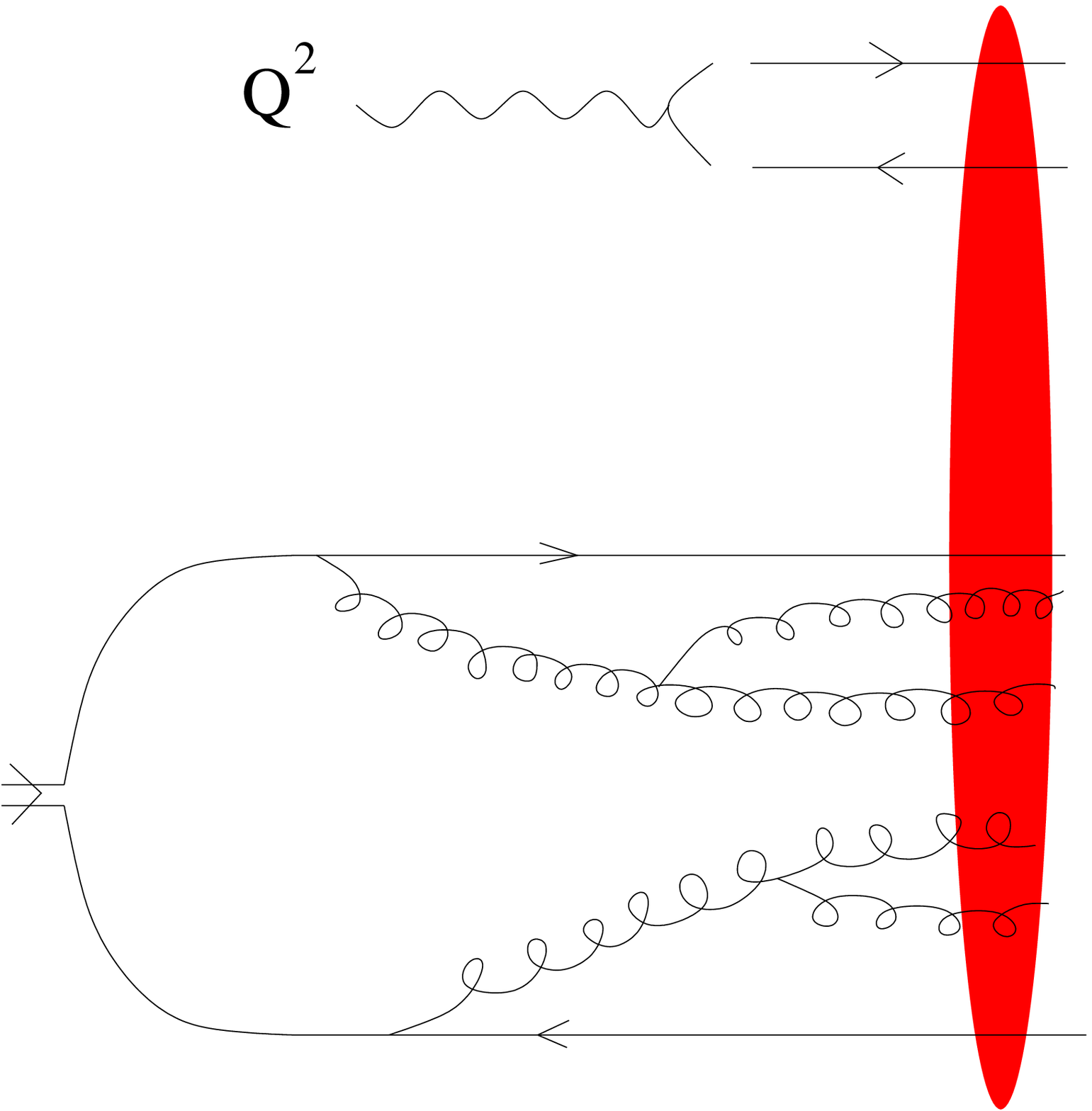,width=5cm}
\phantom{spacespa}
&\epsfig{file=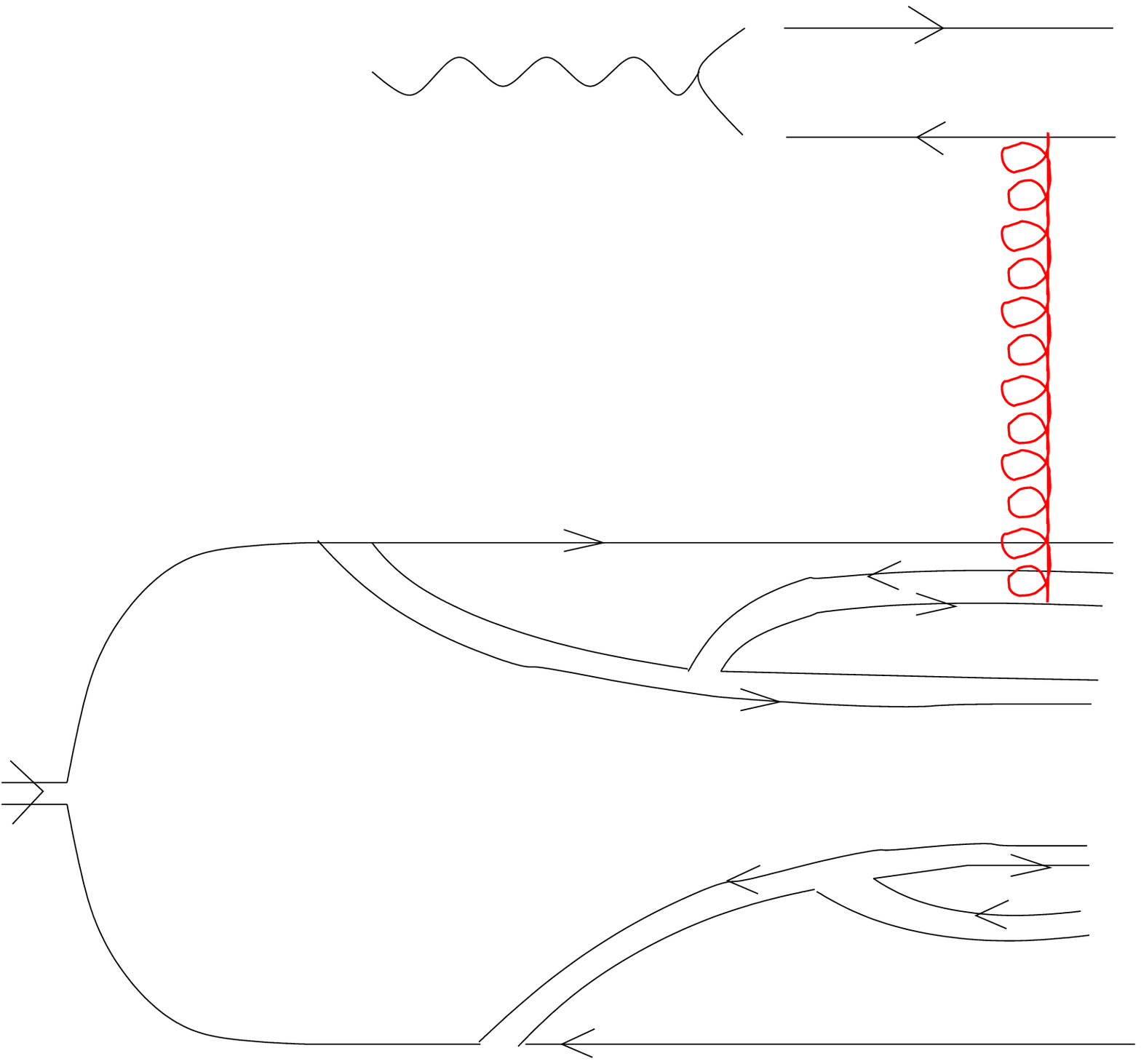,width=5cm}\\
(a) & (b)
\end{tabular}
\end{center}
\caption{\label{fig:basicscat}
{\em (a)}
The scattering of a virtual photon probe off a particular
fluctuation of an evolved target made of a quark and an antiquark
in its bare state.
The photon necessarily goes through a quark-antiquark pair at high
enough energies, when the target is dominated by dense gluonic states.
(What is represented in this figure is actually the inelastic amplitude, which
is a cut of the total cross section or of the forward elastic amplitude).
{\em (b)} In the dipole model, the probe and the 
target may be represented by a set of color dipoles, 
and the interaction proceeds through gluon exchanges.
The curly vertical line represents the lowest-order interaction between
pairs of dipoles, that is to say, the exchange of a gluon
(or a two-gluon exchange if one is speaking 
of the forward elastic amplitude).
}
\end{figure}

For definiteness, let us consider the scattering of a hadronic probe off a given target,
in the restframe of the probe and at a fixed impact parameter, that is to
say, at a fixed distance between the probe and the center of the target
in the two-dimensional plane transverse to the collision axis.
In the parton model, the target interacts
through one of its quantum fluctuations, made of a high occupancy Fock state
if the energy of the reaction is sufficiently high (see Fig.~\ref{fig:basicscat}a).
As will be understood below,
the probe effectively ``counts'' the partons in the current Fock state
of the target whose transverse momenta $k$ (or sizes $r\sim 1/k$)
are of the order of the one that characterizes the probe: 
Roughly speaking, the amplitude for the scattering off this particular
partonic configuration is proportional to the number of such partons.

The observable that is maybe the most sensitive to quantum fluctuations
of a hadron is the cross section for the interaction of a virtual photon
with a hadronic target such as a proton or a nucleus.
The virtual photon is emitted by an electron (or a positron).
What is interesting with this process, called ``deep-inelastic scattering'',
is that the kinematics of the photon is fully controlled
by the measurement of the scattered electron.
The photon can be considered a hadronic object since it interacts through
its fluctuations into a quark-antiquark state. The latter form
color dipoles since although both the quark and the antiquark carry color
charge, the overall object is color neutral due to the color neutrality
of the photon. The probability distribution of these fluctuations may
be computed in quantum electrodynamics (QED). 
Subsequently, the dipole interacts with the target
by exchanging gluons. The dipole-target cross section factorizes
at high energy.
One typical event is depicted in Fig.~\ref{fig:basicscat}a.

Dipole models \cite{Nikolaev:1990ja,Nikolaev:1991et}
have become more and more popular among phenomenologists
since knowing the dipole cross section enables one to compute different
kinds of observables. 
Like parton densities, the latter is a universal quantity, that
may be extracted from one process and used to predict other observables.
Different phenomenological models may 
be tried for the dipole cross section.
QCD evolution equations may even be derived, as we shall discover below.
An accurate recent study of the foundations of dipole
models may be found in Ref.~\cite{Ewerz:2004vf,Ewerz:2006vd}.

In QCD, the state of a hadronic object, encoded in a set of wave functions, is built up
from successive splittings of partons starting from the valence structure.
This is visible in the example of Fig.~\ref{fig:basicscat}a:
The quark and the antiquark that build up, in this example, the target in its
asymptotic state each emit a gluon, which themselves emit, 
later on in the evolution, other gluons.
As one increases the rapidity $y$ by boosting the target,
the opening of the phase space for parton splittings makes the probability
for high occupation numbers larger.
Indeed, the probability to find a gluon that carries a fraction $z$ (up to $dz$)
of the momentum of its parent parton (which may be a quark or a gluon)
is of order $\alpha_s N_c dz/z$ for small $z$.
There is a logarithmic singularity in $z$, meaning that 
emissions of very soft gluons (small $z$) are favored if they
are allowed by the kinematics.
The splitting probability is of order 1 when the total rapidity of the scattering 
$y=\log 1/x$ is increased
by roughly $1/\bar\alpha$, where the convenient notation 
$\bar\alpha=\alpha_s N_c/\pi$ has been introduced.
Only splittings of a quark or of a gluon into a gluon 
exhibit the $1/z$ singularity.
Therefore, at large rapidities, gluons eventually
dominate the partonic content of the hadrons.

The parton model in its basic form, where the fundamental objects of
the theory (quarks and gluons) are directly considered,
is not so easy to handle in the high-energy regime.
One may considerably simplify the problem by going to the limit of
 large number of colors ($N_c$), 
in which a gluon may be seen as a zero-size quark-antiquark pair.
Then, color-neutral objects become collections of color dipoles, whose endpoints consist
in ``half gluons'' (see Fig.~\ref{fig:basicscat}b).
There is only one type of objects in the theory, dipoles, which simplifies very much
the picture.
Furthermore, going to {\em transverse coordinate space} 
(instead of momentum space, usually used in the DGLAP formalism) 
by trading the transverse momenta of the gluons for the sizes of the dipoles
(through an appropriate Fourier transform)
brings another considerable simplification.
Indeed, the splittings that contribute to amplitudes in the high-energy limit
are the soft ones, for 
which the emitted gluons take only a small fraction of the momentum
of their parent (the latter being very large). Therefore,
the positions of the gluons (and thus of the
edges of the dipoles) in the plane transverse to the collision axis are not
modified by subsequent evolution once the gluons have been created.
Thus, the evolution of each dipole proceeds through completely {\em independent}
splittings to new dipoles.

We will now see how this picture translates into a QCD evolution equation
for scattering amplitudes, first in the regime in which there are no nonlinear
effects. In a second step, 
we will try and understand how to incorporate the latter.

\subsubsection{BFKL equation from the dipole model}

\begin{figure}
\begin{center}
\epsfig{file=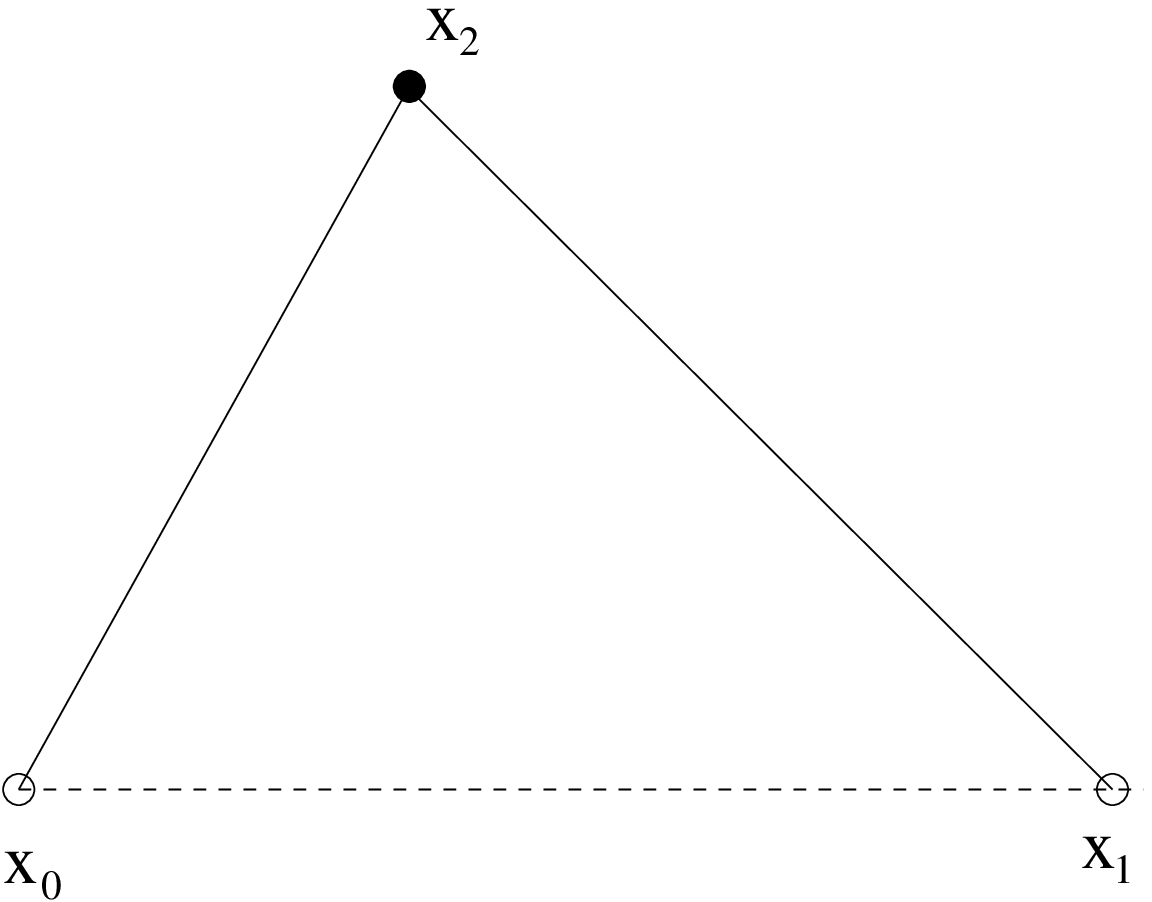,width=6cm}\\
(a)\\
\epsfig{file=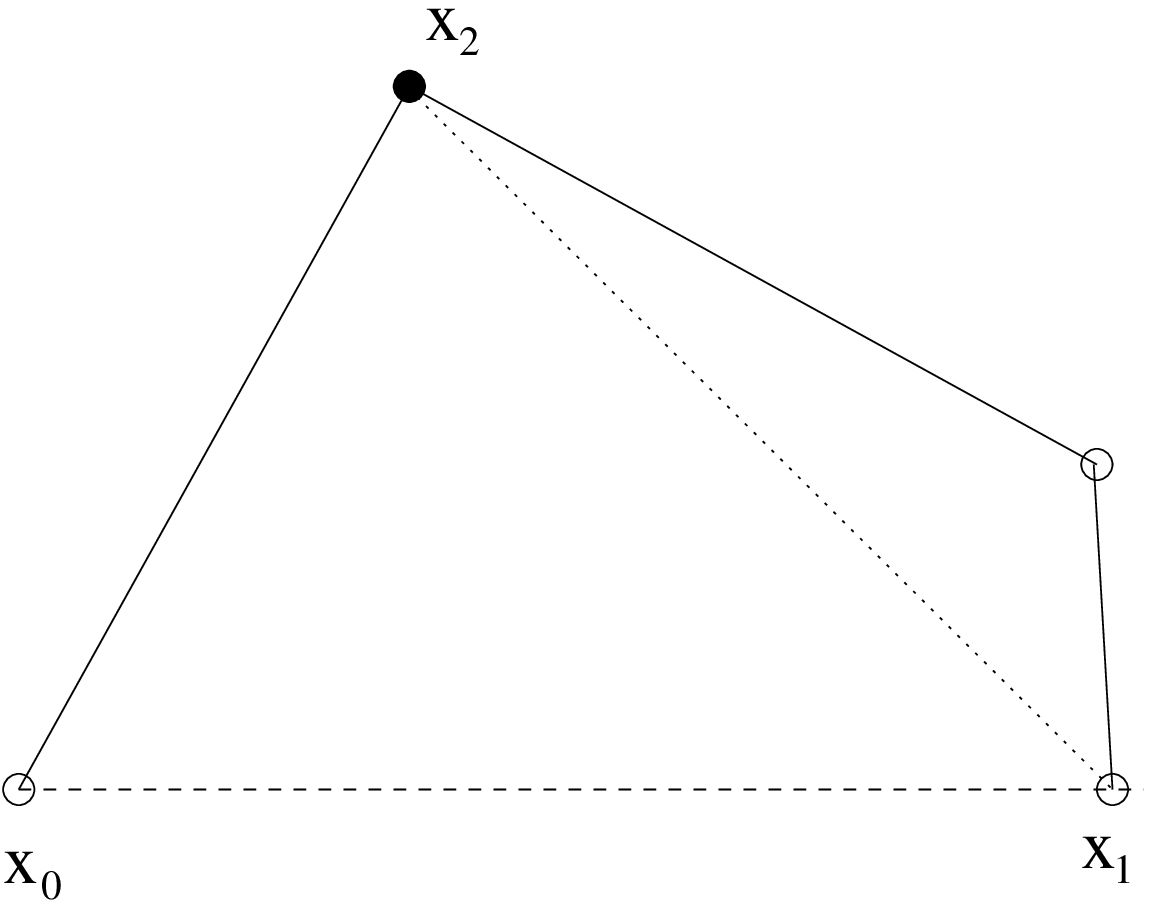,width=6cm}\\
(b)\\
\epsfig{file=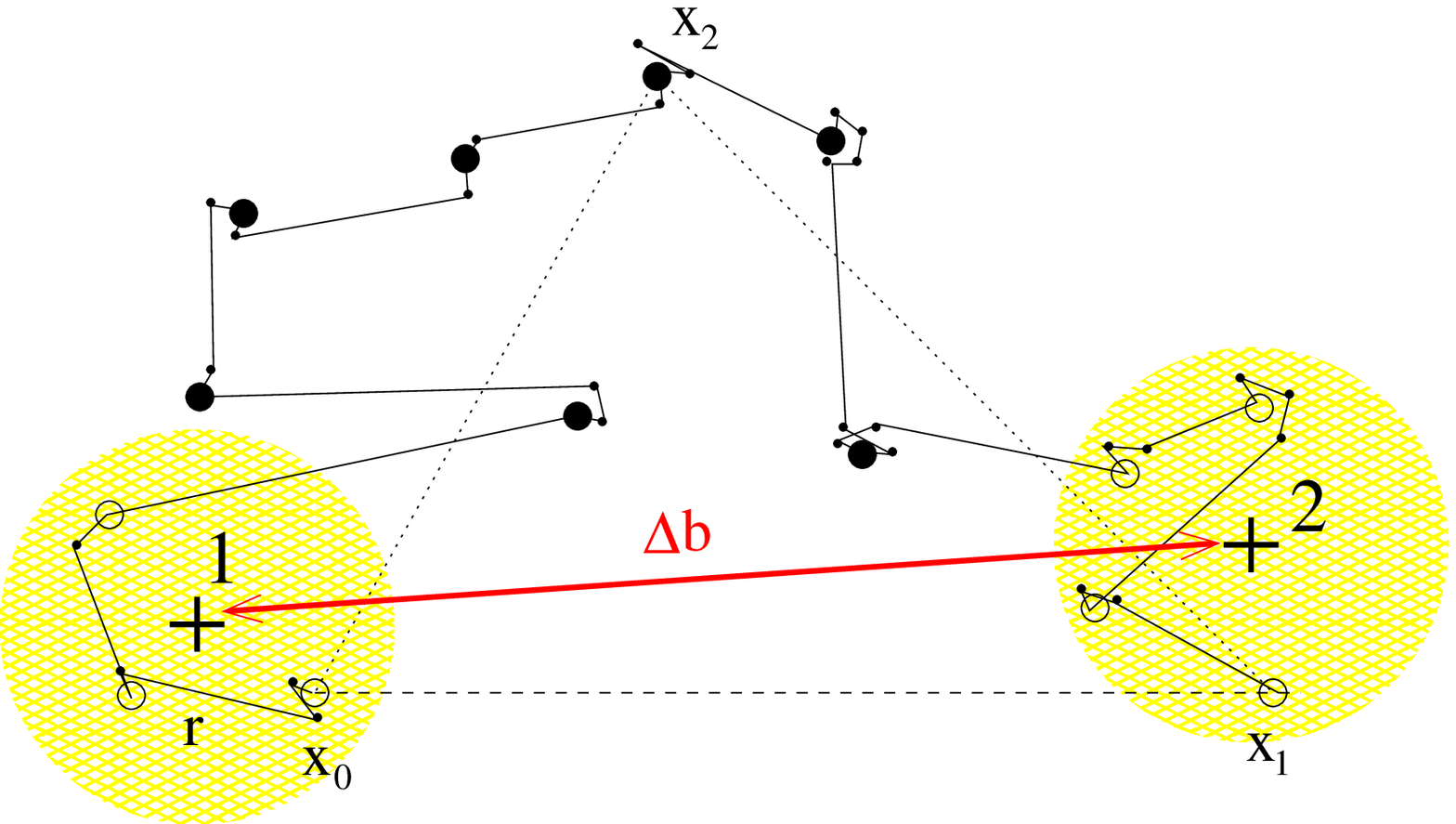,width=10cm}\\
(c)
\end{center}
\caption{\label{fig:scheme}Schematic picture of
a realization of the dipole evolution after the first
two steps of the evolution ({\it (a)} and {\it (b)}), and after some
larger rapidity evolution {\it (c)}.
In the first step {\it (a)}, 
the initial dipole $(x_0,x_1)$ (denoted by a dashed line)
splits to the new dipoles 
$(x_0,x_2)$ and $(x_2,x_1)$ (full lines).
The points represent the edges of each dipole,
that is to say, the position of the gluons.
In the next step {\it (b)}, the dipole $(x_2,x_1)$ itself splits
in two new dipoles. The splitting process proceeds {\it (c)} until
the maximum rapidity is reached.
Many very small dipoles are produced in the vicinity of 
each of these endpoints, due to the infrared singularity visible
in Eq.~(\ref{splitting}) (Only a fraction of them is represented).
The zones 1 and 2 in {\it (c)}, separated by the transverse 
distance $\Delta b$, would
evolve quasi-independently after the stage
depicted in this figure.
}
\end{figure}

The building up of the states of each hadron is specified by providing the
splitting rate of a dipole whose endpoints have transverse coordinates
$(x_0,x_1)$ into two dipoles $(x_0,x_2)$ and $(x_1,x_2)$ as the result of a gluon
emission at position $x_2$.
It is computed in perturbative QCD and reads \cite{Mueller:1993rr}
\begin{equation}
\frac{dP}{d(\bar\alpha y)}((x_{0},x_1)\rightarrow (x_{0},x_2),(x_{2},x_1))
=\frac{|x_{0}-x_1|^2}{|x_{0}-x_2|^2 |x_{1}-x_2|^2}\frac{d^2 x_2}{2\pi}.
\label{splitting}
\end{equation}
Dipole splittings are independent. After some rapidity evolution
starting from
a primordial dipole, one gets a chain of dipoles such as the one
depicted in Fig.~\ref{fig:scheme}.

The elementary scattering amplitude for one projectile dipole $(x_0,x_1)$
off a target dipole $(z_0,z_1)$ is independent of the rapidity and 
reads \cite{Mueller:1993rr}
\begin{equation}
T^{\text{el}}((x_0,x_1),(z_0,z_1))=\frac{\pi^2\alpha_s^2}{2}
\log^2\frac{|x_0-z_1|^2|x_1-z_0|^2}{|x_0-z_0|^2|x_1-z_1|^2}.
\label{eq:Tdipole}
\end{equation}
If the target is an evolved state at rapidity $y$, then it
consists instead in a distribution $n(y,(z_0,z_1))$
of dipoles.
The (forward elastic) scattering amplitude $A(y,(x_0,x_1))$ is just given
by the convolution of $n$ and $T^{\text{el}}$, namely
\begin{equation}
A(y,(x_0,x_1))=
\int \frac{d^2 z_0}{2\pi}\frac{d^2 z_1}{2\pi}
T^{\text{el}}((x_0,x_1),(z_0,z_1))n(y,(z_0,z_1)).
\label{eq:BFKLprojectile}
\end{equation}

Let us examine the properties of $T^{\text{el}}$.
To this aim, it is useful to decompose the coordinates of the
dipoles in their size vector $r_a=x_0-x_1$ (resp. $r_b=z_0-z_1$)
and impact parameter $b_a=\frac{x_0+x_1}{2}$ (resp. $b_b=\frac{z_0+z_1}{2}$).
In the limit in which the relative impact parameters of the dipoles 
$b=b_a-b_b$ is very large compared to their sizes, 
we get the simplified expression
\begin{equation}
T^{\text{el}}(r_a,r_b,b)\underset{|r_a|,|r_b|\ll|b|}{\sim}
 {\alpha_s^2}\frac{r_a^2 r_b^2}{b^4},
\label{eq:Tdipoleapprox1}
\end{equation}
and thus the scattering amplitude decays fast as a function of the relative
impact parameter.
If instead the relative impact parameter is small (of the order of the size of the smallest
dipole), we get
\begin{equation}
T^{\text{el}}(r_a,r_b,b)\underset{|r_a|,|r_b|\sim|b|}{\sim} {\alpha_s^2}
\frac{r_<^2}{r_>^2},
\label{eq:Tdipoleapprox2}
\end{equation}
where $r_<=\min(|r_a|,|r_b|)$, $r_>=\max(|r_a|,|r_b|)$, and
the integration over the angles has been performed.

Equation~(\ref{eq:Tdipoleapprox1}) means that the dipole interaction 
is local in impact parameter: It vanishes as soon as the relative
distance of the dipoles is a few steps in units of their size.
Eq.~(\ref{eq:Tdipoleapprox2}) shows that only dipoles whose sizes
are of the same order of magnitude interact.
These properties are natural in quantum mechanics.
Thus the amplitude $A$ in Eq.~(\ref{eq:BFKLprojectile}) effectively
``counts'' the dipoles of size of the order of $|x_{01}|$
at the impact parameter $\frac{x_0+x_1}{2}$ (up to $|x_{01}|$), 
with a weight factor
$\alpha_s^2$.

An evolution equation for the
amplitude $A$ 
with the rapidity of the scattering can be established.
It is enough to know how the dipole density in the target evolves when
rapidity is increased, since all the rapidity dependence is contained
in $n$ in the factorization~(\ref{eq:BFKLprojectile}), and such an equation
may easily be worked out with the help of the splitting
rate distribution~(\ref{splitting}). It reads \cite{Mueller:1993rr}
\begin{multline}
\frac{\partial n(y,(x_0,x_1))}{\partial (\bar\alpha y)}=
\int \frac{d^2 x_2}{2\pi}\frac{|x_{01}|^2}{|x_{02}|^2 |x_{12}|^2}
[
n(y,(x_0,x_2))+n(y,(x_2,x_1))\\
-n(y,(x_0,x_1))
],
\label{eq:BFKL}
\end{multline}
where $x_{ab}\equiv x_a-x_b$. The very same equation holds for $A$.
The elementary scattering amplitude $T^\text{el}$ only appears
in the initial condition at $y=0$, which is not shown in Eq.~(\ref{eq:BFKL}).
In a nutshell, the integral kernel encodes the branching diffusion of the dipoles.
The total number of dipoles at a given impact parameter grows exponentially,
and their sizes diffuse. The appropriate variable in which diffusion takes
place is $\log(1/|x_{01}|^2)$. (This is due to the collinear singularities
in Eq.~(\ref{splitting}).)
This equation is nothing but the BFKL equation.
A complete solution to this equation, 
including the impact-parameter dependence, 
is known \cite{Lipatov:1985uk}.

An important property of the amplitude $A$ is that it is boost-invariant.
This property is preserved in the BFKL formulation.
We could have put the evolution in the projectile instead of the
target, or shared it between the projectile and the target: The result
for the scattering amplitude would have been the same.
In a frame in which the target carries $y^\prime$ units of rapidity and
the projectile $y-y^\prime$, the amplitude $A$ reads
\begin{multline}
A(y,(x_0,x_1))=\int \frac{d^2 z_0}{2\pi}\frac{d^2 z_1}{2\pi}
\frac{d^2 z_0^\prime}{2\pi}\frac{d^2 z_1^\prime}{2\pi}
n^\text{projectile}(y-y^\prime,(z_0,z_1)|(x_0,x_1))\\
\times
T^\text{el}((z_0,z_1),(z_0^\prime,z_1^\prime))
n^\text{target}(y^\prime,(z_0^\prime,z_1^\prime)).
\label{eq:BFKLframe}
\end{multline}
$n^\text{projectile}(y-y^\prime,(z_0,z_1)|(x_0,x_1))$
is the density of dipoles $(z_0,z_1)$ found in a dipole of initial
size $(x_0,x_1)$ after evolution over $y-y^\prime$ steps in rapidity.
If $y^\prime=y$, one recovers Eq.~(\ref{eq:BFKLprojectile}).
If $y^\prime=0$, then all the evolution is in the projectile instead.

The amplitude $A$ is related to an interaction probability,
and thus, it must be bounded:
In appropriate normalizations, $A$ has to range between 0 and 1.
But as stated above,
the BFKL equation predicts an exponential rise of $A$
with the rapidity for any dipole size, 
which at large rapidities eventually violates unitarity.
Hence the BFKL equation does not provide a complete account of
high-energy scattering in QCD.


\subsubsection{Unitarity and the Balitsky-Kovchegov equation}

It is clear that one important ingredient that has been
left out in the derivation of the BFKL
equation is the possibility of {\em multiple scatterings}
between the probe and the target.
Several among the $n$ dipoles in Eq.~(\ref{eq:BFKLframe})
may actually interact with the dipoles in the other hadron
simultaneously. The only reason why such interactions
may not take place is that $T^\text{el}\sim\alpha_s^2$ (see Eq.~(\ref{eq:Tdipole})),
and thus the probability for two simultaneous
scatterings is of order $\alpha_s^4$, which
is suppressed. But this argument holds only as long as the dipole number densities
are of order $1$. If $n\sim 1/\alpha_s^2$
(which is also the point above which the unitarity of $A$ is no longer
preserved in the BFKL approach), 
then it is clear that multiple scatterings
should occur.

\begin{figure}
\begin{center}
\epsfig{file=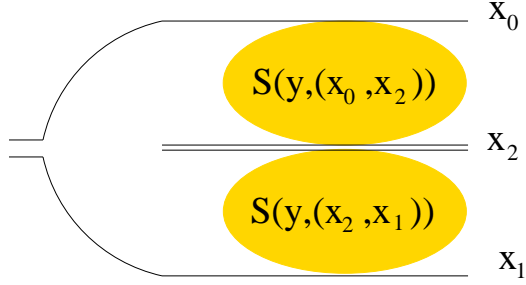,width=0.5\textwidth}
\end{center}
\caption{\label{fig:BK2}Derivation of the Balitsky-Kovchegov equation.}
\end{figure}

\begin{figure}
\begin{center}
\epsfig{file=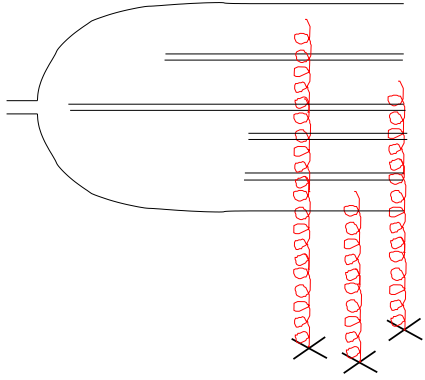,width=0.4\textwidth}
\end{center}
\caption{\label{fig:bk}
Picture of the BK evolution.
All the QCD evolution is put in the probe, which carries the
total rapidity. It develops a high occupancy state of dipoles,
which scatter independently off the target.
}
\end{figure}

In order to try and implement these multiple scatterings, we introduce
the probability that there be no scattering between
a dipole $(x_0,x_1)$ and a given realization of the target with total rapidity $y$,
that we shall denote by
$S(y,(x_0,x_1))$.
Let us start with a system in which the evolution is fully contained in the target.
We increase the total rapidity by boosting the {\em projectile} (initially at rest)
by a small amount $dy$.
Then there are two cases to distinguish, depending
on whether the dipole $(x_0,x_1)$ splits in the rapidity interval $dy$.
In case it splits into two dipoles $(x_0,x_2)$ and $(x_2,x_1)$, the probability that 
the projectile does not interact is just the product of the probabilities
that each of these new dipoles do not interact.
This is because once created, dipoles are supposed to be independent.
In summary:
\begin{equation}
S(y+dy,(x_0,x_1))=\begin{cases}
S(y,(x_0,x_1))\\
\ \ \ \ \ \ \ \ \text{with proba}\ 1-\bar\alpha dy
\int_{x_2} \frac{dP}{d(\bar\alpha y)}(x_{01}\rightarrow x_{02},x_{12})\\
S(y,(x_0,x_2))\times S(y,(x_2,x_1))\\
\ \ \ \ \ \ \ \ \text{with proba}\
\bar\alpha dy \frac{dP}{d(\bar\alpha y)}(x_{01}\rightarrow x_{02},x_{12})
\end{cases}
\label{eq:bk01}
\end{equation}
Taking the average over the realizations of the
target and the limit $dy\rightarrow 0$, we get
\begin{multline}
\frac{\partial}{\partial y}\langle S(y,(x_0,x_1))\rangle=\bar\alpha
\int \frac{d^2 x_2}{2\pi}\frac{x_{01}^2}{x_{02}^2 x_{21}^2}
[
\langle S(y,(x_0,x_2))S(y,(x_2,x_1)) \rangle\\
 -\langle S(y,(x_0,x_1))\rangle
]
\label{eq:bk02}
\end{multline}
(See Fig.~\ref{fig:BK2} for a graphical representation.)
We see that this equation is not closed:
An evolution equation for the correlator 
$\langle S(y,(x_0,x_2))S(y,(x_2,x_1))\rangle$
is required.
However, we may assume that these correlators factorize
\begin{equation}
\langle S(y,(x_0,x_2))S(y,(x_2,x_1))\rangle =
\langle S(y,(x_0,x_2))\rangle
\langle S(y,(x_2,x_1))\rangle.
\label{eq:factocorr}
\end{equation}
This assumption is justified if the dipoles scatter off independent targets,
for example, off the nucleons of a very large nucleus.
Writing $A=1-\langle S\rangle$, we get the following closed equation for $A$:
\begin{multline}
\frac{\partial}{\partial y} A(y,(x_0,x_1))=\bar\alpha
\int \frac{d^2 x_2}{2\pi}\frac{x_{01}^2}{x_{02}^2 x_{21}^2}
[
A(y,(x_0,x_2))+A(y,(x_2,x_1))\\
-A(y,(x_0,x_1))-A(y,(x_0,x_2))A(y,(x_2,x_1))
],
\label{eq:BKA}
\end{multline}
which is the Balitsky-Kovchegov (BK) equation \cite{Kovchegov:1999yj,Kovchegov:1999ua}.
Note that if one neglects the nonlinear term, 
one gets back the BFKL equation~(\ref{eq:BFKL}) (written for $A$ instead of $n$).
A graphical representation of this equation is given in Fig.~\ref{fig:bk}.

It is not difficult to see analytically that the BK equation preserves the 
unitarity of $A$: When $A$ becomes of the order of 1, then the nonlinear term
gets comparable to the linear terms in magnitude, and slows down
the evolution of $A$ with $y$, which otherwise would be exponential.

\begin{figure}
\begin{center}
\begin{minipage}[c]{0.4\textwidth}
\centerline{\epsfig{file=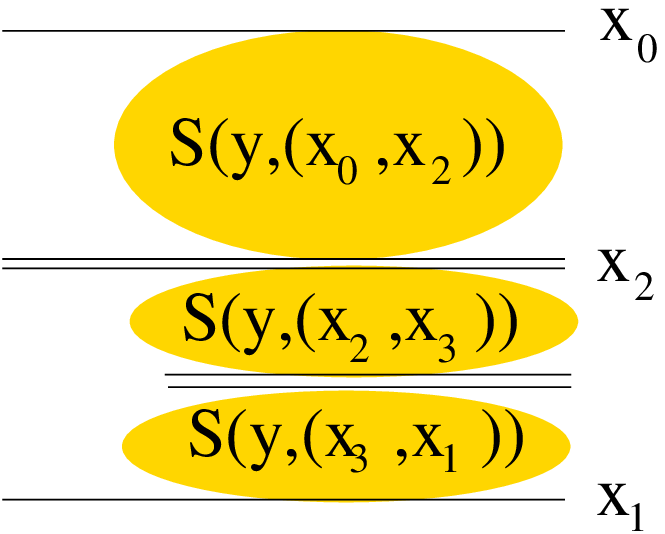,width=0.95\textwidth}}
\centerline{(a)}
\end{minipage}
\hskip 0.5cm
\begin{minipage}[c]{0.4\textwidth}
\centerline{\epsfig{file=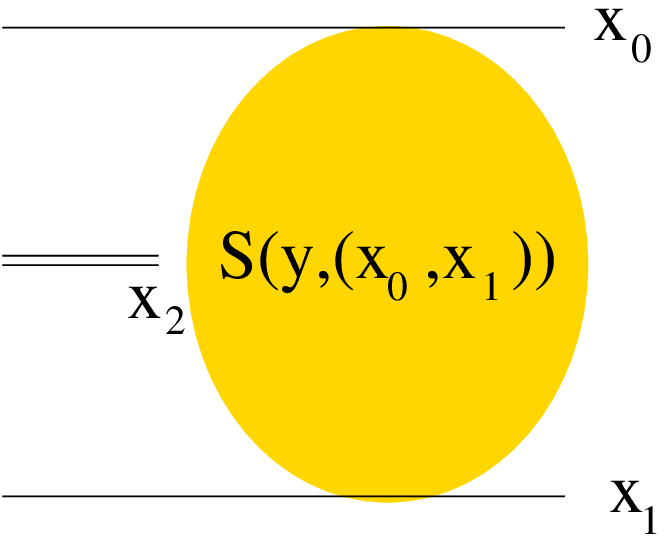,width=0.95\textwidth}}
\centerline{(b)}
\end{minipage}
\end{center}
\caption{\label{fig:JIMWLK2}(a) Contribution to the 
B-JIMWLK equation for the 2-point correlator 
restricted to dipoles ($x_2^\prime$
is taken equal to $x_2$ in this figure). (b) A graph that would
also contribute to the 2-point correlator and
that is missing in the B-JIMWLK formalism.}
\end{figure}

Let us go back to Eqs.~(\ref{eq:bk01}),(\ref{eq:bk02}) and instead of assuming the
factorization of the correlators~(\ref{eq:factocorr}), work out
an equation for the two-point correlator $\langle SS\rangle$. From
the same calculation as before, we get
\begin{multline}
\frac{\partial}{\partial y}\langle S_{02}S_{2^\prime 1}\rangle=
\bar\alpha\int \frac{d^2x_3}{2\pi}\frac{x_{02}^2}{x_{03}^2x_{32}^2}
\left(\langle S_{03}S_{32}S_{2^\prime 1}\rangle-\langle S_{02}S_{2^\prime 1}\rangle\right)\\
+\bar\alpha\int \frac{d^2x_3}{2\pi}\frac{x_{12^\prime}^2}{x_{13}^2x_{32^\prime}^2}
\left(\langle S_{2^\prime3}S_{31}S_{02}\rangle-\langle S_{02}S_{2^\prime 1}\rangle\right),
\label{eq:balitskydipoles}
\end{multline}
where we have introduced the notation $S_{ab}\equiv S(y,(x_a,x_b))$.
(See Fig.~\ref{fig:JIMWLK2}a for the corresponding graphical representation.)

This equation calls for a new equation for the 3-point correlators, and so on.
The obtained hierarchy is nothing but the Balitsky hierarchy \cite{Balitsky:1995ub}
(see also Ref.~\cite{Balitsky:1998kc,Balitsky:1998ya})
restricted to dipoles.
We refer the reader to~\cite{Janik:2004ts,Janik:2004ve} (see also Ref.~\cite{Levin:2004yd})
for the detailed relationship of
this equation to the B-JIMWLK formalism.
Note in particular that the factorized correlators~(\ref{eq:factocorr}) is a 
solution of the whole hierarchy, and is actually a good approximation
to the solution of the full B-JIMWLK equations.
This statement was first made after the results of the numerical solution 
to the JIMWLK equation worked out
in Ref.~\cite{Rummukainen:2003ns}.

We may wonder why there are no terms involving one-point functions 
in the right handside of the previous
equation. Actually, such terms would correspond to graphs like the one 
of Fig.~\ref{fig:JIMWLK2}b, in which, for example, two dipoles merge.
They are expected to occur if saturation is properly taken into account.
While the restriction of the Balitsky equation to dipoles does not
drastically change the solution for the scattering amplitudes,
such terms would instead have a large effect, as we shall discover in 
the following. 
In the next section, we will explain why such terms are actually required
for physical reasons.

\subsubsection{Saturation}

The BK equation may be well-suited for the ideal case
in which the target is a nucleus made of an infinity of independent nucleons.
But it is not quite relevant to describe the scattering
of more elementary objects such as two dipoles (or two virtual photons,
to be more physical).

\begin{figure}
\begin{center}
\begin{tabular}{cc}
\epsfig{file=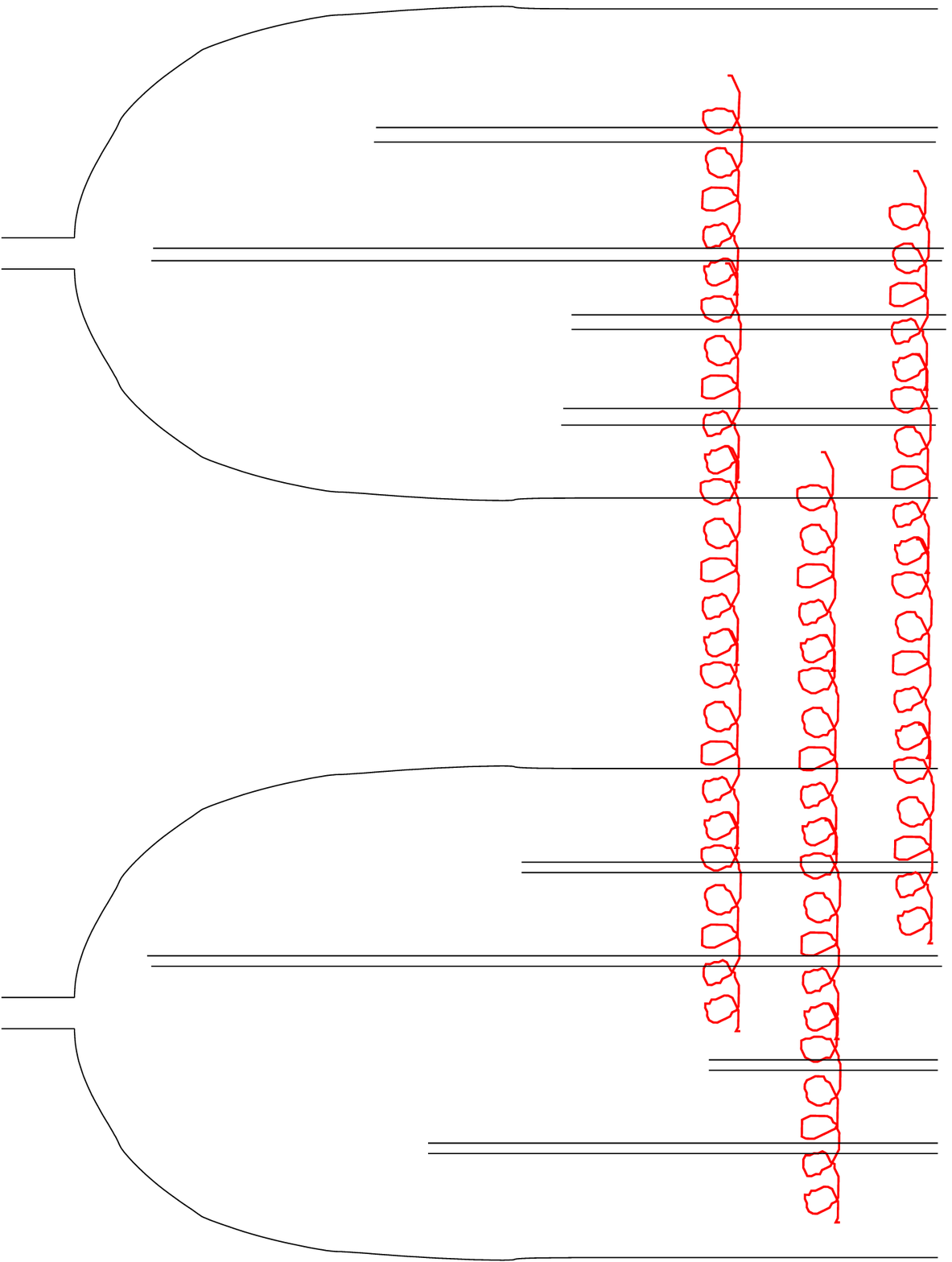,width=0.4\textwidth}
&
\hskip 0.1\textwidth
\epsfig{file=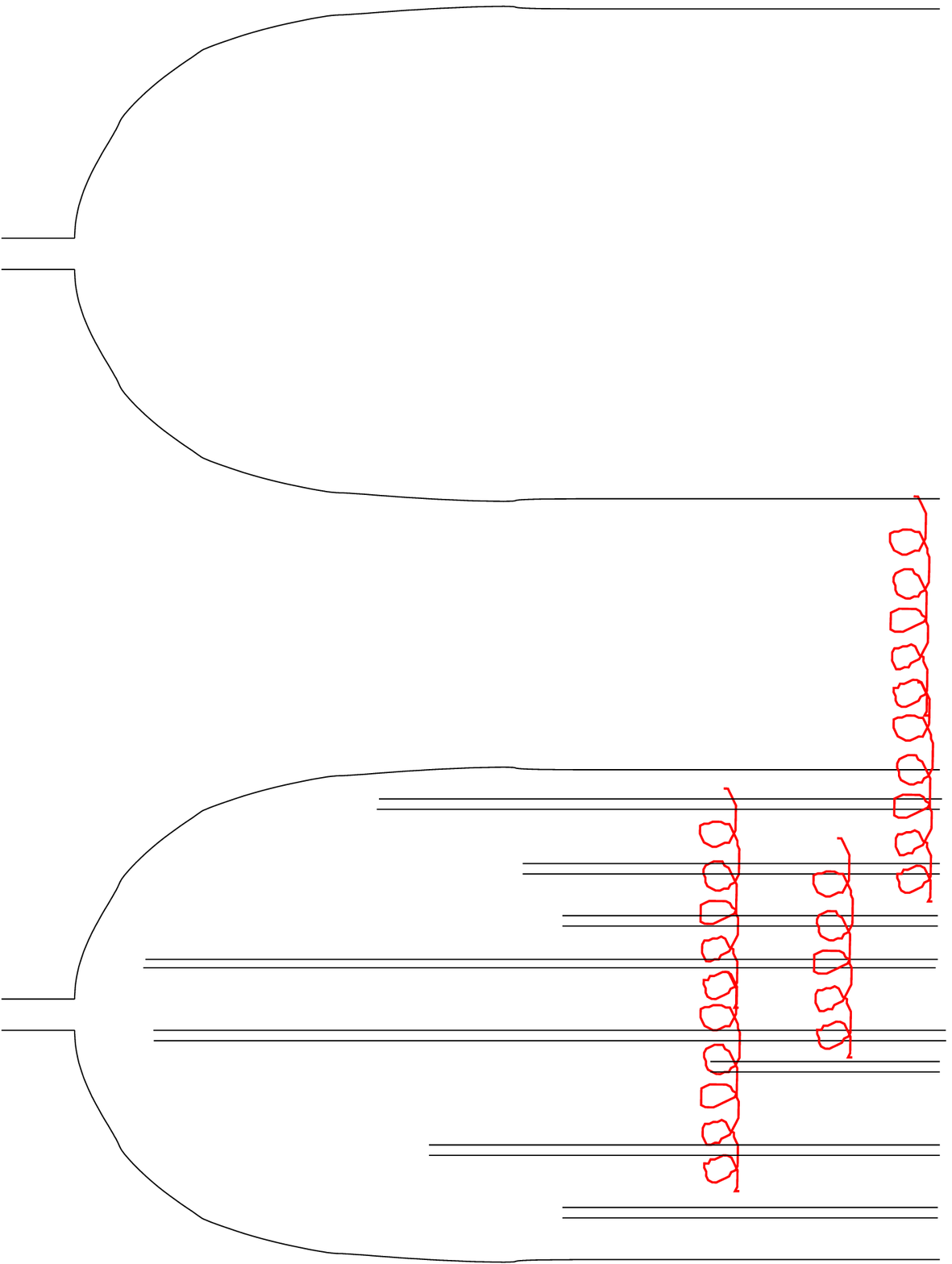,width=0.4\textwidth}\\
(a) &
\hskip 0.1\textwidth
(b)
\end{tabular}
\end{center}
\caption{\label{fig:com}
(a) Scattering in the dipole model in the center-of-mass frame.
The evolution is shared between the target and the probe.
The amplitude is unitarized through the multiple scatterings
occuring between the two evolved wave functions.
(b) Boost of the previous graph to the restframe of the projectile.
There is now twice as much evolution in the target
and the nonlinear effects should occur inside its wavefunction, in the course
of the evolution.
They may take the form of ``internal'' rescatterings (as depicted), or
dipole merging...
}
\end{figure}

Indeed, 
following Chen and Mueller~\cite{Chen:1995pa} (see also Ref.~\cite{Kovchegov:1997dm}),
let us consider dipole-dipole scattering in the center-of-mass frame,
where the rapidity evolution is equally shared between the projectile and the target
(see Fig.~\ref{fig:com}a).
Then at the time of the interaction, the targets are dipoles that stem from the branching
of a unique primordial dipole. 
Obviously, the assumption of statistical independence of the targets,
which was needed for the factorization~(\ref{eq:factocorr}) to hold, is no longer
justified.

So far, we have seen that nonlinear effects which go beyond the factorization
formula~(\ref{eq:BFKLframe}) are necessary to preserve unitarity
as soon as $n\sim 1/\alpha_s^2$. 
This came out of an analysis of Eq.~(\ref{eq:BFKLframe})
in the restframe of the target. The rapidity $y_\text{BFKL}$ at which the system
reaches this number of dipoles and hence at which the BFKL approach breaks
down may be found from the form
of the typical growth of $n$ with $y$, namely $n(y)\sim e^{\bar\alpha y}$.
Parametrically,
\be
y_\text{BFKL}\sim\frac{1}{\bar\alpha}\log\frac{1}{\alpha_s^2}.
\ee
Now we may go to the center-of-mass frame, where Eq.~(\ref{eq:BFKLframe})
with $y^\prime=y/2$ would describe the amplitude in the absence of
nonlinear effects.
There, the typical number of dipoles in the projectile and in the
target are well below $1/\alpha_s^2$:
$n(y_\text{BFKL}/2)\sim 1/\alpha_s$.
We actually see that the evolution of the dipoles 
in each of theses systems
remains linear until $y=2 y_\text{BFKL}$.
In that rapidity interval, nonlinear effects consist in the
simultaneous scatterings of several dipoles from the
target and the projectile but the evolution of $n$ still obeys
the BFKL equation (see Fig.~\ref{fig:com}a).
Now, performing a boost to the projectile restframe, 
the evolution goes into the target. 
Formula~(\ref{eq:BFKLprojectile}) should then apply for the amplitude $A$.
But if the evolution of the target were kept linear,
then the amplitude would not be 
unitarity because the number of dipoles would
be larger than $1/\alpha_s^2$.
Hence, through some nonlinear mechanism, 
which was represented by multiple scatterings between linearly evolving
objets in the center-of-mass frame,
the dipole number density 
has to be kept effectively lower than $1/\alpha_s^2$ in order to preserve 
unitarity (see Fig.~\ref{fig:com}b).
This is called {\em parton saturation}.
The precise saturation mechanism has not been formulated in QCD.
It could be dipole recombinations due to gluon fusion, 
multiple scatterings inside
the target which slow down the production of new dipoles \cite{Mueller:1996te} 
(as in Fig.~\ref{fig:com}b), 
``dipole swing''
as was proposed more recently \cite{Avsar:2005iz,Avsar:2006jy}, or any
other mechanism.

Hence, unitarity of the scattering amplitudes together
with boost-invariance seem to require the saturation of the
density of partons.

A pedagogical review of saturation 
and the discussion of the relationship between saturation and unitarity
may be found in Ref.~\cite{Mueller:2001fv}.
Original papers include Refs.~\cite{Mueller:1989st,Mueller:1999wm}.

\subsubsection{The Pomeron language}

So far, we have presented in detail a $s$-channel picture of hadronic interactions, 
and it is in this
formalism that we will understand most easily the link with reaction-diffusion
processes. In the $s$-channel formulation, 
all the QCD evolution happens in the form
of quantum fluctuations of the interacting hadrons. However, a picture
maybe more familiar to the reader is a $t$-channel picture,
where the rapidity evolution is put in the $t$-channel, while the projectile
and target are in their bare states.
This picture directly stems from the usual Lorentz-invariant
formulation of quantum field theory, while the dipole picture
(or the parton model) is derived in the framework of time-ordered perturbation
theory.

\begin{figure}
\begin{center}
\epsfig{file=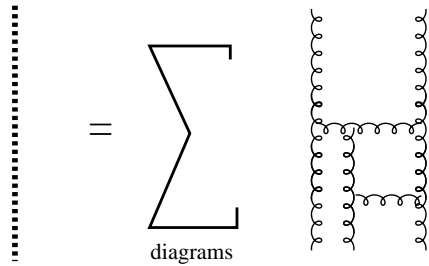,width=0.4\textwidth}
\end{center}
\caption{\label{fig:pomeron}The BFKL Pomeron is a sum of $t$-channel
gluon Feynman diagrams.}
\end{figure}

Classes of Feynam diagrams can be grouped into ``Pomerons'' 
(or Reggeized gluons, see Fig.~\ref{fig:pomeron}),
in terms of which scattering processes may be analyzed. (A pedagogical
review on how to derive the BFKL equation in such a formalism is available
from Ref.~\cite{Forshaw:1997dc}).

\begin{figure}
\begin{center}
   \begin{minipage}[c]{.45\textwidth}
\begin{center}
\epsfig{file=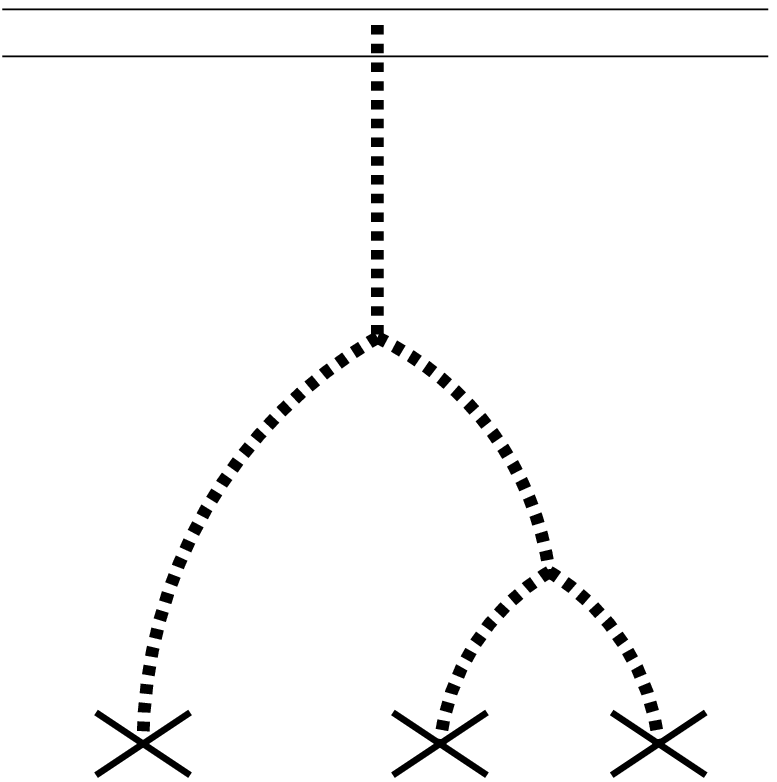,width=0.5\textwidth}
\end{center}
   \end{minipage}
   \begin{minipage}[c]{.45\textwidth}
\begin{center}
\epsfig{file=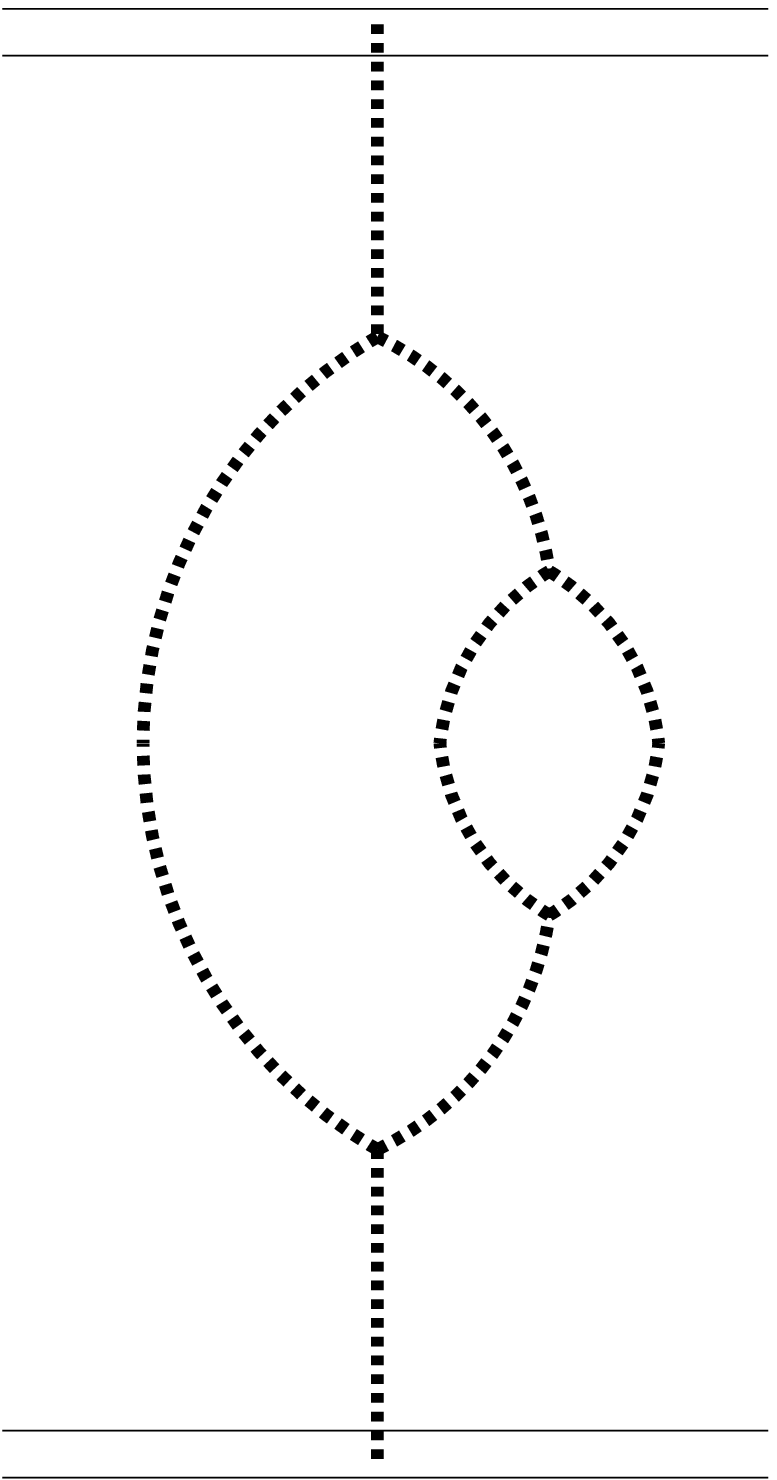,width=0.5\textwidth}
\end{center}
   \end{minipage}
\vskip 0.4cm
   \begin{minipage}[b]{.45\textwidth}
\begin{center}
(a)
\end{center}
   \end{minipage}
   \begin{minipage}[b]{.45\textwidth}
\begin{center}
(b)
\end{center}
   \end{minipage}
\end{center}
\caption{\label{fig:bkmppomeron}(a) Example of a diagram contributing to the BK
equation in the $t$-channel representation (see Fig.~\ref{fig:bk}).
The dashed lines represent Pomerons. The rapidity is proportional to the length
of the Pomeron lines in the $t$-channel.
(b) Pomeron representation of
a class of diagrams to which Fig.~\ref{fig:com}a belongs.
}
\end{figure}

An effective action containing Pomeron 
fields and vertices may be constructed.
In these terms, the $s$-channel diagrams of Figs.~\ref{fig:bk},\ref{fig:com}a may
be translated in terms of the diagrams of Fig.~\ref{fig:bkmppomeron}.
The effective action formalism was initially developped in 
Refs.~\cite{Kirschner:1994gd,Kirschner:1994xi,Lipatov:1995pn}.
More recently, there has been some progress 
in the definition of the effective action
\cite{Antonov:2004hh},
some of it with the help of the correspondence with
statistical physics processes \cite{Blaizot:2005vf,Hatta:2005rn}.

We will not expand on this formulation in the present review,
because it is difficult to see the analogy with statistical
processes in this framework. A $s$-channel picture is much more natural.
However, a full solution of high-energy QCD may require to go back to that kind
of calculation and compute accurately the $1\rightarrow n$ Pomeron vertices.
This program was formulated some time ago 
\cite{Bartels:1992ym,Bartels:1993ih}, and there is continuing
progress in this direction (see e.g. \cite{Bartels:1999aw,Ewerz:2003an}).


\subsection{Analogy with reaction-diffusion processes}

We are now in position to draw the relationship between high-energy QCD
and reaction-diffusion processes.
In the first section below, we will show that the BK equation
is, in some limit, an equation that also appears in the context
of statistical physics.
Second, we will exhibit a particular reaction-diffusion model,
and show in the final section how this model
is related in a more general way to scattering in QCD.

\subsubsection{The BK equation and the FKPP equation}

Let us first show at the technical level 
that under some well-controlled approximations, the BK equation~(\ref{eq:BKA})
may be mapped exactly to a parabolic nonlinear partial differential equation.
This observation was first made in Ref.~\cite{Munier:2003vc}.

To simplify, we will look for impact-parameter independent solutions:
$A(y,(x_0,x_1))$ is supposed to depend on $y$ and $x_{01}$ only,
not on ${x_0+x_1}$.
We switch to momentum space through the Fourier transformation
\begin{equation}
A(y,k)=\int \frac{d^2 x_{01}}{2\pi x_{01}^2}e^{i kx_{01}} A(y,x_{01}).
\end{equation}
This transformation greatly simplifies the BK equation 
\cite{Kovchegov:1999yj,Kovchegov:1999ua}.
It now reads
\begin{equation}
\partial_{\bar\alpha y} A(y,k)=\chi(-\partial_{\log k^2})A(y,k)-A^2(y,k).
\end{equation}
The first term in the right handside, which is a linear term, is actually an 
integral kernel, obtained by Fourier transformation of the BFKL kernel
(first three terms in the right handside of Eq.~(\ref{eq:BKA})).
It is most easily expressed in Mellin space: $k^{-2\gamma}$ is the set
of its eigenfunctions, with the corresponding eigenvalues
\begin{equation}
\chi(\gamma)=2\psi(1)-\psi(\gamma)-\psi(1-\gamma).
\end{equation}
This kernel may be expanded around some real $\gamma=\gamma_0$, fixed 
between 0 and 1. Keeping the terms up to $\mathcal{O}((\gamma-\gamma_0)^2)$ is the
well-known diffusive approximation, which is a good approximation for
large rapidities.
Introducing the notations $\chi_0=\chi(\gamma_0)$, $\chi^\prime_0=\chi^\prime(\gamma_0)$
and $\chi^{\prime\prime}_0=\chi^{\prime\prime}(\gamma_0)$,
the BK equation reads, within this approximation
\be
\partial_{\bar\alpha y}A
={\scriptstyle\frac{\chi_0^{\prime\prime}}{2}}\partial_{\log k^2}^2 A
+(\gamma_0 \chi_0^{\prime\prime}-\chi^\prime_0)\partial_{\log k^2} A
+(\chi_0-\gamma_0 \chi_0^{\prime}+{\scriptstyle\frac{\gamma_0^2\chi_0^{\prime\prime}}{2}})A
-A^2.
\ee
Through some linear change of variable $(\bar\alpha y,\log k^2)\rightarrow (t,x)$,
\be
\begin{split}
\bar\alpha y&= 
\frac{t}{\chi_0-\gamma_0 \chi_0^{\prime}+{\scriptstyle\frac{\gamma_0^2\chi_0^{\prime\prime}}{2}}}\\
\log k^2    &=
\sqrt{\frac{\chi_0^{\prime\prime}}{2(\chi_0-\gamma_0 \chi_0^{\prime})+{{\gamma_0^2\chi_0^{\prime\prime}}}}}x
+\frac{\gamma_0 \chi_0^{\prime\prime}-\chi^\prime_0}
{\chi_0-\gamma_0 \chi_0^{\prime}+{\scriptstyle\frac{\gamma_0^2\chi_0^{\prime\prime}}{2}}}t,
\end{split}
\ee
one may get rid of the first-order partial derivative in the right handside.
We then find that the new function
\begin{equation}
u(t,x)=\frac{A(y(t),\log k^2(t,x))}
{\chi_0-\gamma_0 \chi_0^{\prime}+{\scriptstyle\frac{\gamma_0^2\chi_0^{\prime\prime}}{2}}}
\end{equation}
obeys the equation
\begin{equation}
\frac{\partial u(t,x)}{\partial t}=
\frac{\partial^2 u(t,x)}{\partial x^2}
+u(t,x)-u^2(t,x),
\label{eq:BFKLtoFKPP}
\end{equation}
which is the Fisher \cite{Fisher} and Kolmogorov-Petrovsky-Piscounov \cite{KPP} 
(FKPP) equation.
This equation was first written down as a model for
gene propagation in a population in the large population size limit.
But it turns out to apply directly or indirectly
to many different physical situations, such as
reaction-diffusion processes, but also directed percolation, 
and even mean-field spin glasses \cite{DerridaSpohn}.
A recent comprehensive review on the known mathematics and the
phenomenological implications of the FKPP equation can be found
in Ref.~\cite{vansaarloos-2003-386}.

As a side remark, we note that if $\gamma_0$ is chosen such as
$\chi(\gamma_0)=\gamma_0\chi^\prime(\gamma_0)$, then the mapping
drastically simplifies. Actually, this choice has a physical
meaning, as we will discover in Sec.~\ref{sec:reviewtraveling}
when we try and solve the BK equation.

Beyond the exact mapping~(\ref{eq:BFKLtoFKPP}) 
between an approximate form
of the BK and the FKPP equations,
the full BK equation is said to be in the {\em universality
class} of the FKPP equation. All equations in this universality
class share some common properties, as will be understood below.
The exact form of the equation is unessential.
As a matter of fact, recently, 
it has been checked explicitely that the BFKL equation with next-to-leading
order contributions to the linear evolution kernel 
(but keeping the QCD coupling fixed)
is also in the same universality
class. A mapping to a partial differential equation (which involves
higher-order derivatives in the rapidity variable) was
exhibited \cite{Enberg:2006aq}.
What defines physically the universality class of the FKPP equation is
a branching diffusion process with some saturation mechanism. The details
seem unimportant.

In the next section, we shall give a concrete 
example of a reaction-diffusion process:
We will see how the FKPP equation appears as a fluctuationless 
(or ``mean-field'') limit of some stochastic reaction-diffusion process.
In Ref.~\cite{Munier:2003vc}, it had not been realized that 
the analogy of QCD with such processes is in fact much
deeper than the formal mapping between the BK equation and the FKPP equation
that we have just outlined.
But this is actually the case, as we shall shortly argue.

\subsubsection{\label{sec:reactiondiffusionexample}
Reaction-diffusion processes: An example}

We consider a reaction-diffusion model, which was
introduced in Ref.~\cite{Enberg:2005cb}.
Particles are evolving in discrete time 
on a one-dimensional lattice. At each timestep, a particle may jump
to the nearest position on the left or on the right with respective
probabilities $p_l$ and $p_r$, and may split into two particles with
probability $\lambda$. We also allow that each of the $n(t,x)$ particles
on site $x$ at time $t$ to die with probability
$\lambda n(t,x)/N$.

\begin{figure}
\begin{center}
\epsfig{file=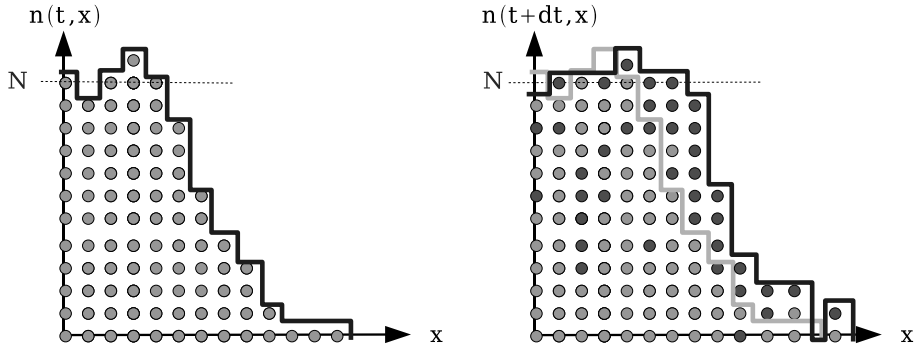,width=0.9\textwidth}
\end{center}
\caption{\label{fig:model1d}
Picture of a realization of the system of particles at two successive times.
In the bins in which the number of particles is of order $N$, 
some particles disappear, others are created by splittings, but overall the
number of particles is conserved up to
fluctuations of order $\sqrt{N}$.
In the bins in which $n$ is small compared to $N$, the dynamics
is driven by branching diffusion. As a result, $n(t,x)$ looks like a noisy
wave front moving to the right.
}
\end{figure}

From these rules, we may guess what a realization of this evolution may
look like at large times.
The particles branch and diffuse (they undergo a linear evolution)
until their number $n$ becomes of the order of $N$, at which point
the probability that they ``die'' starts to be sizable,
in such a way that their number never 
exceeds $N$ by a large amount, on any site.
But if the initial condition is spread on a finite number of lattice sites,
the linear branching-diffusion process may always proceed towards 
larger values of $|x|$, where there were no particles in the beginning of
the evolution.
Hence a realization will look like a front connecting an ensemble
of lattice sites where a quasi-stationary
state in which the number of particles is $N$ (up to fluctuations)
has been reached,
to an ensemble of empty sites (towards $|x|\rightarrow\infty$). 
This front moves with time as the branching diffusion process proceeds.
The position of the front $X(t)$ may be defined in different ways,
leading asymptotically to equivalent determinations, up to a constant.
For example, one may define $X(t)$ 
as the rightmost bin in which there are more 
than $N/2$ particles, or, alternatively, as the total number of particles
in the realization whose positions are greater than 0, scaled by $1/N$.
A realization and its time evolution is sketched in Fig.~\ref{fig:model1d}.

Between times $t$ and $t+\Delta t$, $n_l(t,x)$ particles 
out of $n(t,x)$ move to the left and
$n_r(t,x)$ of them move to the right. 
Furthermore, $n_+(t,x)$ particles are replaced by their
two offspring at $x$, and $n_-(t,x)$ particles disappear.
Hence the total variation in the number of particles on site $x$ reads
\begin{subequations}\label{stocha}
\begin{multline}
n(t+\Delta t,x)-n(t,x)
=-n_l(t,x)-n_r(t,x)-n_-(t,x)\\
+n_+(t,x)+n_l(t,x+\Delta x)+n_r(t,x-\Delta x).
\end{multline}
The numbers describing a timestep at position $x$ have 
a multinomial distribution:
\begin{multline}
P(\{n_l,n_r,n_+,n_-\})=
\frac{n!}{n_l!n_r!n_+!n_-!\Delta n!}
p_l^{n_l}p_r^{n_r}\\
\times\lambda^{n_+}(\lambda n/N)^{n_-}
(1\!-\!p_l\!-\!p_r\!-\!\lambda \!-\!\lambda n/N)^{\Delta n},
\end{multline}
\end{subequations}
where $\Delta n=n-n_l-n_r-n_+-n_-$, and all quantities 
in the previous equation
are understood at site $x$ and time $t$.
The evolution of $u\equiv n/N$ is obviously stochastic. One could
write the following equation:
\be
u(t+\Delta t,x)=\langle u(t\!+\!\Delta t,x)\rangle
+\sqrt{\langle u^2(t\!+\!\Delta t,x)\rangle
-\langle u(t\!+\!\Delta t,x)\rangle^2}
\,\nu(t+\Delta t,x)
\label{eq:stochamodel}
\ee
where the averages are understood over the time step that takes
the system from $t$ to $t+\Delta t$. They are conditioned to
the value of $u$ at time $t$.
$\nu$ is a noise, i.e. a random function.
The equation was written in such a way that it has zero mean and
unit variance.
Note that the noise is updated at time $t+\Delta t$ in this equation.

One can compute the mean evolution of $u\equiv n/N$ in one step of time
which appears in the right handside of Eq.~(\ref{eq:stochamodel})
from Eq.~(\ref{stocha}).
It reads
\begin{multline}
\langle u(t\!+\!\Delta t,x)|\{u(t,x)\}\rangle\!=\!u(t,x)\!+\!
p_l[u(t,x\!+\!\Delta x)\!-\!u(t,x)]\\
+\!p_r[u(t,x\!-\!\Delta x)\!-\!u(t,x)]\!+\!\lambda u(t,x)[1\!-\!u(t,x)].
\label{mod_mf}
\end{multline}
The mean evolution of the variance of $u$ that appears in 
Eq.~(\ref{eq:stochamodel}) may also be computed. The precise form of the
result is more complicated, but roughly speaking, the variance 
of $u$ after evolution over a unit of time is
of the order of $u/N$ for small $u\sim 1/N$. 
This is related to the fact
that the noise has a statistical origin: Having
$n$ particles on the average
in a system means that each realization typically
consists in $n\pm\sqrt{n}$ particles.

When $N$ is infinitely large, one can replace the $u$'s in Eq.~(\ref{mod_mf})
by their averages: This would be a mean field approximation.
Obviously, the noise term drops out, and the equation becomes
deterministic.
Note that if we appropriately take
the limits $\Delta x\rightarrow 0$ and $\Delta t\rightarrow 0$,
setting
\begin{equation}
\lambda=\Delta t,\ \ p_R=p_L=\frac{\Delta t}{(\Delta x)^2},
\end{equation}
the obtained mean-field equation is nothing but the
FKPP equation~(\ref{eq:BFKLtoFKPP}).
For the numerical simulations of this model that we will perform in 
Sec.~\ref{sec:reviewtraveling},
we will keep $\Delta t$ and $\Delta x$ finite,
which is usually more convenient for computer implementation.

Thus we have seen that the evolution of
reaction-diffusion systems
is governed by a stochastic equation~(\ref{eq:stochamodel}) 
whose continuous limit ($\Delta t\rightarrow 0,\ \Delta x\rightarrow 0$) and
mean-field limit ($N\gg 1$) is a partial differential equation
of the form of (exactly actually, in our simple case study) the FKPP equation.
We shall now argue that partons in high-energy QCD
form such a system.

\subsubsection{Universality class of high-energy QCD}

Let us come back to the QCD dipole model.
We have seen that rapidity 
evolution of the hadron wavefunctions
proceeds through a branching diffusion
process of dipoles. Let us denote by $T(y,r)$ the scattering amplitude
of the probe dipole off one particular realization of the target at rapidity $y$
and at a given impact parameter.
This means that we imagine for a while that we may freeze the target
in one particular realization after the rapidity evolution $y$, 
and probe the latter with
projectiles of all possible sizes.
Of course, this is not doable in an actual experiment, not even in principle.
But it is very important for the statistical picture to go through such
a ``gedanken observable''.
The amplitude $A$, which is related to the measurable total cross section,
is nothing but the average of $T$ over all possible
realizations of the fluctuations of the target, namely
\begin{equation}
A(y,r)=\langle T(y,r)\rangle.
\end{equation}

\begin{figure}
\begin{center}
\epsfig{file=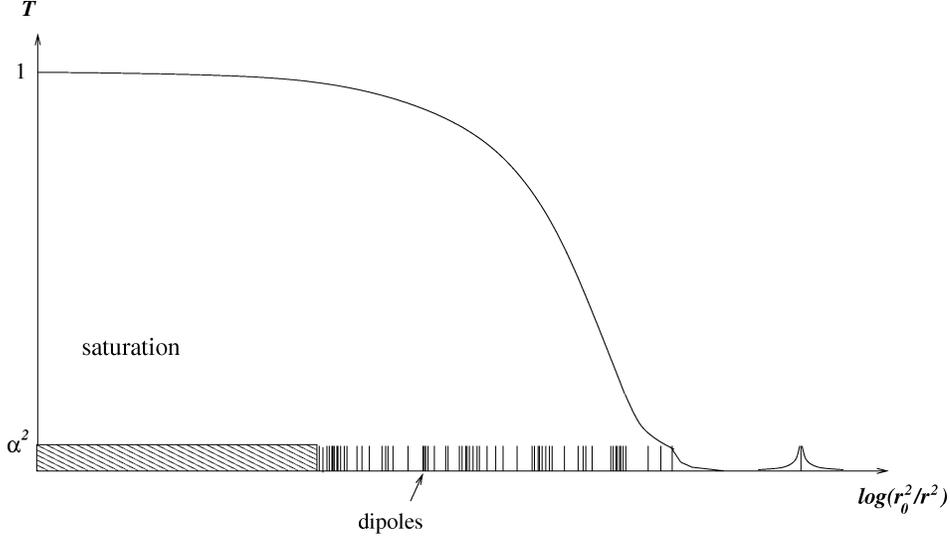,width=0.9\textwidth}
\end{center}
\caption{\label{fig:T}Sketch of the scattering amplitude $T$ 
of a dipole of size $r$ off a frozen
partonic configuration. The small lines on the axis denote the
dipoles ordered by their logarithmic sizes.
Up to fluctuations, $T$ looks like a wave front.
}
\end{figure}

The branching diffusion of the dipoles essentially occurs in the $\log(1/r^2)$ variable.
The scattering amplitude is roughly equal to the number of dipoles
in a given bin of (logarithmic) dipole size, multiplied by $\alpha_s^2$.
From unitarity arguments and consistency with boost-invariance, 
we have seen that the branching diffusion process
should (at least) slow down in a given bin as soon as the number of
objects in that very bin is of the order of $N=1/\alpha_s^2$, in such
a way that effectively, the number of dipoles in each bin is limited to $N$.
A typical realization of $T$ is sketched in Fig.~\ref{fig:T}.
As in the case of the reaction-diffusion process, 
from similar arguments,
it necessarily looks like a front.
The position of the front, defined to be the value $r_s$ of $r$ for which
$T$ is equal to some fixed number, say $\frac12$,
is related to the saturation scale defined
in the Introduction: $r_s=1/Q_s(y)$.

We now see that there is a very close analogy between what we are describing
for QCD here and the model that we were introducing in the previous section.
So in particular, one might be able to formulate interaction processes in QCD 
with the help of a stochastic
nonlinear evolution equation for the ``gedanken'' amplitude $T$.
We already know the mean-field limit that one should get, when $N$ is very large:
This is the BK equation, as was rigorously proven above.
Thus we know the equivalent of the term $\langle u(t+\Delta t,x)\rangle$
in Eq.~(\ref{eq:stochamodel}).
The noise term is not known, but since it is of statistical origin, we know that
it must be of the order of the square root of the number of dipoles normalized to $N$, 
that is to say, of order $\sqrt{T/N}$.
We may write an equation of the form
\be
\partial_{\bar\alpha y} T(y,k)=\chi(-\partial_{\log k^2})T(y,k)
-T^2(y,k)+\alpha_s\sqrt{2T(y,k)}\,\nu(y,k),
\label{eq:stochaqcd}
\ee
where $\nu$ is a noise, uncorrelated in rapidity and transverse momentum,
with zero mean and unit variance. (The factor of 2 under the square
root is essentially arbitrary).
This equation is to be compared to the following one:
\be
\partial_t u(t,x)=\partial_x^2 u(t,x)+u(t,x)-u^2(t,x)
+\sqrt{\frac{2u(t,x)}{N}}\,\nu(t,x),
\label{eq:RFT}
\ee
which is the so-called ``Reggeon field theory'' equation
when the noise $\nu$ is exactly a normal Gaussian white noise, that is
to say, of zero mean and whose non-vanishing cumulant reads
\be
\langle \nu(t,x)\nu(t^\prime,x^\prime)\rangle=
\delta(t-t^\prime)\delta(x-x^\prime).
\ee
It is a stochastic extension of Eq.~(\ref{eq:BFKLtoFKPP}).
If the noise term were of the form 
\be
\sqrt{\frac{2u(t,x)(1-u(t,x))}{N}}\,\nu(t,x)
\ee 
instead,
then this equation would be what is usually referred to as the
{\em stochastic Fisher-Kolmogorov-Petrovsky-Piscounov equation}.
The sFKPP equation and the physics that it represents is
reviewed in Ref.~\cite{panja-2004-393}.

Taking averages over events converts this equation into
a hierarchy of coupled equations, which has a lot in common in its structure
with the Balitsky hierarchy~(\ref{eq:balitskydipoles}).
(Actually, there are some extra
terms compared to the Balitsky hierarchy, which
were first found from the analogy with reaction-diffusion processes,
and which precisely represent nonlinear effects inside the wavefunctions.
A detailed study may be found in Ref.~\cite{Iancu:2004iy}).
We will perform explicit calculations in this
spirit within simpler models in Sec.~\ref{sec:zerodimensional} below.

Based on these considerations, 
we may establish a dictionary between QCD and reaction-diffusion processes.
The correspondence is summarized in Tab.~\ref{tab:dictionary}.

\begin{table}
\begin{center}
\begin{tabular}{c  c}
\hline\hline
     Reaction-diffusion & QCD \\
\hline
     Occupation fraction $u(t,x)$  & 
\begin{minipage}[t]{7cm}
Scattering amplitude for the probe
off a frozen realization of the target $T(k,y)$
\end{minipage}
\\
     Average occupation fraction $\langle u(t,x)\rangle$ 
& Physical scattering amplitude $A=\langle T\rangle$\\
     Space variable $x$ & $\log(k^2/\Lambda^2)$ or $\log(1/r^2\Lambda^2)$\\
     Time variable $t$   & Rapidity $\bar\alpha y$\\
     Average maximum density of particles $N$ & $1/\alpha_s^2$ \\
     Position of the front $X(t)$ &
Saturation scale $\log(Q_s^2(y)/\Lambda^2)$\\
\begin{minipage}[t]{7cm}
Branching-diffusion kernel $\omega(-\partial_x)$\\
($\omega(-\partial_x)=\partial_x^2+1$ in the
FKPP case)
\end{minipage}&
\begin{minipage}[t]{7cm}
BFKL kernel $\chi(-\partial_{\log k^2})$\\
or its equivalent in coordinate space
\end{minipage}\\
\hline\hline\\
\end{tabular}
\end{center}
\caption{\label{tab:dictionary}Dictionary between QCD and the reaction-diffusion
model for the main physical quantities.
$\Lambda$ is a typical hadronic scale.}
\end{table}

The mechanism for saturation of the parton densities 
(i.e. of the dipole number density) is
not known for sure in QCD. There are also important differences between
the reaction-diffusion model introduced above and QCD
that lie in the ``counting rule'' of the particles (provided by
the form of $T^\text{el}$ in the QCD case, see Eq.~\ref{eq:Tdipole}).
But from the general analysis of processes described by equations in the
universality class of the stochastic FKPP equation and the underlying evolution
mechanisms presented in Sec.~\ref{sec:reviewtraveling}, 
we will understand that most of the observables have universal properties in 
appropriate limits,
which do not depend on the details of the mechanism at work.
We draw the reader's attention to Refs.~\cite{Mueller:2005ut,Iancu:2005nj}, 
where a precise stochastic
equation was searched for in QCD. Some of the problems one may face
with the use and the very interpretation of
such equations were studied in Ref.~\cite{Iancu:2005dx}.

The way in which we view high energy QCD
is actually not particularly original: It is nothing but the QCD dipole model,
which was implemented numerically in the form of a Monte Carlo
event generator by Salam \cite{Salam:1995zd,Salam:1995uy,Salam:1996nb} 
(see also \cite{Avsar:2005iz} for another more recent implementation).
He also devised and implemented a saturation mechanism \cite{Mueller:1996te}
that went beyond the original dipole model pictured in Fig.~\ref{fig:com}a,
but which is necessary, as was argued before.

Before discussing more deeply the physical content of equations of the
form of Eq.~(\ref{eq:stochaqcd}),
we shall first study a model in which spatial dimensions are left out,
that we will be able to formulate in different ways.


\section{\label{sec:zerodimensional}Zero-dimensional model}

In the previous section, we have understood that scattering at high energy
in QCD may be viewed as a branching-diffusion process supplemented
by a saturation mechanism. We have exhibited a simple toy model
with these characteristics, whose dynamics is represented by an equation
of the type~(\ref{eq:RFT}).

Unfortunately, even that toy model is too difficult to solve analytically.
We shall study a still simplified model, where there is no diffusion 
mechanism: Realizations are completely specified by the number of particles
that the system contains at a given time. 
Of course, in this case, a saturation scale cannot be defined, which limits
the relevance of this model for QCD. However, we will be able to
formulate this model in many different ways, and to draw parallels
with QCD.
 
We start by defining precisely the model. Then, two approaches to the
computation of the moments of the number of particles are presented.
The first set of methods relies on field theory (Sec.~\ref{sec:fieldtheory}).
The second method relies on a statistical approach (Sec.~\ref{sec:statisticalmethods})
and will be extended in a phenomenological way to models with a spatial dimension
in Sec.~\ref{sec:reviewtraveling}.
We shall then draw the relation to a scattering-like formulation 
(Sec.~\ref{sec:relscattering}).
Finally (Sec.~\ref{sec:alternative}), some variants of the basic model are reviewed.

\subsection{Definition}

Let us consider a simple model in which the system is characterized
by its number $n_t$ of particles at each time $t$.
Between times $t$ and $t+dt$, each particle has a probability $dt$ to split
in two particles. For each pair of particles, there is a probability $dt/N$
that they merge into one.
We may summarize these rules in the following form:
\begin{equation}
n_{t+dt}=\left\{
\begin{aligned}
n_t\!+\!1 & \text{  proba  } n_tdt\\
n_t\!-\!1 & \text{  proba  } \frac{n_t (n_t\!-\!1)dt}{N}\\
n_t   & \text{  proba  } 1\!-\!n_tdt -\frac{n_t(n_t\!-\!1)dt}{N}.
\end{aligned}
\right.
\label{rule0}
\end{equation}
From this, one can easily derive an equation for the 
time evolution of the
probability $P(n,t)$
of having exactly
$n$ particles in the system at time $t$:
\begin{equation}
\frac{\partial P}{\partial t}(n,t)=
(n-1)P(n-1,t)+\frac{n(n+1)}{N}P(n+1,t)-\left(n+\frac{n(n-1)}{N}\right)P(n,t).
\label{master0d}
\end{equation}
This is the {\em master equation} for the Markovian process under consideration.
The two first terms with a positive sign represent the process of going
from one state containing $n$ particles 
to an adjacent one containing $n+1$ or $n-1$ particles respectively,
while the last term
simply corrects the probability to keep it unitary.

By multiplying both sides of this equation by $n$ and summing over $n$,
we get an evolution equation for the average number of 
particles $\langle n_t\rangle$:
\begin{equation}
\frac{d\langle n_t\rangle}{dt}=\langle n_t\rangle
-\frac{1}{N}\langle n_t(n_t-1)\rangle
\label{evolutionnnaive}
\end{equation}
Unfortunately, this equation is not closed, and one would have to establish an
equation for $\langle n_t(n_t-1) \rangle$, which would involve 3-point correlators
of $n_t$, and so on, ending up with an infinite hierarchy of equations, exactly
like in Sec.~\ref{sec:schannel} for QCD (see Eq.~(\ref{eq:balitskydipoles})).

This illustrates the difficulties one has to face
before one can get an analytical expression for $\langle n_t\rangle$,
even in such a simple model.

\subsection{\label{sec:fieldtheory}``Field theory'' approach}

In the next subsections, 
we will follow different routes to get analytical results
on the moments of the number of particles in the system at a given time $t$. 
The first one will be similar
to the $s$-channel picture of QCD (see Sec.~\ref{sec:schannel}), 
since it will consist
in computing  the time (rapidity in QCD) evolution of realizations of the system.
The second one will be closer to the $t$-channel picture of QCD. 
We will see how ``Pomerons'' may appear in these simple systems.
We will then examine a formulation in terms of a stochastic nonlinear
partial differential equation, which is nothing but the sFKPP equation
in which the space variable ($x$) has been discarded.

\subsubsection{\label{sec:0dfockstate}Particle Fock states and their weights}

Statistical problems were first formulated as field theories
by Doi \cite{Doi} and Peliti \cite{Peliti}. 
Different authors have used these methods (see Ref.~\cite{tauber-2007-716} for a review).
We shall start by following the presentation given in
Ref.~\cite{PhysRevE.59.3893}.

We would like to interpret the master equation~(\ref{master0d})
 as a quasi-Hamiltonian evolution
equation of the type of the ones that appear in quantum mechanics.
To this aim, we need to introduce the basis of states $|n\rangle$ of
fixed number $n$ of particles.
We define the ladder operators $a$ and $a^\dagger$ 
by their action on these states:
\begin{equation}
a|n\rangle=n|n-1\rangle,\ a^\dagger|n\rangle=|n+1\rangle
\end{equation}
and which obey the commutation relation
\begin{equation}
[a,a^\dagger]=1.
\end{equation}
The $n$-particle state may be constructed from the vacuum 
(zero-particle) state
by repeated application of the ladder operator:
\begin{equation}
|n\rangle=\left(a^\dagger\right)^n|0\rangle.
\end{equation}
The normalization is not standard with respect to what is usually
 taken in quantum mechanics.
In particular, the orthogonal basis $|n\rangle$ 
is normalized in such a way that
$\langle m|n\rangle=n!\delta_{m,n}$.
This implies that the completeness relation reads
\be
\sum_n \frac{1}{n!}|n\rangle\langle n|=1.
\ee
We also introduce the state vector of the system at a time $t$
as a sum over all possible Fock states weighted by their probabilities:
\begin{equation}
|\phi(t)\rangle=\sum_n P(n,t)|n\rangle.
\end{equation}
It is straightforward to see that the master equation~(\ref{master0d}) is
then mapped to the Schr\"odinger-type equation
\begin{equation}
\frac{\partial}{\partial t}|\phi(t)\rangle=-{\cal H}|\phi(t)\rangle,
\end{equation}
where $\mathcal{H}$ is the ``Hamiltonian'' operator
\begin{equation}
{\cal H}=(1-a^\dagger)a^\dagger a-\frac{1}{N}(1-a^\dagger)a^\dagger a^2.
\label{hamiltonian0d}
\end{equation}
The first term represents the splitting of particles, while the second one,
proportional to $1/N$,
represents the recombination.
We may rewrite ${\cal H}$ as
\begin{equation}
{\cal H}={\cal H}_0 +{\cal H}_1,
\end{equation}
where
\be
{\cal H}_0 =a^\dagger a
\ee
is the ``free'' Hamiltonian whose eigenstates are the Fock states.
We now go to the interaction picture by introducing the time-dependent
Hamiltonian
\begin{equation}
{\cal H}_I(t)=e^{{\cal H}_0 t} {\cal H}_1 e^{-{\cal H}_0 t}
\end{equation}
and the states $|\phi\rangle_I=e^{{\cal H}_0 t}|\phi\rangle$.
The solution of the evolution reads
\begin{equation}
\begin{split}
|\phi\rangle_I
&=T \exp\left(-\int_0^t dt^\prime {\cal H}_I(t^\prime)\right)|\phi_0\rangle_I\\
&=|\phi_0\rangle_I-\int_0^t dt^\prime {\cal H}_I(t^\prime)|\phi_0\rangle_I
+\int_0^t dt^\prime \int_0^{t^\prime} dt^{\prime\prime} 
{\cal H}_I(t^\prime){\cal H}_I(t^{\prime\prime})|\phi_0\rangle_I+\cdots
\end{split}
\label{Hevolsol}
\end{equation}
We may then compute the weights of the successive Fock states
by applying this formula.
Let us show how it works in detail by computing the state of a single
particle evolved from time 0 to time $t$, in the limit $N=\infty$ in which
there are no recombinations.
We follow the usual method to deal with
such problems in field theory. 
We insert 
repeatedly
complete basis of eigenstates of ${\cal H}_0$ into Eq.~(\ref{Hevolsol}), 
namely
\begin{equation}
|\phi\rangle_I=|1\rangle
-\int_0^t dt^\prime \sum_{n_1}
\frac{1}{n_1!}
|n_1\rangle\langle n_1|
{\cal H}_I(t^\prime)
|1\rangle+\cdots
\end{equation}
(We have kept the first two terms in Eq.~(\ref{Hevolsol}) explicitely).
Using the expression for ${\cal H}_I(t)$ 
as a function of ${\cal H}_0$ and ${\cal H}_1$,
together with the knowledge
that the Fock states are eigenstates of ${\cal H}_0$, we get
\begin{equation}
|\phi\rangle=e^{-t}|1\rangle-\sum_{n_1} e^{-n_1 t}
\int_0^t dt^\prime e^{n_1 t^\prime-t^\prime}
\frac{1}{n_1!}
|n_1\rangle\langle n_1|{\cal H}_1|1\rangle+\cdots
\end{equation}
Inserting the expression
for ${\cal H}_1$, one sees that
in the infinite-$N$ limit, there is only one possible
transition, $1\rightarrow 2$.
Performing the integration over $t^\prime$
and computing in the same manner the higher orders,
one finally gets the expansion
\begin{equation}
|\phi\rangle=e^{-t}|1\rangle+e^{-t}(1-e^{-t})|2\rangle+\cdots
+e^{-t}(1-e^{-t})^{n-1}|n\rangle+\cdots
\end{equation}
from which one can read the probabilities of the successive Fock states.
This expansion is similar to the expansion in dipole Fock states
introduced in Sec.~\ref{sec:schannel}: The $n$-particle states
correspond to $n$-dipole states in QCD,
and their weights are computed by applying successive splittings to the system,
whose rates are given by Eq.~(\ref{splitting}). (They are just unity in the case of
the zero-dimensional model.)

We see that this method is well-suited to compute the probabilities of the lowest-lying
Fock-states, and their successive corrections
at finite $N$. 
But in general we are rather interested in averages such as $\langle n^k\rangle$,
for which the weights of all Fock states are needed.
We will develop a slightly different (but equivalent) formalism below, 
that will enable us
to get these averages in a much more 
straightforward way.

\subsubsection{\label{sec:pomeronfieldtheory}Pomeron field theory}

Let us introduce the generating function of the factorial
moments of the distribution of the number of particles
\begin{equation}
Z(z,t)=\sum_n (1+z)^n P(n,t).
\end{equation}
The evolution equation obeyed by $Z$ can easily be derived
from the master equation~(\ref{master0d}):
\begin{equation}
\frac{\partial Z}{\partial t}=z(1+z)\left(\frac{\partial Z}{\partial z}-
\frac{1}{N}\frac{\partial^2 Z}{\partial z^2}\right).
\end{equation}
We may represent this equation in a second-quantized formalism
by introducing the operators
\begin{equation}
b^\dagger=z,\ b=\frac{\partial}{\partial z}=\bar z
\end{equation}
acting on the set of states $|Z\rangle$ consisting in the analytic
functions of $z$.
Then we may write
\begin{equation}
\frac{\partial Z}{\partial t}=-{\cal H}^{\mathbb{P}}Z,
\end{equation}
where
\begin{equation}
{\cal H}^{\mathbb{P}}={\cal H}_0^{\mathbb{P}}+{\cal H}_1^{\mathbb{P}},\ \text{with}\
{\cal H}_0^{\mathbb{P}}=-b^\dagger b,\ {\cal H}_1^{\mathbb{P}}
=-b^\dagger b^\dagger b+\frac{1}{N}
b^\dagger(1+b^\dagger)b^2.
\label{eq:hamiltonianpft}
\end{equation}
A basis for the states is
\begin{equation}
|k\rangle=z^k,\ \langle k|={\bar z^k}
\end{equation}
which is orthogonal with respect to the scalar product
\begin{equation}
\langle Z_1|Z_2\rangle=\int \frac{dzd\bar z}{2i\pi}e^{-|z|^2} 
\bar Z_1(z,\bar z) Z_2(z,\bar z),
\end{equation}
and obeys the normalization condition $\langle k|l\rangle=k! \delta_{k,l}$.
We shall call these states ``$k$-Pomeron'' states, by analogy with high-energy QCD.
We may apply exactly the same formalism as before, since the operators $b$, $b^\dagger$
have the same properties as the $a$, $a^\dagger$.

From the definition of the scalar product,
it is not difficult to see that the $k$-th factorial moment of $n$ may be obtained
by a mere contraction of the state vector $|Z\rangle$, computed by
solving the Hamiltonian evolution,
with a $k$-Pomeron state. The following identity holds:
\begin{equation}
\langle k|Z\rangle=
\left\langle\frac{n_t!}{(n_t-k)!}\right\rangle,
\end{equation}
where the average in the right handside goes over the realizations of 
the system.
As for the initial condition, starting the evolution with one particle
means taking as an initial condition the superposition $|0\rangle+|1\rangle$
of zero- and one-Pomeron states respectively. 
The zero-Pomeron state does not contribute to the
evolution, hence a one-Pomeron state is like a one-particle state.

\begin{figure}
\begin{center}
\epsfig{file=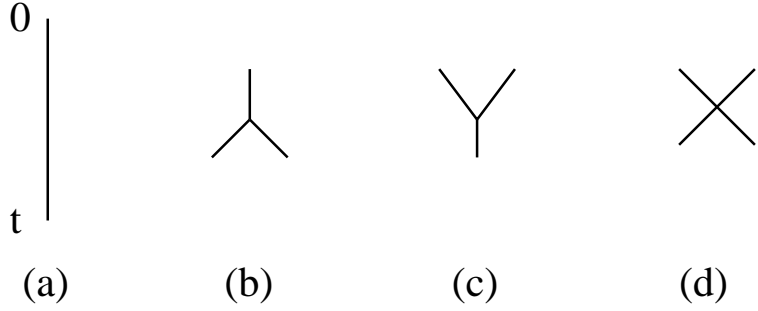,width=10cm}
\end{center}
\caption{\label{fig:feynman0d}
Propagator and vertices for the Pomeron field theory.
Time flows from the top to the bottom.
}
\end{figure}

In order to simplify the systematic computation
of these moments, we may 
use a diagrammatic method and establish Feynman rules.
To this aim, we write the contribution of the graphs with $l$-vertices
(corresponding to the term of order $l$ in the expansion of 
Eq.~(\ref{Hevolsol})), starting with a one-Pomeron state:
\be
\langle k|Z\rangle\supset (-1)^l
\int_0^t dt_1 \int_0^{t_1}dt_2\cdots\int_0^{t_{l-1}}dt_l
\sum_{n_1,\cdots,n_l}
\langle k|n_l\rangle
\frac{1}{n_l!}
\langle n_l |{\cal H}_I^{\mathbb{P}}|n_{l-1}\rangle
\cdots
\frac{1}{n_1!}
\langle n_1 |{\cal H}_I^{\mathbb{P}}|1\rangle.
\ee
Each matrix element that appears in this equation is associated
to a vertex, and propagators connect these vertices.
We read off the expression for the Hamiltonian~(\ref{eq:hamiltonianpft})
that there is one propagator and three vertices in the theory:
one splitting ($1\rightarrow 2$), one recombination
($2\rightarrow 1$) and a $2\rightarrow 2$ elastic diffusion vertices.

The method to compute the 1 to $k$ Pomeron transition amplitude is standard.
First, one draws all possible diagrams for this transition that contain
$l$ vertices, including all possible permutations. (Note that a splitting
may occur in $k$ different ways, if $k$ is the number of
Pomerons before the splitting; A recombination instead
may occur in $k(k-1)/2$ ways).
Then, the propagators (Fig.~\ref{fig:feynman0d}a) are replaced by 
\be
\langle 1|e^{-t {\cal H}_0^\mathbb{P}}|1\rangle = e^t,
\label{eq:PFTpropagator}
\ee
(where $t$ is the time
interval that they span) in such a way that
the $n$-Pomeron state propagates as
$\langle n|e^{-t {\cal H}_0^\mathbb{P}}|n\rangle=e^{nt}$. 
Intermediate times are integrated out.
As for the vertices (Figs.~\ref{fig:feynman0d}b-d), 
the following factors have to be applied:
\begin{equation}
1\rightarrow 2:\ -1;\ \
2\rightarrow 1:\ \frac{2}{N};\ \
2\rightarrow 2:\ \frac{2}{N}.
\end{equation}
In addition, there is a $(-1)^{\#\text{vertices}}$ factor.
Finally, an overall $k!$ factor leads to the expression for the factorial 
moment $\langle n_t(n_t-1)\cdots(n_t-k+1)\rangle$.

The lowest-order diagram for the average particle number, 
consisting in a simple propagator, reads $\langle n_t\rangle=e^t$.
We now understand that this method leads to a more straightfoward computation
of the moments of the number of particles 
than the one based on the
computation of the probabilities of successive Fock states, for
a single Pomeron already resums an infinity of particle Fock states.
The Pomeron in this case is exactly like the BFKL Pomeron introduced 
in Sec.~\ref{sec:schannel}, which leads to
an exponential increase of the scattering amplitudes with the rapidity 
(Eq.~(\ref{eq:PFTpropagator})).

\begin{figure}
\begin{center}
\epsfig{file=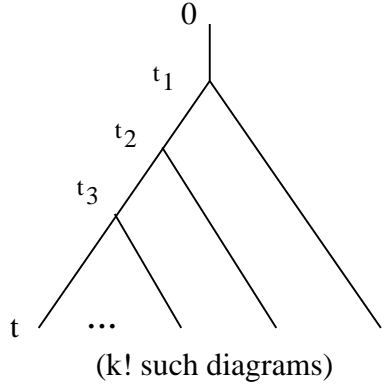,width=5cm}
\end{center}
\caption{\label{fig:diag0dtree}Diagrams contributing to
the one Pomeron $\rightarrow$ $k$-Pomeron transition, which gives the moments
$\langle n_t(n_t-1)\cdots(n_t-k+1)\rangle$ at leading order in a $1/N$ expansion.
}
\end{figure}

We now move on to the computation of higher-order diagrams.
First, let us recover simple results by taking the infinite-$N$ limit.
We consider the diagrams in Fig.~\ref{fig:diag0dtree}, which are 
the only ones that survive for $N=\infty$ in the evaluation of the
moment $\langle n_t(n_t-1)\cdots(n_t-k+1)\rangle$.
Using the Feynman rules, we get for each individual diagram
\begin{equation}
(-1)^k\times(-1)^k\times 
e^{kt}\int_0^t dt_1e^{-t_1}\int_{t_1}^{t}dt_2 e^{-t_2}\cdots\int_{t_{k-1}}^t dt_k e^{-t_k}
=\frac{1}{k!}\left(1-e^{-t}\right)^{k-1}.
\end{equation}
There are $k!$ such diagrams (corresponding to all possible permutations
of the Pomerons), and there is an extra overall $k!$ factor 
to be added in order to get
the relevant factorial moment:
\begin{equation}
\langle n_t(n_t-1)\cdots (n_t-k+1)\rangle=k! e^{kt}\left(1-e^{-t}\right)^{k-1}.
\end{equation}

\begin{figure}
\begin{center}
\epsfig{file=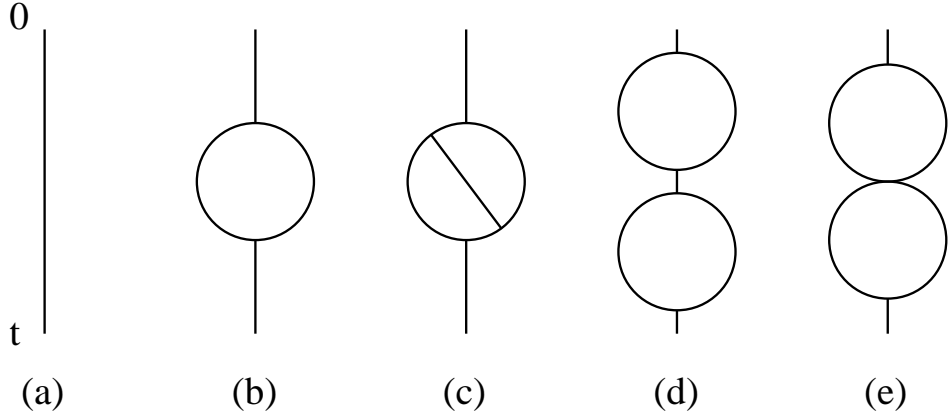,width=0.9\textwidth}
\end{center}
\caption{\label{fig:diag0dloops}Diagrams up to order $1/N^2$ contributing to
the average of the number of particles in the system after an evolution over the
time interval $t$.}
\end{figure}

Next, we would like to perform the computation of the one-Pomeron $\rightarrow$ 
one-Pomeron transition (which provides the value of $\langle n_t\rangle$)
within the full theory, including the recombinations.
Some of the lowest-order diagrams are shown in Fig.~\ref{fig:diag0dloops}.
A straightforward application of the Feynman rules 
edicted above
leads to the following results
for the graphs 
that are depicted in Fig.~\ref{fig:diag0dloops}:
\begin{equation}
\begin{split}
\left.\langle n_t\rangle\right|_\text{tree, Fig.~\ref{fig:diag0dloops}a}&=e^{t}\\
\left.\langle n_t\rangle\right|_\text{1 loop, Fig.~\ref{fig:diag0dloops}b}
&=-2!\frac{e^{2t}}{N}\left(1-e^{-t}(1+t)\right)\\
\left.\langle n_t\rangle\right|_\text{2 loops, Fig.~\ref{fig:diag0dloops}c}
&=3!\frac{e^{3t}}{N^2}\left(1+4e^{-t}(1-t)-e^{-2t}(2t+5)\right)\\
\left.\langle n_t\rangle\right|_\text{2 loops, Fig.~\ref{fig:diag0dloops}d}&
=4 \frac{e^{2t}}{N^2}\left(t-3+e^{-t}\left({\scriptstyle\frac{t^2}{2}}+2t+3\right)\right)\\
\left.\langle n_t\rangle\right|_\text{2 loops, Fig.~\ref{fig:diag0dloops}e}&
=4 \frac{e^{2t}}{N^2}\left(t-2+e^{-t}\left(t+2\right)\right)
\end{split}
\end{equation}
We may classify these different contributions according to their order
in $e^t/N$: The leading terms for large $t$ and $e^t/N\sim 1$
are always of the form $N(e^t/N)^{1+\#\text{loops}}$.
It turns out that we may compute easily these dominant terms at any number
of loops. They stem from the graphs in which all splittings
occur before all recombinations. These terms build up a series that reads
\begin{equation}
\langle n_t\rangle=\sum_{k=1}^{\infty}(-1)^{k-1}k!\frac{e^{kt}}{N^{k-1}}.
\end{equation}
This series is factorially divergent, but is easy to resum
with the help of the Borel transformation.
Indeed, using the identity
\begin{equation}
k!=\int_0^{+\infty}db\, b^k e^{-b},
\end{equation}
then exchanging the integration over $b$ and the sum over the number of Pomerons $k$,
one gets
\begin{equation}
\langle n_t\rangle=N^2 e^{-t}\int_0^{+\infty}db\frac{1}{1+\frac{1}{b}}e^{-Ne^{-t}b}
=N\left(1-N e^{Ne^{-t}}\Gamma(0,N{e^{-t}})\right),
\label{eq:finalntpomeron}
\end{equation}
where $\Gamma$ is the incomplete Gamma function.

This result was obtained for the first time using a diagrammatic method
in Ref.~\cite{Shoshi:2005pf}. The authors of that paper
also computed the
next-to-leading order, that is to say, 
the terms of relative order $1/N$ after
the resummation has been performed.
The equivalent of the diffractive processes known in QCD
were also investigated by the same authors
in Ref.~\cite{Shoshi:2006eb}.
More results were obtained on that kind of models
by another group in Ref.~\cite{Levin:2007yv,Kozlov:2006cu}, 
using different techniques, which go beyond the perturbative approach.
Remarkably, the latter calculations can be applied to some extent
to QCD \cite{Kozlov:2007xc,Levin:2007wc}.

\subsubsection{Stochastic evolution equations}

The model may also be formulated in the form of a stochastic evolution
equation for the number of particles $n_t$ it contains at each time $t$.
The most straightforward way of doing this would be to
first compute the mean and variance of $n_{t+dt}$ given
$n_t$, with the help of the master equation~(\ref{master0d}).
This would enable one to write the time evolution
of $n_t$ in terms of a drift and of a noise of zero mean
and normalized variance, namely:
\begin{equation}
\frac{dn_t}{dt}=n_t-\frac{n_t(n_t-1)}{N}+
\sqrt{n_t+\frac{n_t(n_t-1)}{N}}\nu_{t+dt},
\label{evoln}
\end{equation}
where $\nu$ is such that $\langle \nu_t\rangle=0$ and
$\langle\nu_t\nu_{t^\prime}\rangle=\delta(t-t^\prime)$.
This equation is similar to Eqs.~(\ref{eq:stochaqcd}) and~(\ref{eq:RFT}), except for
it does not have a spatial dimension where some diffusion could take place.
The noise term is of order $\sqrt{n}$, as it should according to
the argumentation of Sec.~\ref{sec:schannel}.
Note that the distribution of $\nu$
depends on $n_t$ and
is not a Gaussian.
This last point is easy to understand: The evolution
of $\nu_t$ is intrinsically discontinuous, since
it stems from a rescaling of
$n_t$, which is an integer at all times.
A Brownian evolution (i.e. with a Gaussian noise) would
necessarily be continuous.
For completeness, we write the statistics of  $\nu_{t+dt}$, which is
easy to derive from the evolution of $n$:
\begin{equation}
\nu_{t+dt}=
\begin{cases}
\phantom{-}\frac{1}{\sigma\, dt}-\frac{\Delta}{\sigma} & \text{proba}\
{n_t\, dt}\\
\phantom{-\frac{1}{\sigma\, dt}}-\frac{\Delta}{\sigma} & \text{proba}\
1-n_t\, dt-\frac{n_t(n_t-1)}{N}dt
\\
-\frac{1}{\sigma\,dt}-\frac{\Delta}{\sigma} & \text{proba}\
\frac{n_t(n_t-1)}{N}dt,
\end{cases}
\end{equation}
where
$\Delta=n_t-\frac{n_t(n_t-1)}{N}$ and
$\sigma=\sqrt{n_t+\frac{n_t(n_t-1)}{N}}$.
There are jumps induced by the terms proportional to $1/dt$.

This formulation is not of great interest, neither for analytical
calculations nor for numerical simulations, since it is much
easier to just implement the rules that define the model in 
the first place (Eq.~(\ref{rule0})) 
in the form of a Monte Carlo event generator.

There is a better way to arrive at a stochastic evolution equation for
this model, although it is a bit more abstract. (It is actually equivalent
to the Pomeron field theory formulated before.)
Instead of following states with a definite number of particles like
above, we may introduce coherent states
\begin{equation}
|z\rangle=e^{-z+z a^\dagger}|0\rangle,
\end{equation}
where $z$ is a complex number.
For real positive values of $z$, the state $|z\rangle$ is nothing but a Poissonian
state, which is a superposition of $|k\rangle$-particle states, where
the weight of each state follows the 
Poisson law
of parameter $z$.
For the simplicity of the argument, let us restrict ourselves
to these states.
By applying the Hamiltonian ${\cal H}$ 
(defined in Eq.~(\ref{hamiltonian0d}))
to a Poissonian state $|z_t\rangle$,
one gets a new state $|\phi_{t+dt}\rangle$:
\begin{equation}
|\phi_{t+dt}\rangle=|z_t\rangle-dt\,{\cal H}|z_t\rangle.
\label{evolcoherent0d}
\end{equation}
Of course, that new state is not itself
at Poissonian state in general, but
may be written as a superposition of such states.
One writes
\begin{equation}
|\phi_{t+dt}\rangle=\int dz\,f(z)|z\rangle=\int dz\,f(z)\sum_n e^{-z}\frac{z^n}{n!}|n\rangle.
\label{decompcoherent0d}
\end{equation}
The idea is to interpret the weight function $f(z)$ as the probability to observe
a given Poissonian state $|z\rangle$.
Hence the evolution is viewed as a stochastic path
\begin{equation}
\cdots\rightarrow z_{t-dt}
\rightarrow z_t\rightarrow z_{t+dt}\rightarrow z_{t+2dt}\rightarrow\cdots
\end{equation}
with well-defined transition rates from one Poissonian state to the next one.
Inserting the explicit expression for the Hamiltonian~(\ref{hamiltonian0d}) and the
decomposition~(\ref{decompcoherent0d}) in
Eq.~(\ref{evolcoherent0d}), one gets for each Fock state $|n\rangle$
\begin{multline}
\int dz\,e^{-z}f(z)\frac{z^n}{n!}=e^{-z_t}\frac{z_t^{n}}{n!}\\
-dt\,e^{-z_t}
\left[
\frac{z_t^n}{(n-1)!}-\frac{z_t^{n-1}}{(n-2)!}-\frac{1}{N}
\left(
\frac{z_t^{n+1}}{(n-1)!}-\frac{z_t^n}{(n-2)!}
\right)
\right].
\end{multline}
Finally, this equation is easy to invert for $f(z)$ by integrating over $n$
with the weight
$\int \frac{dn}{2i\pi}z_{t+dt}^{-n-1}$, along
an appropriate contour in the complex plane.
After some straightforward algebra, we get
\begin{multline}
f(z_{t+dt})=\delta(z_{t+dt}-z_t)+dt\left(z_{t}-\frac{z_t^2}{N}\right)\delta^\prime(z_{t+dt}-z_t)\\
+\frac12\left[
2dt\left(z_t-\frac{z_t^2}{N}\right)\delta^{\prime\prime}(z_{t+dt}-z_t)
\right].
\end{multline}
This is a Gaussian centered at $z_t+dt(z_t-\frac{z_t^2}{N})$
of variance $2dt(z_t-\frac{z_t^2}{N})$.
Introducing a normal Gaussian noise $\nu_t$ which satisfies
\begin{equation}
\langle \nu_t\rangle=0\ \ \text{and}\ \ 
\langle \nu_t\nu_{t^\prime}\rangle=\delta(t-t^\prime),
\end{equation}
we may write
\begin{equation}
\boxed{
\frac{dz_t}{dt}=z_{t}-\frac{z_t^2}{N}
+\sqrt{2\left(z_{t}-\frac{z_t^2}{N}\right)}\nu_{t+dt}
}
\label{eq:stochastic0d}
\end{equation}
where the noise is taken at time $t+dt$, and hence, 
this equation is to be interpreted in the Ito sense.
If $z_{t=0}$ is a real number between 0 and $N$, then
the equation keeps it in this range.
But one may consider more general coherent states,
with complex $z_t$.

This equation is suitable for numerical simulations:
One may discretize the time in small steps $\Delta t\ll 1$
in which case $\nu_t$ is distributed as
\be
p(\nu_t)=\frac{1}{\sqrt{2\pi \Delta t}}\exp\left(
-\frac{\nu_t^2}{2\Delta t}
\right).
\ee
(In many cases, one has to use more sophisticated methods, see e.g. 
Ref.~\cite{gardiner}).
Analytical manipulations of this equation
using Ito's calculus are also quite easy. We are going to give
an example of such a calculation below, avoiding unnecessary
formalism. (We refer the reader to \cite{gardiner} for a textbook
on a more mathematical treatment of stochastic processes.)

We may transform the stochastic 
equation~(\ref{eq:stochastic0d}) 
to a hierarchy of equations for the factorial moments
of the number of particles, using the relation
\begin{equation}
\langle z_t^k\rangle=\langle n_t(n_t-1)\cdots(n_t-k+1)\rangle
\equiv n_t^{(k)}.
\label{correspfactorials}
\end{equation}
First, let us write Eq.~(\ref{eq:stochastic0d}) in a discretized form:
\begin{equation}
{z_{t+dt}}=z_t+dt\left(z_{t}-\frac{z_t^2}{N}\right)
+dt\sqrt{2\left(z_{t}-\frac{z_t^2}{N}\right)}\nu_{t+dt}.
\end{equation}
We then take the $k$-th power of the left and the right handside, and
we average the result over realizations. 
Expanding in powers of $dt$ for small $dt$, we get
\begin{multline}
\left\langle z_{t+dt}^k\right\rangle=\langle z_t^k\rangle
+dt\,k
\left\langle z_{t}^k-\frac{z_t^{k+1}}{N}\right\rangle
+dt\,k \left\langle z_t^{k-1}\sqrt{2\left(z_{t}-\frac{z_t^2}{N}\right)}
\right\rangle
\langle \nu_{t+dt}\rangle\\
+dt^2\,\frac{k(k-1)}{2} \left\langle
2\left(z_{t}^{k-1}-\frac{z_t^{k}}{N}\right)
\right\rangle
\langle\nu_{t+dt}^2\rangle
+\cdots
\end{multline}
We have factorized the average over the noise over the time intervals $[t,t+dt]$ and
$[0,t]$, since the noise $\nu$ is uncorrelated in time.
The term proportional to $dt$ vanishes thanks to the fact that $\nu_{t+dt}$ averages
to zero. One may think that the next term could be neglected for it
is apparently proportional to $dt^2$.
Actually, it
gives a contribution of order $dt$,
because for discretized $t$,
$\langle \nu_{t+dt}^2\rangle=1/dt$.
The dots stand for terms of order $dt^2$ at least. 
Using Eq.~(\ref{correspfactorials}) to identify the
factorial moments of $n$, we eventually get
\begin{equation}
\frac{dn_t^{(k)}}{dt}=k\left(n_t^{(k)}-\frac{n_t^{(k+1)}}{N}\right)+k(k-1)
\left(n_t^{(k-1)}-\frac{n_t^{(k)}}{N}\right)
\end{equation}
This equation is similar to the (modified) Balitsky hierarchy in high-energy QCD.
Let us write explicitely the equations for the first two moments:
\be
\begin{split}
\frac{d\langle n_t\rangle}{dt}&=\langle n_t\rangle-\frac{1}{N}
\langle n_t(n_t-1)\rangle,\\
\frac{d\langle n_t(n_t-1)\rangle}{dt}&=2\left(1-\frac{1}{N}\right)
\langle n_t(n_t-1)\rangle
-\frac{2}{N}{\langle n_t(n_t-1)(n_t-2)\rangle}+2\langle n_t\rangle.
\end{split}
\ee
We note the similarity in structure with Eq.~(\ref{eq:balitskydipoles}), except for
the term $2\langle n\rangle$ in the right handside. This term stems precisely
from the particle recombinations, and was absent in the B-JIMWLK formalism.

Finally, let us mention that for a more rigorous and general derivation
of this stochastic formulation, one may use a path integral formalism
obtained from the Hamiltonian~(\ref{hamiltonian0d}), see Ref.~\cite{tauber-2007-716}.


\subsection{\label{sec:statisticalmethods}Statistical methods}

\begin{figure}
\begin{center}
\epsfig{file=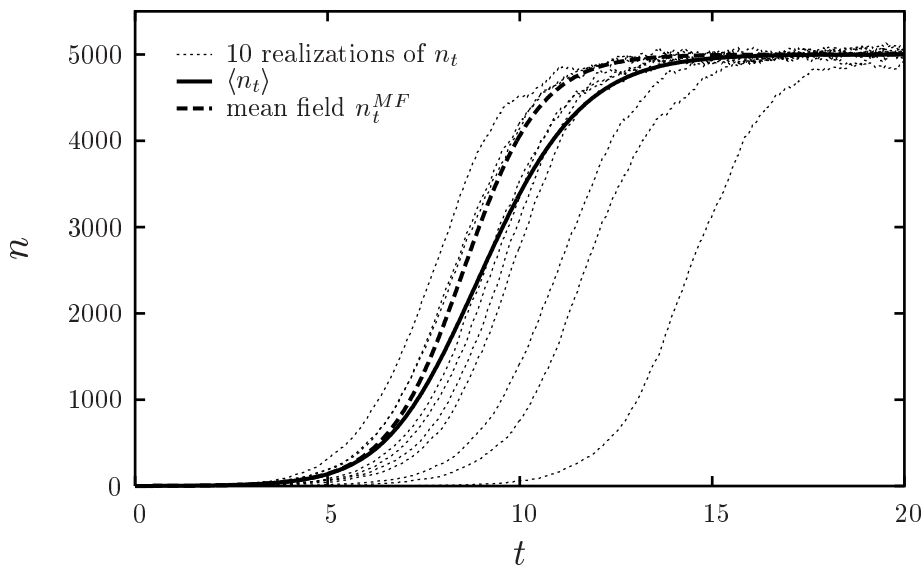,width=0.9\textwidth}
\end{center}
\caption{\label{fig:realizations0d}
[From Ref.~\cite{Munier:2006um}]
Ten different realizations of the stochastic evolution of the 
zero-dimensional model (dotted lines; $N=5\times 10^3$).
All realizations look the same, up to a shift in time. They are all parallel to
the solution to the mean-field equation~(\ref{eq:MFequation}) (dashed line).
Note the significant difference between the latter and the average of the particle
number over the realizations (full line).
}
\end{figure}

The field theory methods presented above provide a systematics to solve the evolution
of the system to arbitrary orders in $1/N$, at least theoretically.
(In practice, identifying and resumming the relevant diagrams becomes
increasingly difficult).
However, it would look quite unreasonable
to get into such an involved formalism
if one were only interested in computing
the lowest order in a large-$N$ expansion.
Indeed, as we shall demonstrate it below,
in the case of this simple model, an intuitive and economical
calculation leads to the right answer \cite{Munier:2006um}.
We work it out here because this line of reasoning is at
the basis of the solution to more complicated models, closer to QCD, that we shall
address in the next section (Sec.~\ref{sec:reviewtraveling}).

As before, we denote by $n_t$ the value of the number
of particles in a given realization of the system.
We further introduce
$p_{\bar n}(\bar t)$ the distribution of the times
at which the number of particles in the system reaches some given
value $\bar n$,
and $\langle n_t|\bar n,{\bar t}\rangle$ the conditional average number of
particles at time $t$ given that there were $\bar n$ particles in the system
at time $\bar t$.
One may write the following factorization formula:
\begin{equation}
\langle n_t\rangle=\int_0^\infty d\bar t p_{\bar n}(\bar t)
\langle n_t|\bar n,{\bar t}\rangle.
\label{facto0d}
\end{equation}
This formula holds exactly for any value of $\bar n$. In particular, if $N$ is large enough,
one may choose $\bar n$ such that $1\ll \bar n\ll N$.

Looking at a few realizations generated numerically
(Fig.~\ref{fig:realizations0d}), 
one sees that the curves that represent $n_t$ look like the solution
to the mean-field equation obtained by neglecting the noise term in Eq.~(\ref{evoln}),
up to a translation of the origin of times by some random $t_0$.
(The curves look also slightly noisy around the average trend, but the noise
would still be much weaker for larger values of $N$.)
This suggests that once there are enough particles in the system 
(for $n_t>\bar n\gg 1$), 
the evolution
becomes essentially deterministic and in that stage of the evolution, the noise
can safely be discarded.
Thus stochasticity only manifests itself in the initial stages of the evolution,
but in a crucial way.
Indeed, as one can see in Fig.~\ref{fig:realizations0d}, after averaging, 
$\langle n_t\rangle$
differs significantly from the mean-field result, and this difference stems from
rare realizations in which the particle number stays low for a long time.
Therefore, in individual realizations, stochasticity should accurately be
taken into account as long as $n_t<\bar n$. Fortunately, when the number
of particles in the system is small compared to the parameter
$N$ that fixes the typical maximum number of particles
in a realization, the stochastic evolution is essentially governed
by a linear equation.

Thanks to this discussion,
we may assume that the evolution is linear as long as there are 
less than $\bar n$ particles
in the system and deterministic when $n_t>\bar n$.
It is then enough to compute $p_{\bar n}(\bar t)$ 
for an evolution without recombinations,
and $\langle n_t|\bar n,{\bar t}\rangle$ for an evolution without noise.
The second quantity is most easily computed by replacing 
$n_t$ in Eq.~(\ref{evolutionnnaive}) by the average quantity
$n_t^\text{MF}$ 
(or equivalently by discarding the noise term in Eq.~(\ref{evoln})) 
and
neglecting the term $n_t^\text{MF}/N$ (which is small compared to the
term $n_t^\text{MF}$).
One gets a closed equation for $n_t^\text{MF}$ in the form
\begin{equation}
\frac{dn_t^\text{MF}}{dt}=n_t^\text{MF}-\frac{(n_t^\text{MF})^2}{N}
\label{eq:MFequation}
\end{equation}
which is solved by
\begin{equation}
n_{t}^\text{MF}=\frac{N}{1+\frac{N}{\bar n}e^{-(t-\bar t)}}
=\langle n_t|\bar n,{\bar t}\rangle
\label{eq:MFsolution}
\end{equation}
where the initial condition 
at time $t=\bar t$
has been chosen in such a way that
$n_{\bar t}=\bar n$.

As for the distribution $p_{\bar n}(\bar t)$ for the 
waiting times $\bar t$ to observe
$\bar n$ particles in the system, its derivation is a bit more subtle.
Since the evolution is taken linear
until there are $\bar n$ particles in the system, 
the number of particles increases
with time in any given realization.
Then the following relation is true
\begin{equation}
p_{\bar n}(\bar t)=\left.\frac{d}{dt}\right|_{t=\bar t}
\sum_{n=\bar n}^\infty P(n,t)
\end{equation}
where $P(n,t)$ solves the master equation~(\ref{master0d}) 
in which terms of order $1/N$ are
discarded. This relation only holds 
because the probability that $n$ be larger than $\bar n$
reads
\begin{equation}
\text{Prob}(n\geq \bar n,t)=\sum_{n=\bar n}^\infty P(n,t)
\end{equation}
thanks to the fact that $n$ never decreases in realizations when nonlinear
effects are neglected.

We could solve the simplified equation for $P(n,t)$, but for the sake of
presenting a method that may be more general, 
we shall follow a slightly different route
and establish first an equation that gives $p_{\bar n}(\bar t)$ 
more directly.

Let us introduce $Q(n,t)$ the probability that the number of
particles remain strictly less than $\bar n$ for any time in $[0,t]$,
starting with a system of $n$ particles at time 0.
Then we obviously have
\begin{equation}
\int_t^\infty d\bar t\, p_{\bar n}(\bar t)=Q(1,t).
\end{equation}
which by simple derivation of $Q(1,t)$ with respect to $t$
gives the relevant distribution.
We now establish an evolution equation for $Q$.
Recall that the evolution equation for $P$ was obtained by
considering the variation in the number of particles in the system
between times $t$ and $t+dt$.
Here we consider the beginning of the time evolution, between times $0$ and $dt$.
The probability that the system does not exceed $\bar n$ particles
up to time $t+dt$ starting with $n$ particles at time $t=0$, 
$Q(n,t+dt)$, is the
probability $ndt$ that the system gains 
a particle between times $0$ and $dt$ multiplied by
$Q(n+1,t)$, plus a unitarity-preserving term.
In this way, after having taken the limit $dt\rightarrow 0$,
we get
\begin{equation}
\frac{\partial Q(n,t)}{\partial t}=
n\left(Q(n+1,t)-Q(n,t)\right).
\end{equation}
This equation is valid when we neglect recombination
processes, which is the relevant approximation here.
In order to find a solution, 
we introduce the generating function 
for the moments of $n$:
\begin{equation}
G(u,t)=\sum_{n=0}^\infty u^n Q(n,t).
\end{equation}
The evolution of $Q$ implies
\begin{equation}
\frac{\partial G}{\partial t}=(1-u)\frac{\partial G}{\partial u}-\frac{1}{u} G.
\label{eq:dGdt}
\end{equation}
This equation may be solved by the method of characteristics well-known for example
in fluid mechanics,
but also in QCD where it is commonly used 
to solve the renormalization group equation.
We provide all details of the derivation of the solution in our simple case
since it is not used so often in the particular 
field of high-energy QCD.

The method consists in promoting the independent variable 
$u$ to a function of time:
 $u\rightarrow u(t)$.
One then writes the total time derivative of $G$ as
\begin{equation}
\frac{dG(u(t),t)}{dt}=\frac{\partial G(u(t),t)}{\partial t}
+\frac{du(t)}{dt}\frac{\partial G(u(t),t)}{\partial u}.
\end{equation}
Identifying the right handside of 
this equation to Eq.~(\ref{eq:dGdt}), one gets
\begin{equation}
\frac{dG(u(t),t)}{dt}=-\frac{1}{u} G,
\label{chara1}
\end{equation}
provided that $u(t)$ solves
\begin{equation}
\frac{du(t)}{dt}=u-1.
\label{chara2}
\end{equation}
This equation is easily integrated:
\begin{equation}
u(t)=1+(u_0-1)e^t,
\label{chara3}
\end{equation}
where the initial condition $u_0=u(0)$ is taken at zero time.
The backward solution is also needed:
\begin{equation}
u_0=1+(u(t)-1)e^{-t}.
\label{chara4}
\end{equation}
Next, one integrates the ordinary differential 
equation~(\ref{chara1})
\begin{equation}
G(u(t),t)=G(u_0,0)\exp\left(-\int_0^t dt^\prime \frac{1}{u(t^\prime)}\right)
\end{equation}
Replacing $u(t^\prime)$ by its value given by Eq.~(\ref{chara3}) 
under the integration sign, then replacing $u_0$ by its expression as a function
of $u$ and of $t$ (Eq.~(\ref{chara4})) one gets
\begin{equation}
G(u,t)=G(1+(u-1)e^{-t},0)\frac{u}{u-1+e^t}.
\label{chara5}
\end{equation}
Finally, the initial condition for $G$ stems from the fact
that $Q(n,0)=0$ for $n\geq\bar n$ and $Q(n,0)=1$ 
for $n<\bar n$. Therefore,
\begin{equation}
G(u,0)=\sum_{n=0}^{\bar n -1} u^n=\frac{1-u^{\bar n}}{1-u^{\phantom{n}}}.
\end{equation}
Inserting this result into Eq.~(\ref{chara5}), we get
\begin{equation}
G(u,t)=\frac{u}{1-u}\frac{1-(1-(1-u)e^{-t})^{\bar n}}{1-e^{-t}(1-u)}.
\end{equation}
$Q(1,\bar t)$ is easily obtained from $G$, by a simple integration:
\begin{equation}
Q(1,\bar t)=\int\frac{du}{2i\pi}\frac{G(u,\bar t)}{u^2},
\end{equation}
where the integration runs over an appropriate contour in the complex
$u$ plane. We get from the Cauchy theorem
\begin{equation}
Q(1,\bar t)=1+\left(1-e^{-\bar t}\right)^{\bar n-1},
\end{equation}
and
\begin{equation}
p_{\bar n}(\bar t)=-\frac{dQ(1,\bar t)}{d\bar t}=
(\bar n-1)e^{-\bar t}(1-e^{-\bar t})^{\bar n-2}.
\end{equation}
In the limits $\bar n \gg 1$ and $\bar t\gg 1$ which are relevant here,
the distribution simplifies to
\begin{equation}
p_{\bar n}(\bar t)\simeq
\bar n e^{-\bar t-\bar n e^{-\bar t}}.
\label{eq:gumbel}
\end{equation}
This is a Gumbel distribution.

Plugging Eqs.~(\ref{eq:gumbel}) and~(\ref{eq:MFsolution}) into Eq.~(\ref{facto0d}), 
we get for the average number of particles after $t$ units of time 
of evolution:
\begin{equation}
\langle n_t\rangle=N\int_0^\infty d\bar t\frac{\bar n e^{-\bar t-\bar n e^{-\bar t}}}
{1+\frac{N}{\bar n}e^{-(t-\bar t)}}.
\end{equation}
Because the Gumbel distribution is strongly damped for $\bar t<0$, the lower
integration
boundary may safely be extended to $-\infty$. Indeed, it is easy to see that
a conservative upper bound for the contribution of the domain $]-\infty,0]$ to the
integral is $e^{-\bar n}$, which is very small in the limit $\bar n\gg 1$.
Finally, we perform the change of variable $b=\bar n e^{-\bar t} \frac{e^t}{N}$ to 
arrive at the form
\begin{equation}
\langle n_t\rangle=N^2 e^{-t}\int_0^\infty db\frac{1}{1+\frac{1}{b}}e^{-Ne^{-t}b}.
\label{eq:finalntstat}
\end{equation}
It can be checked that it is exactly the form found through
the diagrammatic approach to Pomeron field theory (compare Eq.~(\ref{eq:finalntstat}) 
to Eq.~(\ref{eq:finalntpomeron})).

The factorization in Eq.~(\ref{facto0d})
and the convenient approximations that it subsequently allows
are actually very important.
Indeed, we realized that we may write the average number of particles
at time $t$, whose expression would {\em a priori} be given
by the solution of a {\em nonlinear stochastic differential equation},
by solving two much simpler problems.
The key observation was the following. When the number of particles
in the system is low compared to the maximum average number of particles
$N$ allowed by the reaction process, then the nonlinearity is
not important, but the noise term is instead crucial.
On the other hand, when the number of particles is large compared to 1,
then the noise may be discarded, but
the nonlinearity of the evolution equation, which corresponds
to recombinations of particles, must be treated accurately.
From this method, one gets an expression for $\langle n_t\rangle$ 
up to relative
corrections of order $1/N$.

When we address the problem of reaction-diffusion with one spatial dimension, we will
rely on the very same observation.
It is essentially the latter which will enable us to find analytical results
also in that case.


\subsection{\label{sec:relscattering}
Relation to high energy scattering and the parton model approach}

So far, we have focussed on the factorial moments of the number $n$ of
particles in the system. We have seen how they may be computed from ``Pomeron'' 
diagrams, which are quite similar to the diagrams that appear in effective
formulations of high-energy QCD.
However, the relation to scattering amplitudes, which are the observables
in QCD, may not be clear to the reader at this stage.
In particular, we do not understand yet what would correspond to boost invariance
of the QCD amplitudes.
The aim of this section is to clarify these points.

Let us consider a realization of the system of particles, 
evolved up to time $t$
(at which it contains $n_t$ particles),
that we may call the projectile.
A convenient formalism to compute the weights of
Fock states was presented in Sec.~\ref{sec:0dfockstate}.
We imagine that at time $t$, 
it scatters off a target consisting of a single particle, and
can have at most one exchange with the target, which ``costs''  a factor $1/N$.
All the particles in the system have an equal probability to scatter.
Hence the probability that the system scatters reads $T=n_t/N$.
The average of $T$ over events is the average particle number
normalized to $N$.

This way of viewing the evolution of the system makes it obviously 
very similar to the
QCD dipole model introduced in Sec.~\ref{sec:schannel}, 
provided one identifies the number of particles to the number
of dipoles and the time to the rapidity variable. 
The average of $T$ over
realizations is the elastic scattering amplitude.

From this analogy, there is a property similar 
to boost invariance that should hold.
Instead of putting all the evolution in the projectile, we may share it between
the projectile and the target. Let us call $n_{t^\prime}$ the number of particles
in the projectile at the time of the interaction, and $m_{t-t^\prime}$ the number
of particles in the target. The total evolution time is the same as before.
To establish the expression for $T$ in this frame, it is easier to
work with the probability $S=1-T$ that there is no interaction.
If any number of interactions were allowed between each pair of particles
from the projectile and the target, then one would simply write 
$S=\exp(-n_{t^\prime}m_{t-t^\prime}/N)$.
But since the number of interactions should be limited to one per particle,
one has to decrease $n$ and $m$ for each new power of $1/N$, i.e. for each
additional rescattering:
\begin{multline}
S=1-\frac{1}{N}nm+\frac1{2!}\frac{1}{N^2}[n(n-1)][m(m-1)]\\
-\frac{1}{3!}\frac{1}{N^3}
[n(n-1)(n-2)][m(m-1)(m-2)]\cdots
\label{eq:forminteraction}
\end{multline}
where the time dependences are understood in order to help the reading.
This is like a ``normal ordering'' of the expression 
to which we would arrive by assuming
any number of exchanges.
Note that $S$ is not necessarily positive in a given event,
and hence one looses the probabilistic interpretation
once one has performed the normal ordering.

\begin{figure}
\begin{center}
\epsfig{file=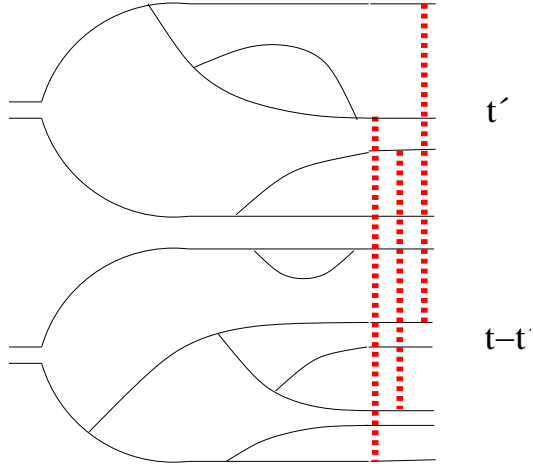,width=0.5\textwidth}
\end{center}
\caption{\label{fig:scat0d}Representation of the scattering of two
systems of particles. The systems evolve in time from the
left to the right.
The horizontal lines represent the particles, and the vertical
dashed lines the interactions between the systems. Each of
the elementary scatterings comes with a power of $1/N$.
Note the strong similarity with the QCD diagram in Fig.~\ref{fig:com}a, except
that in the present case, 
recombinations are included in the evolution 
of each of the systems.}
\end{figure}

Taking the average over realizations, one gets
\begin{equation}
\langle S\rangle =\sum_{k=0}^\infty 
\left\langle \frac{n!}{(n-k)!}\right\rangle_{t^\prime} 
\left\langle \frac{m!}{(m-k)!}\right\rangle_{t-t^\prime}
\frac{(-1)^k}{k! N^k}.
\end{equation}
If $t^\prime=t-t^\prime$, the first two factors 
in each term of the series are of course identical after averaging.
The sum runs over the number of actual exchanges between the probe and the target.
A realization of the evolution, which would correspond to an event in QCD,
is represented in Fig.~\ref{fig:scat0d}.
Note that the figure is very similar to Fig.~\ref{fig:com}a, except
that particle mergings are allowed, while they have not been properly
formulated in QCD yet.

Now this expression should be independent of $t^\prime$.
It is not difficult to check that this is indeed true by taking
the derivative of $\langle S\rangle$ with respect to $t^\prime$.
Expressing the averages of the factorial moments of the number
of particles with the help of the probability distributions
$P(n,t^\prime)$ and $P(m,t-t^\prime)$ respectively,
each term of the sum over $k$ and $m,n$ reads
\begin{equation}
\left.\frac{d\langle S\rangle}{dt^\prime}\right|_{n,m,k\ \text{fixed}}
=(\dot P_n P_m-P_n \dot P_m)\frac{n!}{(n-k)!}
\frac{m!}{(m-k)!}\frac{(-1)^k}{k!N^k}.
\end{equation}
The time dependence is understood,
and we introduced the notation $P_n=P(n,\cdot)$ and $\dot P_n=\partial_t P(n,\cdot)$ 
to get a more
compact expression. The time variabe that should be used for each factor
is unambiguous since it is in one-to-one correspondence with 
the particle number index.
We may use the master equation~(\ref{master0d})
 to express the time derivatives:
\begin{equation}
\dot P_n P_m-P_n \dot P_m=\left[(n-1)P_{n-1}+\frac{n(n+1)}{N}P_{n+1}-
\left(n+\frac{n(n-1)}{N}\right)P_n\right]P_m 
- [n\leftrightarrow m].
\end{equation}
Recalling that there are sums over $m$, $n$ and $k$ which go from 0 to $\infty$,
one may shift first the indices $m$ and $n$ in order to factorize
$P_nP_m$ in each term. The factors $1/N$ may then be absorbed by shifting
$k$ for the relevant terms. Then cancellations occur between the
terms of both squared brackets in such a way that 
once the summations over $n$, $m$ and $k$ have been performed,
the global result is 0.
This proves the independence of $\langle S\rangle$ 
upon $t^\prime$, that is, ``boost invariance'' in a relativistic
quantum field theory
language.
Of course, boost invariance is a consequence of some subtle interplay
between the form of the interaction~(\ref{eq:forminteraction}) and
the form of the evolution encoded in the master equation~(\ref{master0d}).
Had we not normal ordered the expression for $S$ in Eq.~(\ref{eq:forminteraction}),
boost invariance would not have hold as we shall check shortly.

We have seen that we may formulate scattering amplitudes
in the zero-dimensional toy model, exactly in the same way as in QCD.
We have seen in particular how crucial it is to include
particle mergings consistently with the form of the interaction
between the states of the projectile and of the target at the time of the interaction, 
in order to get a boost-invariant amplitude.


\subsection{\label{sec:alternative}Alternative models in 0 dimensions}

For the sake of completeness, we shall now construct some variants of
the zero-dimensional model introduced above,
since the latter were also discussed in the literature.
We review two of the most popular models.

\subsubsection{Allowing for multiple scatterings between pairs of particles}

Instead of assuming that there is at most one single exchange between each pairs
of partons, one may allow for any number of exchanges. Then the definition
of $S$ is modified as follows:
\begin{equation}
\langle S\rangle=\left\langle e^{-\frac{nm}{N}}\right\rangle
=\sum_{n,m\geq 1}P(n,t^\prime) P(m,t-t^\prime) e^{-\frac{mn}{N}}.
\end{equation}
One sees immediately that if the probabilities $P$
satisfy the master equation~(\ref{master0d}), then this expression
cannot be boost-invariant (i.e. independent of $t^\prime$).
Indeed, if Eq.~(\ref{master0d}) holds, then
\begin{equation}
P(n,t\rightarrow\infty)=\delta_{n,N}\ \text{and}\ 
P(n,t=0)=\delta_{n,1}.
\end{equation}
It follows that in the frame in which
the projectile is at rest,
\begin{equation}
\langle S\rangle_{t^\prime=0,t\rightarrow\infty}=e^{-1}
\end{equation}
while in the center-of-mass frame (if the projectile and the target
share an equal fraction of the evolution),
\begin{equation}
\langle S\rangle_{t^\prime=\frac{t}{2},t\rightarrow\infty}=e^{-{N}}
\end{equation}
which is very different.
Actually, in this model, the average 
number of particles cannot saturate to a fixed
value $N$.
It would not be compatible with boost invariance.

In order to preserve boost-invariance, one has to modify the master equation.
We may write most generally
\begin{equation}
\dot P_n=\sum_{k\ne 0}(\alpha^k_{n-k} P_{n-k}-\alpha^k_{n}P_n).
\label{eq:mastermodif}
\end{equation}
The coefficients $\alpha^k_n$ are the transition rates from
a $(n-k)$-particle state to a $n$-particle state.
We determine the $\alpha^k_n$ from the
boost-invariance requirement. 
Actually, only one coefficient $\alpha^{k=1}_n$
is needed in the case of this model.

Using the same method as the one employed for checking the boost invariance
in the previous model, we write
\begin{equation}
\frac{d\langle S\rangle}{dt^\prime}
=\sum_{n,m}(\dot P_n P_m-P_n \dot P_m)\left\langle e^{-\frac{mn}{N}}\right\rangle,
\end{equation}
and express $\dot P_n, \dot P_m$ with the help 
of the master equation~(\ref{eq:mastermodif}).
Requiring that the sum over $n$ and $m$ vanishes
leads to the rates
\begin{equation}
\alpha^1_{n}=N\left(1-e^{-n/N}\right),
\end{equation}
where the overall constant is determined from the rate
in the unsaturated version of the model, which should hold for
values of $n\ll N$.
This model was first proposed by Mueller and Salam \cite{Mueller:1996te}.

We see that the saturation mechanism is quite different than in the previous model.
Indeed, the average number of particles in the system keeps growing, but at a rate
that slows down and depends on the number of particles in the system itself.
Unitarity of the scattering probability $T$ is ensured first 
by multiple scatterings rather than 
by the saturation of the number of particles to a constant number $N$ 
(up to statistical fluctuations).

This model was studied in detail in Ref.~\cite{Blaizot:2006wp}. The conclusions
drawn in there is that the saturation mechanism implied by the above model
is likely to be quite close to the one at work in QCD.
We could get analytical results for this model using one of the methods
presented above.
In particular, the statistical method outlined in Sec.~\ref{sec:statisticalmethods}
would apply and lead in a straightforward way to the expression for
$\langle n\rangle$, up to corrections of relative order $1/N$.

\subsubsection{Reggeon field theory}

Starting from the field theory formulation in Sec.~\ref{sec:pomeronfieldtheory}, 
we may discard the 4-Pomeron vertex (term $(b^\dagger)^2b^2/N$ in Eq.~(\ref{eq:hamiltonianpft})).
The new Hamiltonian then reads
\begin{equation}
{\cal H}^{RFT}=-b^\dagger b-(b^\dagger)^2 b+\frac{1}{N}b^\dagger b^2.
\label{hamiltonianRFT}
\end{equation}
The stochastic formulation reads
\begin{equation}
\frac{dz_t}{dt}=z_t-\frac{z_t^2}{N}+\sqrt{2z_t}\,\nu_{t+dt}
\end{equation}
(Compare to Eq.~(\ref{eq:stochastic0d}).)
This is the zero-dimensional version of the stochastic equation
defining the so-called Reggeon field theory, which was intensely 
studied in the 70's as a pre-QCD model for hadronic interactions.

This model has peculiar properties if one insists on interpreting it
as a particle model. Indeed,
the Hamiltonian~(\ref{hamiltonianRFT}) corresponds to 
a generating function for the factorial moments of the number
$n$ of particles in the system at a given time $t$
that satisfies the partial differential equation
\be
\frac{\partial Z(z,t)}{\partial t}=z(1+z)\frac{\partial Z(z,t)}{\partial z}
-\frac{z}{N}\frac{\partial^2 Z(z,t)}{\partial z^2}
\ee
and the corresponding master equation, obeyed by
the probability $P(n,t)$ to find
$n$ particles in the system at time $t$, writes
\begin{multline}
\frac{\partial P(n,t)}{\partial t}=-n P(n,t)+(n-1)P(n-1,t)\\
+\frac{1}{N}(n+1)(n+2)P(n+2,t)
-\frac{1}{N}n(n+1)P(n+1,t).
\end{multline}
One can read off this equation the rates for particle creation/disappearance.
One has a $1\rightarrow 2$ splitting, with rate $dt$; 
a $2\rightarrow 0$ annihilation
with rate
$dt/N$; and a $2\rightarrow 1$ recombination with rate $-dt/N$.
This is a negative number, and of course, it is unacceptable for a physical 
probability not to take its values between $0$ and $1$.
But we should not reject {\em a priori} negative probabilities as a formal calculation
tool, as long as the physical probabilities
are well-defined.
However, a Monte-Carlo code based on these negative rates turns out
to be extremely unstable, and thus of no practical use.

Note that the statistical approach teaches us that
in the $N\gg 1$ limit, the moments of the number of particles
in the system should not be 
very different than for the model with 3 and 4-Pomeron vertices, 
since it is essentially the form of the fluctuations
in the dilute regime that determine the moments at all times.

A detailed study of the special properties of this model as well
as a comparison with reaction-diffusion-like models may be found 
in Ref.~\cite{Bondarenko:2006rh}.


\section{
\label{sec:reviewtraveling}
Review of general results on stochastic traveling-wave equations}

In Sec.~\ref{sec:schannel}, we have shown the relevance of the stochastic
FKPP equation for high-energy QCD. The latter represents (classical)
particle models that undergo a branching-diffusion process in one dimension,
supplemented by a saturation mechanism.
Sec.~\ref{sec:zerodimensional} was dedicated to a detailed study,
from different points of view, of simplified models obtained from the
former ones by switching off diffusion.
We now go back to the study of one-dimensional models.
We proceed by steps: First, we shall address the deterministic FKPP equation
(which is equivalent to the BK equation in QCD)
(Sec.~\ref{sec:deterministicFKPP}). 
Second, we shall introduce fluctuations
to get solutions for equations 
in the universality class of the sFKPP equation
(Sec.~\ref{sec:combining} and~\ref{sec:beyond}).

\subsection{\label{sec:deterministicFKPP}Deterministic case: the FKPP equation}

We address the simplest reaction-diffusion equation, namely
the FKPP equation
\begin{equation}
\partial_t u=\partial_x^2 u+u-u^2.
\label{eq:FKPP}
\end{equation}
This equation 
was found to describe scattering in QCD under some assumptions,
see Sec.~\ref{sec:schannel}.

It is a mathematical theorem \cite{Bramson} that this equation admits 
{\em traveling waves} as solutions, that is to say, 
soliton-like solutions such
that
\begin{equation}
u(t,x)=u(x-vt)
\label{eq:travelsolution}
\end{equation}
where $v$ is the velocity of the wave.
$u$ is a front that smoothly connects 1 (for $x\rightarrow -\infty$)
to 0 (for $x\rightarrow +\infty$).
The velocities of the traveling waves and their shapes for large $x$
are also known mathematically. 
Starting with some given initial condition which itself
is not necessarily a traveling wave such as Eq.~(\ref{eq:travelsolution}), 
the solution converges at large times to a stationary wave front.
The front velocity
during this phase may also be predicted asymptotically.
We informally review these results in this section.

\subsubsection{\label{sec:generalanalysis}General analysis and wave velocity}

The FKPP equation~(\ref{eq:FKPP}) encodes 
a diffusion in space (through the
term $\partial_x^2 u$ in the right handside), 
a growth (term $u$), and a saturation of this growth
(term $-u^2$). It admits two fixed points:
the constant functions $u(t,x)=0$ and $u(t,x)=1$.
A linear stability analysis shows that $0$ is unstable, while $1$ is stable.
Indeed, thanks to the growth term $u$ in the right handside, a small perturbation
$u(t,x)=\varepsilon\ll 1$
grows exponentially with time. On the other hand, a perturbation near 1 
of the form $u(t,x)=1-\varepsilon$ goes back to the fixed point $1$ through evolution.
Hence the FKPP equation describes the transition from an
unstable to a stable state.
Therefore, we expect that the linear part of the equation drives the
motion of the traveling wave, since the role of the nonlinear
term is just to stabilize the fixed point $u=1$.

We shall cast the linear part of the equation in a more general
form:
\begin{equation}
\partial_t u(t,x)=\omega(-\partial_x)u(t,x),
\label{eq:linear}
\end{equation}
where $\omega(-\partial_x)$ is a branching diffusion kernel.
It may be an integral or differential operator.
An appropriate kernel is, in practice, an operator
such that the ``phase velocity'' $v(\gamma)=\omega(\gamma)/\gamma$ (see below)
has a minimum in its domain of analyticity.
The FKPP equation corresponds to the choice 
$\omega(-\partial_x)=\partial_x^2+1$.

Let us follow the wave front in the vicinity of a specific value of $u$.
To this aim, we define a new coordinate $x_{\text{WF}}$
such that
\begin{equation}
x=x_{\text{WF}}+vt.
\end{equation}
The solution of the linearized equation~(\ref{eq:linear})
writes most generally
\begin{equation}
u(t,x)=\int_{\mathcal{C}} \frac{d\gamma}{2i\pi}u_0(\gamma)
\exp\left(-\gamma(x_{\text{WF}}+vt)+\omega(\gamma)t
\right),
\label{eq:sollinear}
\end{equation}
where $\omega(\gamma)$ is the Mellin transform of the linear kernel
$\omega(-\partial_x)$ 
(and thus $\gamma$ corresponds to $-\partial_x$),
and defines the dispersion relation of the
linearized equation. $u_0(\gamma)$ is the Mellin transform of the initial
condition $u(t=0,x)$.
Let us assume that the initial condition is a function
smoothly connecting 1 at $x=-\infty$ to 0 at $x=+\infty$, with
asymptotic decay of the form $u(t=0,x)\sim e^{-\gamma_0 x}$.
Then $u_0(\gamma)$ has singularities on the real negative axis, and on
the positive axis starting from $\gamma=\gamma_0$ and extending towards $+\infty$.
Let us take a concrete example: If $u(0,x\leq 0)=1$ and $u(0,x>0)=e^{-\gamma_0 x}$,
then $u_0(\gamma)=1/\gamma+1/(\gamma_0-\gamma)$.
The integration contour $\mathcal{C}$ should go parallel to the imaginary
axis in the complex $\gamma$-plane and cross
the interval $[0,\gamma_0]$.

Each partial wave of wave number $\gamma$ has a phase velocity
\be
v_\phi(\gamma)=\frac{\omega(\gamma)}{\gamma},
\ee
whose expression is found by imposing that the exponential 
factor in the integrand of Eq.~(\ref{eq:sollinear})
be time-independent for $v=v_\phi(\gamma)$.

\begin{figure}
\begin{center}
\epsfig{file=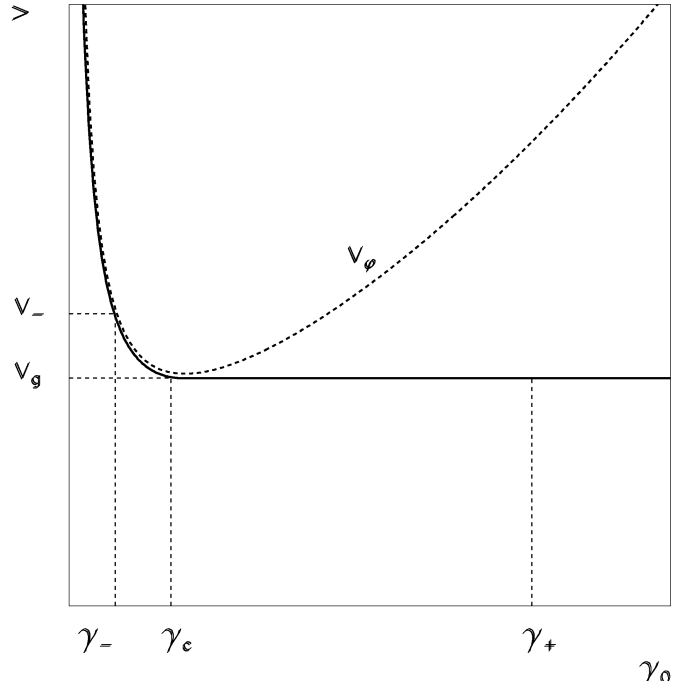,width=10cm}
\end{center}
\caption{\label{fig:velocitytw.eps}
Front velocity as a function of its asymptotic
decay rate $\gamma_0$ (dashed curve). 
It has a minimum at $\gamma=\gamma_c$. 
The full line represents the actual velocity 
that would be selected starting with
an initial condition decaying as $e^{-\gamma_0 x}$ for large $x$.
If $\gamma_0=\gamma_-<\gamma_c$ 
(initial condition less steep than $\gamma_c$), then
the asymptotic velocity is the phase velocity of a front 
which has the same asymptotics as the initial condition.
For any $\gamma_0=\gamma_+>\gamma_c$, the velocity of the front is the minimum 
of the phase velocity $v_\phi(\gamma)$.
}
\end{figure}

We are interested in the large-time behavior of $u(t,x)$.
The integrand in Eq.~(\ref{eq:sollinear}) admits a saddle point
at a value $\gamma_c$ of the integration variable
such that
\begin{equation}
\omega^\prime(\gamma_c)=v,
\end{equation}
that is to say, when $v$ coincides with the group velocity of the wave packet.
But the large-time solution is not necessarily given by the saddle point:
This depends on the initial condition $u_0(\gamma)$.
In order to understand this point, let us
 work out in detail the simple example of initial condition
quoted above.
The integral has two contributions for large $t$:
\begin{equation}
u(t,x)=e^{-\gamma_0(x_{\text{WF}}+vt)+\omega(\gamma_0)t}+
\kappa
e^{-\gamma_c(x_{\text{WF}}+vt)+\omega(\gamma_c)t},
\label{eq:solsaddle}
\end{equation}
up to a relative $\mathcal{O}(1)$ factor $\kappa$.
The time invariance of $u(t,x)$ in the frame of the wave
may only be achieved by tuning $v$
to one of the following two values:
\be
\begin{split}
\mathit{(i)}\ \ & v_0=\frac{\omega(\gamma_0)}{\gamma_0}\\
\mathit{(ii)}\ \ & v_c=\frac{\omega(\gamma_c)}{\gamma_c}=\omega^\prime(\gamma_c)
\end{split}
\ee
In the second case, $v$ coincides with 
the minimum of the phase velocity $\omega(\gamma)/\gamma$ and in particular,
$v_c\leq v_0$.
The relevant value of $v$ depends on the shape of the initial condition:
\begin{itemize}
\item If $\gamma_0<\gamma_c$, i.e. the decay of the initial
condition is less steep than the decay of the wave from the saddle-point,
then one has to pick the first choice {\it (i)} for the velocity.
Indeed, this is the only one for which the first term in Eq.~(\ref{eq:solsaddle})
is time-independent, and the second term vanishes at large time.
Due to the fact that $v_c<v_0$, choice {\it (ii)} would make the first term
in Eq.~(\ref{eq:solsaddle}) blow up exponentially, $u\sim e^{\gamma_0 (v_0-v_c)t}$.

\item If instead $\gamma_0>\gamma_c$, then it is the second
choice {\it (ii)} that has to be made. The saddle point dominates,
and the wave velocity at large time is independent of the initial condition.

\end{itemize}
Fig.~\ref{fig:velocitytw.eps} summarizes these two cases.

The limiting case $\gamma_0=\gamma_c$ requires a special treatment.
Since it is not relevant for the physics of QCD traveling waves
(only the case $\gamma_0>\gamma_c$ is actually relevant),
we refer the interested reader to the review paper of 
Ref.~\cite{vansaarloos-2003-386}
for a complete treatment also of that case.

There exists a rigorous mathematical proof of these solutions
in the case of the straight FKPP equation \cite{Bramson}.
These results are largely confirmed in numerical simulations
for various other branching diffusion
kernels, including the ones of interest for QCD 
(see e.g.~\cite{Albacete:2005ef,Enberg:2005cb}, 
and Ref.~\cite{Levin:2001et,Armesto:2001fa,Albacete:2003iq} for
earlier simulations of the BK equation).

Actually, in QCD as well as in many problems in statistical physics,
the initial condition is localized or has a finite support, and hence,
its large-$x$ decay is always very steep.
Thus for the physical processes of interest in this review, the asymptotic
front velocity, that we will denote by $V_\infty$ for reasons that
will become clear later, reads
\be
V_\infty=v_c=\frac{\omega(\gamma_c)}{\gamma_c}=\omega^\prime(\gamma_c),
\ee
where the last equality defines $\gamma_c$.
Note that in the context of particle physics, this result was already known from 
the work of Gribov, Levin, Ryskin~\cite{Gribov:1984tu}, and was rederived later
in the framework of the BK equation 
\cite{GolecBiernat:2001if,Iancu:2002tr,Mueller:2002zm}.

\begin{figure}
\begin{center}
\epsfig{file=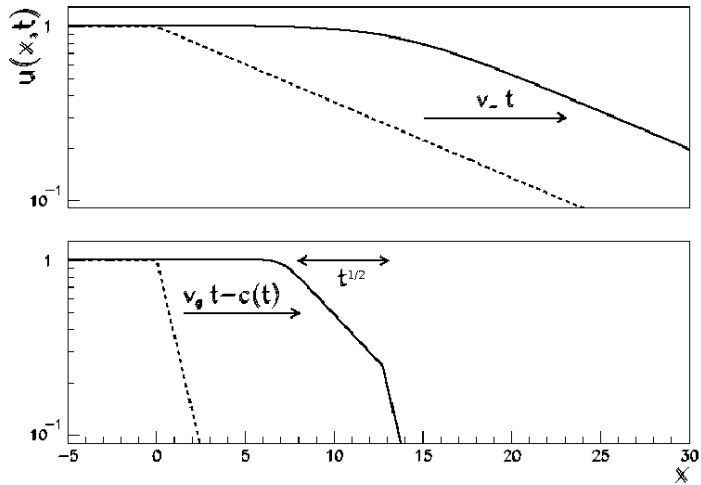,width=0.7\textwidth}
\end{center}
\caption{\label{fig:plot2.eps}
Sketch of the shape of the front according to the large-$x$
behavior of the
initial condition
$u(t=0,x)\sim e^{-\gamma_0 x}$.
{\em Top:} $\gamma_0<\gamma_c$. 
The asymptotic
shape of the initial condition is conserved. The relaxation of the front
is fast.
{\em Bottom:} $\gamma_0>\gamma_c$. The asymptotic shape of the front
is $e^{-\gamma_c x}$, and the velocity for $t=\infty$ is $v_c=\omega(\gamma_c)/\gamma_c$.
The asymptotic shape is reached over a distance $\sqrt{t}$ ahead of the front,
and the velocity at finite time is less than the asymptotic velocity
by $\frac{3}{2\gamma_c t}$.
}
\end{figure}

So far, we have discussed the asymptotic velocity of the solutions
to the FKPP equation as a function of the initial condition.
When the initial condition is steep enough, then the asymptotic front velocity
takes a fixed value which is the minimum of $\omega(\gamma)/\gamma$. 
In the opposite case, the shape of the initial condition is retained
(see Fig.~\ref{fig:plot2.eps}).
We wish to know more detailed properties of the wave front,
such as its shape and the way its velocity approaches
the asymptotic velocity.
There are several methods to arrive at this result.
At the level of principle, they all rely on a matching between a solution
near the fixed point $u=1$, and a solution of the linearized equation
which holds in the tail $u\ll 1$.

\subsubsection{Diffusion equation with a boundary}

We now come back to the original FKPP equation~(\ref{eq:FKPP}).
We have seen that the nonlinearity $-u^2$ has the effect of taming
the growth induced by the linear term $u$, when $u$ gets close to 1.
But nonlinear partial differential equations are
very difficult to address mathematically.
It may be much simpler to address the linear equation 
\begin{equation}
\partial_t u=\partial_x^2u+u
\label{eq:FKPPlinear}
\end{equation}
supplemented with an absorptive
(moving with time) boundary condition that ensures that $u(t,x)$ has
a maximum value of $1$ at any time.
We need to work out the solution of Eq.~(\ref{eq:FKPPlinear})
with this kind of boundary condition.
Here, we reformulate the approach proposed in the QCD context by Mueller
and Triantafyllopoulos \cite{Mueller:2002zm}
(see also Ref.~\cite{Triantafyllopoulos:2002nz} for an account of the next-to-leading order
BFKL kernel).

A solution to Eq.~(\ref{eq:FKPPlinear}) with initial condition
$u(t=0,x)=\delta(x-x_0)$ is given, for positive times, by
\begin{equation}
u(t,x)=\frac{1}{\sqrt{4\pi t}}\exp\left(t-\frac{(x-x_0)^2}{4t}\right).
\label{eq:u0}
\end{equation}
This solution holds if the boundary condition is at spatial infinity.
The solution of the pure diffusion equation, without the growth term,
is of course nothing but $u(t,x)e^{-t}$.
We shall denote it by $u^{\text{PD}}(t,x)$.

\begin{figure}
\begin{center}
\epsfig{file=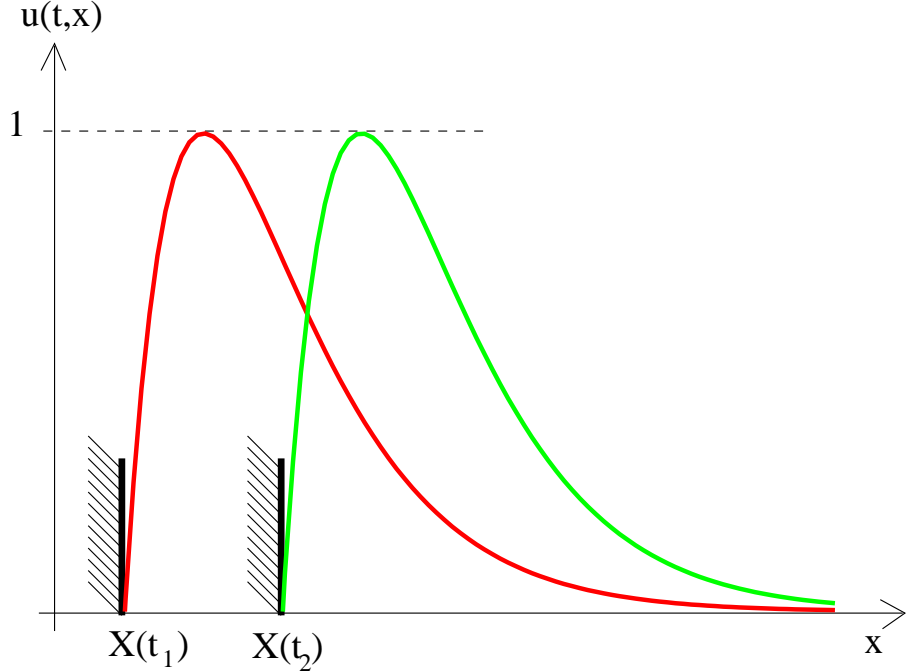,width=12cm}
\end{center}
\caption{\label{fig:cutoff2}Shape of the solution of 
the branching diffusion equation~(\ref{eq:FKPPlinear}) with
a moving cutoff, whose position is adjusted in such a way that the 
maximum of $u(t,x)$ be 1 at all times. The solution is represented at 
two different times $t_1$ and $t_2$, 
showing the soliton-like behavior of the solution.}
\end{figure}

If instead of the boundary condition at infinity 
there is an absorptive barrier
at say $x=X$, i.e. if $u(t,x=X)=0$ for any $t$, then a solution
may be found through a linear combination of the latter solution
with different initial conditions, in such a way as the sum vanishes at $x=X$.
This is known as the method of images.
It is based on the observation that any linear combination
of Eq.~(\ref{eq:u0}) also solves Eq.~(\ref{eq:FKPPlinear}).
From the solution with initial condition $\delta(x-x_0)$,
we subtract the solution of the same equation but 
with initial condition $\delta(x-(2X-x_0))$, in such a way that the solution
vanishes for $x=X$, at any time.
We get
\begin{equation}
u_X(t,x)=\frac{e^t}{\sqrt{4\pi t}}\left(
e^{-\frac{(x-x_0)^2}{4t}}-e^{-\frac{(x-2X+x_0)^2}{4t}}
\right)
\end{equation}
We do not expect the solution to this problem to represent accurately
the solution to the full FKPP equation near the boundary $x\sim X$.
So the region of interest will be ahead of the boundary by a few units, while the
starting point $x_0$ of the evolution is at some finite distance of the boundary:
\begin{equation}
x-X\gg 1\ \ \text{and}\ \ x_0-X\sim 1.
\end{equation}
One may then expand the two Gaussian terms:
\begin{equation}
u_X(t,x)=\frac{x_0-X}{\sqrt{4\pi}}\frac{x-X}{t^{3/2}}
\exp\left(t-\frac{(x-X)^2}{4t}\right).
\end{equation}
The solution to the simple diffusion equation without the growth term, namely 
$\partial_t u=\partial^2_x u$,
is the one that we will actually use in the following.
It would again be the latter solution scaled by $e^{-t}$,
namely
\begin{equation}
u_X^{\text{PD}}(t,x)=\frac{x_0-X}{\sqrt{4\pi}}\frac{x-X}{t^{3/2}}
\exp\left({-\frac{(x-X)^2}{4t}}\right),
\label{eq:uXPD}
\end{equation}
where the superscript PD stands for ``pure diffusion''.
Note that in this equation, $X$ does not depend on time.
We cannot implement in a straightforward way
a time-dependent absorptive boundary. We will
get to such a solution by successive iterations: The main trick
is to go to a frame in which the solution of the branching diffusion
with a boundary is stationary for large times.

Let us start from the solution $u_0$ in Eq.~(\ref{eq:u0}).
The lines $x$ of constant $u_0(t,x)=C$ (without a boundary) are obviously given by
\be
x=x_0+2t-\frac12\log t-\log(C\sqrt{4\pi})
+\text{terms vanishing for $t\rightarrow\infty$}.
\label{eq:cstugaussian}
\end{equation}
(We have selected the rightmost solution $x>x_0$).
Let us change frame by writing $x=x_1+x_0+2t$. Then in this new variable,
$u(t,x)$ in Eq.~(\ref{eq:u0}) reads
\begin{equation}
u(t,x)=e^{-x_1}\left[\frac{e^{\frac{-x_1^2}{4t}}}{\sqrt{4\pi t}}\right]
=e^{-x_1}u^{\text{PD}}(t,x_1),
\label{eq:frontx1}
\end{equation}
where we have factored out the solution of the pure diffusion equation, but
this time, in the moving frame defined by the coordinate $x_1$.
We may implement an absorptive boundary condition, fixed in this new frame, by
replacing $u^{\text{PD}}$ by $u^{\text{PD}}_X$ in Eq.~(\ref{eq:uXPD}). 
Note that $X$ is fixed with respect
to $x_2$, but in the original frame defined by coordinate $x$, it is of order $2t$.
The solution has lines of constant $u$ which solve
\be
x_1=-\frac32\log t-\frac{(x_1-X)^2}{4t}+\log(x_1-X)
-\log\frac{C\sqrt{4\pi}}{x_0-X}.
\ee
The two last terms are subdominant because according to Eq.~(\ref{eq:cstugaussian}),
$x_1-X\sim\log t$, and because $x_0-X$ is a constant.
We further define a new frame
\be
x_2=x_1+\frac32\log t-\log\frac{C\sqrt{4\pi}}{x_0-X}
=x-x_0-2t+\frac32\log t-\log\frac{C\sqrt{4\pi}}{x_0-X}
\label{eq:x2}
\ee
Going back to the expression for $u_0(t,x)$ (see Eq.~(\ref{eq:u0})), we
substitute $x$ by its expression as a function of $x_2$ and get
\begin{multline}
u(t,x)=t^{3/2}e^{-x_2}\frac{C\sqrt{4\pi}}{x_0-X}
\left[\frac{e^{-\frac{x_2^2}{4t}}}{\sqrt{4\pi t}}\right]\\
\times
\exp\left(-\frac{\left(\frac32\log t+\log\frac{C\sqrt{4\pi}}{x_0-X}\right)^2}{4t}
+x_2\frac{\frac32\log t+\log\frac{C\sqrt{4\pi}}{x_0-X}}{2t}\right)
\label{eq:frontx2}
\end{multline}
We replace the expression inside the squared brackets, which is nothing but
$u^{\text{PD}}$, by the solution with a boundary $u_X^{\text{PD}}$.
We check that the value of $x_2$ for which $u(t,x_2)$ is constant
is now a mere constant for large $t$.
Going back to the original frame, we get
\begin{equation}
u(t,x)=Ce^{-X} (x-X(t))e^{-(x-X(t))}\exp\left(-\frac{(x-X(t))^2}{4t}\right),
\label{eq:front0}
\end{equation}
where 
\begin{equation}
X(t)=2t-\frac32\log t+\mathcal{O}(1)
\label{eq:posfront0}
\end{equation}
is the position of the absorptive boundary for large times, 
and thus, the position of the front. The constant 
$X$ is the position of the front in the moving frame.
Setting $X=-1$ and $C=1$, the maximum of $u$ is reached at $x=X(t)$, and 
is indeed equal to 1.

For large $t$ or in the region $x-X(t)\leq \sqrt{t}$ 
which expands with time, 
the Gaussian factor goes to 1, and we see that
$u(t,x)$ only depends on one single variable $x-X(t)$.
This was expected: It is precisely the defining property of traveling waves. 
But in addition
to these asymptotic solutions, we get from this calculation 
the first finite-$t$ correction to the front shape and front velocity.

Actually, the speed of the front is intimately related
to its shape.
At time $t$, it has reached its asymptotic shape
over the distance $\sqrt{t}$
from the saturation point.
This remark will be important in the following.

We have derived the solution of a problem
that was not exactly the initial one,
however, we believe that the shape of the front
in its forward part ($u\ll 1$)
as well as its velocity are quite universal.
Indeed, physically, these properties are completely 
derived from the linear part of the equation.
For this reason, the front is said to be ``pulled''
by its tail.
The nonlinearity only tames the growth of $u$ near $u\sim 1$, 
and so its precise
form should not influence the front position itself, at least
at large enough times.
Thus we expect these solutions to have a broad validity,
only depending on the diffusion kernel, and so,
may be obtainable from our calculation up to
the replacement of the relevant parameters.
For the more general branching diffusion kernel
in Eq.~(\ref{eq:linear}), the velocity of the front would read
\begin{equation}
\boxed{
\frac{dX}{dt}=\frac{\omega(\gamma_c)}{\gamma_c}-\frac{3}{2\gamma_ct}+\cdots
}
\label{eq:velocitygeneral0}
\end{equation}
where $\gamma_c$ solves $\omega(\gamma_c)=\gamma_c\omega^\prime(\gamma_c)$,
as was explained in Sec.~\ref{sec:generalanalysis}.
The front shape in its forward part $x-X(t)\gg 1$ reads
\be
\boxed{
u(t,x)=(x-X(t))e^{-\gamma_c(x-X(t))}
\exp\left(-\frac{(x-X(t))^2}{2\omega^{\prime\prime}(\gamma_c) t}\right),
}
\label{eq:frontgeneral0}
\ee
up to an overall constant.
Fig.~\ref{fig:cutoff2} represents a sketch of the solution 
at two different times.
Note that the asymptotic shape is an exponential decay,
\begin{equation}
u(t,x)\sim e^{-\gamma_c(x-X(t))}.
\label{eq:asymptoticdecay}
\end{equation}
From Eq.~(\ref{eq:frontgeneral0}), this shape extends over a range
\be
L=x-X(t)\sim \sqrt{2\omega^{\prime\prime}(\gamma_c)t}.
\ee
In other words, the time needed for the front to reach its asymptotic
shape over a range $L$ reads
\be
t\sim\frac{L^2}{2\omega^{\prime\prime}(\gamma_c)}.
\label{eq:asymptotictime}
\ee

Through our simple calculation,
we got the lowest order in an expansion of the front shape
and position at large times.
The next corrections to $X(t)$
would be of order 1 (this constant depends
on the way we define the position of the front), 
followed by an algebraic series in $t$ whose
terms all vanish at large $t$.
The first next-to-leading term in the series has been computed 
(see Ref.~\cite{ebert-2000-146}): It turns out to be of
order $1/\sqrt{t}$.
We will not reproduce the calculations that lead to it
because they are rather technical and
there is already a comprehensive review paper available on
the topic \cite{vansaarloos-2003-386}.
But let us write the result for the position and the shape of the front
at that level of accuracy, for the more general branching diffusion
kernel given by Eq.~(\ref{eq:linear}).
To that accuracy, the front position reads \cite{ebert-2000-146}
\begin{equation}
X(t)
=\frac{\omega(\gamma_c)}{\gamma_c} t-\frac{3}{2\gamma_c}\log t
-\frac{3}{\gamma_c^2}
\sqrt{\frac{2\pi}{\omega^{\prime\prime}(\gamma_c)}}\frac{1}{\sqrt{t}}
+{\cal O}(1/t),
\end{equation}
For the simple FKPP case, we recall that $\omega(-\partial_x)=\partial_x^2+1$,
then $\gamma_c=1$ and $\omega(\gamma_c)=2$.
The first two terms in the last equations match the ones found in 
Eq.~(\ref{eq:velocitygeneral0}).
The shape of the front in its forward part has the following form:
\begin{multline}
u(t,x)=C_1 e^{-\gamma_c(x-X(t))}\exp\left({-z^2}\right)
\times\\
\Bigg\{\gamma_c(x-X(t))]+C_2
+\left(3-2C_2+\frac{\gamma_c\omega^{(3)}(\gamma_c)}{\omega^{\prime\prime}(\gamma_c)}
\right)
z^2\\
-\left({\frac23}\frac{\gamma_c\omega^{(3)}(\gamma_c)}{\omega^{\prime\prime}(\gamma_c
)}
+
\frac13{}_2\!F_2\left[1,1;{\scriptstyle\frac52},3;
z^2\right]\right)z^4\\
+6\sqrt{\pi}
\left(1-{}_1\!F_1\left[-{\scriptstyle \frac12},{\scriptstyle \frac32};
z^2
\right]\right)z+{\cal O}(1/\sqrt{t})
\Bigg\},
\label{eq:front1}
\end{multline}
where
\be
z=\frac{x-X(t)}{\sqrt{2 \omega^{\prime\prime}(\gamma_c) t}}
\ee
and ${}_2F_2$, ${}_1F_1$ are generalized hypergeometric functions.
The terms in the first line match with the
result of our calculation (Eq.~(\ref{eq:front0})) for the relevant
value of $\gamma_c$.
These expressions should apply also to QCD, up to the relevant replacements
given in Tab.~\ref{tab:dictionary}.

So far, we have considered equations of the type of Eq.~(\ref{eq:FKPP})
as saturation equations, in the sense that they
describe the diffusive growth of a continuous function $u$ until
it is tamed for $u\sim 1$.
We will see below
that these equations 
may actually be given a different physical interpretation.

\subsubsection{\label{sec:discretebranchingdiffusion}Discrete branching diffusion}

We have investigated the solutions of the FKPP equation
in a mathematical way, without discussing the physics that may lead to such an equation.
The absorptive boundary that we have put
replaces the nonlinear term in the FKPP equation, whose role is to
make sure that $u$ never exceeds the limit $u=1$.
Hence we have thought of this boundary as a way to enforce the {\em saturation}
of some density of particles.
Actually, the FKPP equation~(\ref{eq:FKPP})
may stem from a branching diffusion 
process in which the number of particles is unlimited, and
thus, for which there is no saturation at all.
As a matter of fact, this is 
the way how the BK equation is built in QCD: An exponentially growing 
number of dipoles,
stemming from the rapidity evolution of a hadronic probe, scatters off some
target. The overall interaction probability is unitary because
multiple scatterings are allowed (the interaction probability of $n$ dipoles
is actually of the form $e^{-\alpha_s^2 n}$), but not because
there is a saturation of the number of dipoles in the wavefunction
of the probe. We refer the reader to Fig.~\ref{fig:bk} for a picture
of the process.

\begin{figure}
\begin{center}
\epsfig{file=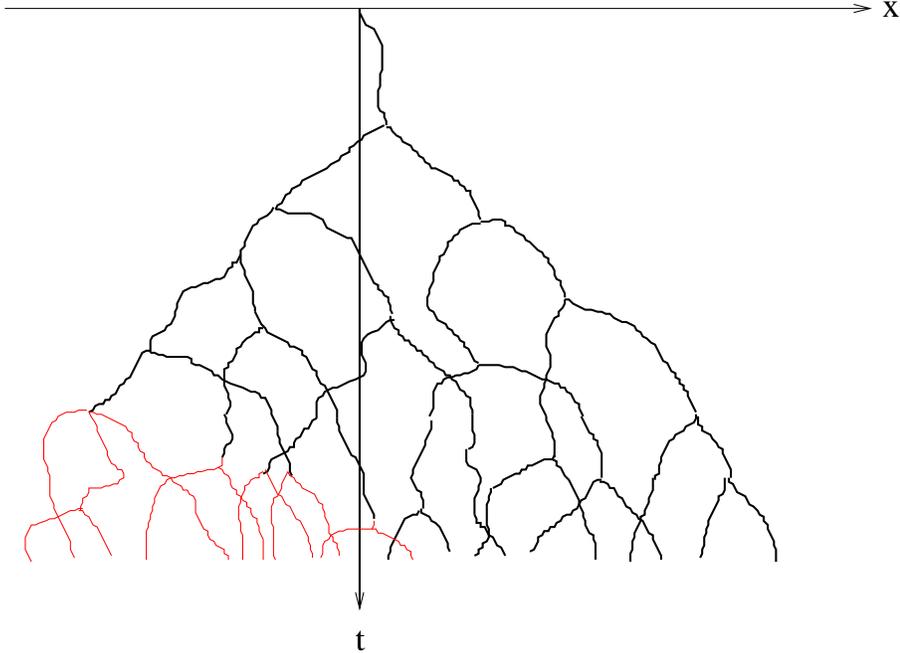,width=12cm}
\end{center}
\caption{\label{fig:branchingdiff}Example of branching diffusion process
on a line (see the text for a mathematical description of the evolution rules).
If the number of individuals is limited by a selection process which,
at each new branching, eliminates the individual  
sitting at the smallest $x$ as soon
as the total number of individuals reaches say $N$ ($N=10$ in this figure),
then only the branches drawn in thick line survive.
}
\end{figure}

To illustrate how the FKPP equation
arises in such a simple model of branching diffusion,
let us consider a set of particles
on a line, each of them being indexed by a continuous variable $x$.
(Such a model was considered for instance in Ref.~\cite{DerridaSpohn}).
We let the system evolve according to the following rules.
During the time interval $dt$, each particle has a probability $dt$
to split in 2 particles. Unless it splits, it moves
of the small random amount $\delta x$, which is a Gaussian
random variable distributed like
\begin{equation}
p(\delta x)=\frac{1}{\sqrt{4\pi dt}}
\exp\left(-\frac{(\delta x)^2}{4 dt}\right).
\label{eq:pgauss}
\end{equation}
Let us consider the number of particles $n(t,x)$ contained in an interval
of given size $\Delta x$ centered around the coordinate $x$. At time $t=0$, the
system is supposed to consist in a single particle sitting at the origin $x=0$.
A sketch of a realization of this model is shown in Fig.~\ref{fig:branchingdiff}.
From the evolution rules, 
we easily get an equation for the average number of particles $\langle n\rangle$:
\begin{equation}
\langle n(t+dt,x)\rangle=dt\,2\langle n\rangle
+(1-dt)\int d(\delta x)p(\delta x)\langle n(t,x-\delta x)\rangle
\end{equation}
which reads, after replacing $p$ by Eq.~(\ref{eq:pgauss}) and after
the limit $dt\rightarrow 0$ has been taken,
\begin{equation}
\frac{\partial\langle n\rangle}{\partial t}=\langle n\rangle
+\frac{\partial^2 \langle n\rangle}{\partial x^2}.
\end{equation}
All the dependence on the size $\Delta x$ 
of the ``bin''
is contained in the initial condition.
It is clear that for large enough times,
the solution to this equation is given by Eq.~(\ref{eq:u0}),
and thus the lines of constant $\langle n\rangle$ 
are given by Eq.~(\ref{eq:cstugaussian}).

Let us now define
\begin{equation}
S(t,x)=e^{-n(t,x)/N}
\end{equation}
where $N$ is some (large) constant.
This definition is reminiscent of the $S$-function, related
to the scattering amplitude, introduced in the discussion
of the BK equation in Sec.~\ref{sec:schannel}.
For large enough $x$, $n(t,x)\ll N$ and thus 
$1-S(t,x)\simeq n(t,x)/N\rightarrow 0$.
For any $x$, 
the exponential makes sure that $S$ ranges between $0$ and $1$.
Thus $S$ (or $1-S$) has the shape of a traveling wave.
Its position $X(t)$ is the value of $x$ for
which $n(t,x)$ is some given constant say of the order of $N$.
Hence, up to fluctuations,
it is given by Eq.~(\ref{eq:cstugaussian}).

On the other hand however, the average of $S$ over events, namely $A=1-\langle S\rangle$
obeys the FKPP equation.
Indeed
\begin{equation}
\langle S(t+dt,x)\rangle=
dt\langle S(t,x)\rangle^2
+(1-dt)\int d(\delta x)p(\delta x)\langle S(t,x-\delta x)\rangle.
\end{equation}
In the limit $dt\rightarrow 0$ and rewriting the equation for $A$, we get
\begin{equation}
\frac{\partial A}{\partial t}=\frac{\partial^2 A}{\partial x^2}+A-A^2.
\end{equation}
Hence $A$ is a traveling wave at large times, and its position $X(t)$
is given by Eq.~(\ref{eq:posfront0}).
It is obviously behind by a term $\log t$ with respect to
the value of $x$ for which the average number of particles
has a given constant value.
Furthermore, 
the probability distribution of the
position of the rightmost particle (or of the $k$-th rightmost particle
for any given $k$)
may also be derived from the FKPP equation.
It turns out that in any event, the average $x$ for which
$n(t,x)$ has a given value, say $n_0$, 
moves with the FKPP velocity
which can be read off from Eq.~(\ref{eq:posfront0}). This is much slower
than the velocity with which $X(t)$ defined in such a way that
$\langle n(t,X(t))\rangle=n_0$ moves.

All this may seem a bit paradoxical.
But actually, it is just related to the fact that 
$\langle e^{-n/N}\rangle$ cannot be
approximated by $e^{-\langle n\rangle/N}$.
We may understand it in the following way.
By taking the average of $n$, we have somewhat forgotten a fundamental
property of $n$: its {\em discreteness}. Indeed, it only takes integer
values, and in particular, the distribution of $n$ in a realization
has a finite support: At any time, there is a value of $x$
to the right of which there are no particles at all. $n$ obeys a stochastic equation.
This is not the case for $\langle n\rangle$, which just obeys
an ordinary branching diffusion equation.

\begin{figure}
\begin{center}
\epsfig{file=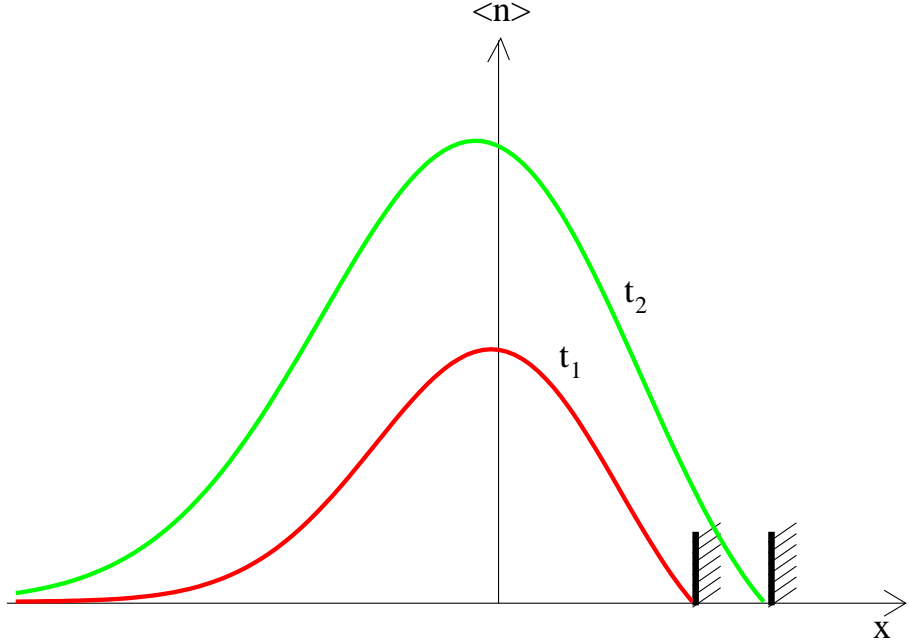,width=12cm}
\end{center}
\caption{\label{fig:cutoff1}Solution of the branching 
diffusion equation~(\ref{eq:absorpboundary})
with a moving  absorptive boundary that forces $\langle n\rangle$ to vanish at some
well-chosen point.
Two different times are represented.}
\end{figure}

In order to recover the effect of
the discreteness of $n$ and compute the velocity,
we may again use the absorptive boundary trick.
Let us solve the linear equation
\begin{equation}
\partial_t\langle n\rangle=\partial_x^2\langle n\rangle+\langle n\rangle
\label{eq:absorpboundary}
\end{equation}
with an absorptive boundary.
The absorptive boundary will be placed in such a way that at a distance of order one to
its left (we will focus on the right-moving wave), $\langle n\rangle=1$
(see Fig.~\ref{fig:cutoff1}).
There is no difference in principle with the boundary calculation that we have performed
before, except that the absorptive boundary is now placed to the right of the front (i.e.
$x_0<X$ in the notations used above).
Thus we find without any further calculation
that the realizations of $n$ move, on the average,
with the FKPP velocity~(\ref{eq:velocitygeneral0}).

\subsection{\label{sec:combining}Combining saturation and discreteness}

We have seen that physically, 
the KPP equation (or the BK equation in QCD) may be interpreted either
as an equation for the growth, diffusion and
saturation of a continuous function, or as
the evolution equation for the average of a bounded function
of a discrete (thus stochastic) branching diffusion process.
For each of these interpretations,
we may find the main features of the solutions by imposing one absorptive
boundary on the linear partial differential equation encoding branching diffusion.
In one case, the boundary is a cutoff that prevents $u$ to be larger than 1:
It represents saturation, i.e. the explicit nonlinearity present in the
FKPP equation. In the other case, the boundary forces 
the function $n$ that represents the number of particles
to vanish quickly when it becomes
less than 1. Formally, it actually models the intrinsic
discreteness of the number $n$ of particles, 
and avoids to address a stochastic
equation directly.

\begin{figure}
\begin{center}
\epsfig{file=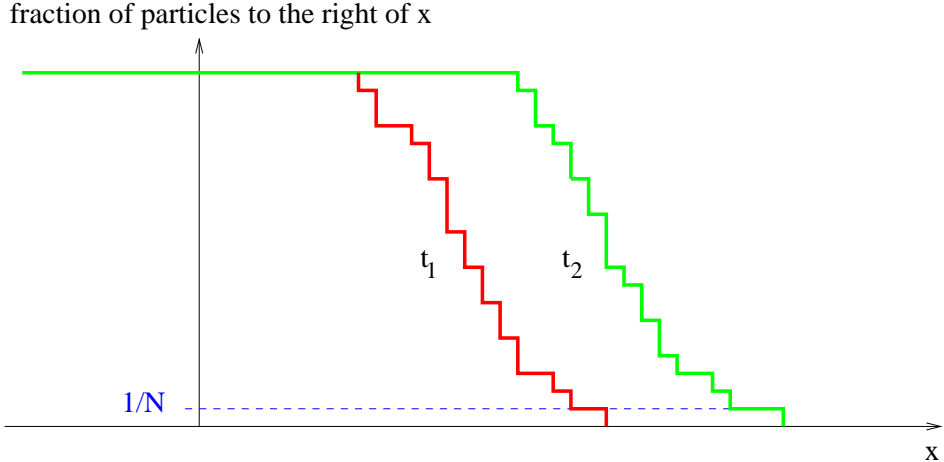,width=0.9\textwidth}
\end{center}
\caption{\label{fig:frontbranching}Branching diffusion model of 
Sec.~\ref{sec:discretebranchingdiffusion} with
selection that limits the total number of particles to $N$.
One sees that the fraction of particles to the right of $x$ looks like
a traveling wave front.
}
\end{figure}

In physical cases such as reaction-diffusion processes for finite $N$,
we define $u(t,x)$ as the number of particles per site 
(or per bin) in $x$ normalized to $N$.
Hence it takes discrete values: $1/N$, $2/N$ etc...
While for large $N$ discreteness is unlikely to play a role in the region $u\sim 1$,
it is expected to be crucial when $u\sim 1/N$.
It is thus natural to impose the two boundaries: one representing
saturation of the particle number, the other one discreteness of the same
quantity.
A model that these two cutoffs may represent is for example, the branching diffusion
model in Sec.~\ref{sec:discretebranchingdiffusion}, but in which the total
number of particles is limited to $N$ by keeping only the $N$ rightmost ones
at each new branching.
It is clear that the function ${\cal U}(t,x)$ defined to be the number of particles to
the right of some position $x$ normalized to the maximum number $N$
is, for large enough times, a front connecting 1 (for $x\rightarrow -\infty$) to 0 
(for $x\rightarrow +\infty$) (see Fig.~\ref{fig:frontbranching}).

Reaction-diffusion problems 
(described by nonlinear stochastic partial differential equations)
were interpreted  
as branching diffusion problems taking place between two absorptive boundaries
for the first time
by Brunet and Derrida in Ref.~\cite{brunet-1997-57} and, independently,
by Mueller and Shoshi, in the case of QCD in Ref.~\cite{Mueller:2004sea}.
Note however that the present interpretation of the cutoffs was only found
in Ref.~\cite{Iancu:2004es} in the context of the QCD parton model.
Mueller and Shoshi considered both cutoffs for reasons related to the
boost-invariance of the QCD amplitude.
The duality of the two boundaries, that is to say of the dense and dilute 
regimes of the traveling wave, was studied more deeply in 
Refs.~\cite{Kovner:2005jc,Kovner:2005en,Kovner:2005uw,Kovner:2005aq,Kovner:2007zu}.

Before moving on to the technical derivation of the shape and position
of the front in this case, let us
figure out what we expect to find.

Starting from the initial condition, the front builds up
and its velocity increases with $t$ (see Eq.~(\ref{eq:velocitygeneral0}))
until it reaches its asymptotic shape, which is a decreasing
exponential $e^{-\gamma_c(x-X(t))}$
that holds for all $x\gg X(t)$.
But if the front is made of discrete particles, then
it has a finite support, and the exponential shape 
may not extend to infinity to the right, since 
$u(t,x)$ has to be either larger than $1/N$, or zero.
It cannot take values that would be a fraction of $1/N$ in realizations,
and thus, we cannot accommodate the shape $e^{-\gamma_c(x-X(t))}$
for arbitrarily large values of $x$, since
it would mean authorizing arbitrarily small positive
values of $u(t,x)$.
From Eq.~(\ref{eq:asymptotictime}) and from the shape of
the asymptotic front~(\ref{eq:asymptoticdecay}), 
the exponential shape sets down to $u=1/N$ at time
\begin{equation}
t_\text{relax}=\frac{c}{2\omega^{\prime\prime}(\gamma_c)}\left(\frac{\log N}{\gamma_c}\right)^2.
\label{eq:relaxtime}
\end{equation}
Beyond, the front cannot develop any longer, and thus, its shape and
velocity remain fixed.
$t_\text{relax}$ is the time that is needed 
for the front to relax from any perturbation,
which is why we have put the subscript ``relax''.

From Eq.~(\ref{eq:velocitygeneral0}) evaluated at $t=t_\text{relax}$, we get 
the new asymptotic velocity, that takes into account the effects of discreteness,
in the form
\be
\frac{dX}{dt}=\frac{\omega(\gamma_c)}{\gamma_c}
-3c\frac{\gamma_c\omega^{\prime\prime}(\gamma_c)}{\log^2 N}.
\ee
The calculation of $c$ requires a proper account of the exact shape of the front.
We shall turn to this calculation now.

As announced, we are now going to solve the 
linear branching 
diffusion equation with two absorptive boundaries: one representing
saturation, the other one discreteness.
\begin{figure}
\begin{center}
\epsfig{file=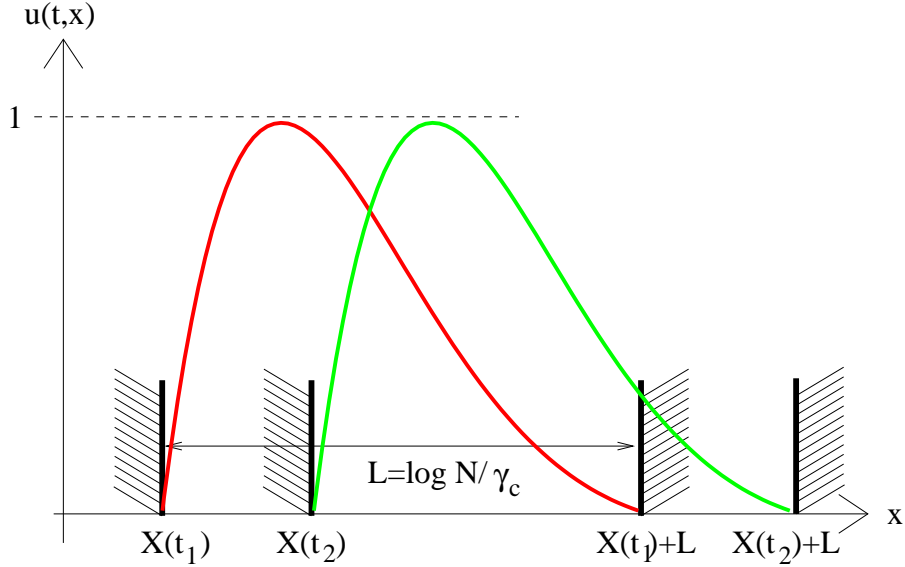,width=12cm}
\end{center}
\caption{\label{fig:cutoff3}Solution to the branching diffusion equation with two
boundaries.}
\end{figure}
First, as in the one-boundary case,
let us solve the simple diffusion equation $\partial_t u=\partial_x^2 u$ 
between two boundaries, at $X$ and $Y$ respectively, that is to say, with the conditions
$u(t,X)=u(t,Y)=0$.
The simplest method in this case is to take the ansatz
\be
u_{X,Y}^{\text{PD}}(t,x)=f(t)g(x).
\ee
Then the diffusion equation reads
\begin{equation}
\frac{f^\prime(t)}{f(t)}=\frac{g^{\prime\prime}(x)}{g(x)}=-\lambda
\end{equation}
$\lambda$ is necessarily a constant, being both a function of $t$ only and of $x$ only.
The equations for $f$ and for $g$ are easily solved. All in all, we get for $u$
\begin{equation}
u_{X,Y}^{\text{PD}}(t,x)=A e^{-\lambda t}\sin\left[\sqrt{\lambda}(x-X_0)\right],
\end{equation}
where $A$ and $X_0$ are constants which we will shortly determine
from the boundary condition.
Note that only positive values of $\lambda$ are physical, since negative ones
would lead to an exponential increase of the solutions.
The boundary conditions at $X$ and $Y$ fix $X_0$ to $X$ and lead to a quantization of $\lambda$:
\begin{equation}
\lambda=\frac{k^2\pi^2}{L^2},
\end{equation}
where $L=Y-X$ is the size of the wave front and $k$ is an integer.
The general solution is a sum of $u_{X,Y}^{\text{PD}}$ over all possible values of $k$, 
with coefficients fixed by the initial condition.
But at large time, thanks to the exponential decay 
of $f(t)$ with $t$, only the mode $k=1$ survives.
The final solution thus reads
\begin{equation}
u_{X,X+L}^{\text{PD}}(t,x)=A \exp\left({-\frac{\pi^2 t}{L^2}}\right)
\sin\frac{\pi(x-X)}{L},
\end{equation}
where the constant $A$ is determined from 
the projection of the initial condition
on the fundamental mode of the ``cavity'' $[X,X+L]$.

We now need to determine the time dependence of $X$. It will follow
from the search of the frame in which the front is stationary in time.

The first step (determination of $x_1$, see Eq.~(\ref{eq:frontx1})) is the same as 
in the one-boundary case.
Starting from Eq.~(\ref{eq:frontx1}), 
we substitute $u^{\text{PD}}(t,x_1)$ with
$u_{X,X+L}^{\text{PD}}(t,x_1)$
and look for the lines of constant $u$.
This leads us to introduce 
\be
x_2=x_1+\frac{\pi^2}{L^2}t.
\ee
Going back to the original variables, we get, for $L$ large,
\be
u(t,x)=A e^{-X} e^{-(x-X(t))}\sin\frac{\pi}{L}(x-X(t)),
\label{eq:2boundfront}
\ee
where 
\be
X(t)=2t-\frac{\pi^2}{L^2}t.
\label{eq:2boundposition}
\ee
We are left with the determination of the size $L$ of the front.
Near the left boundary (at a distance of order 1), 
$u$ should be of order 1, while close to
the right boundary, it should approach $1/N$. We write
\be
u(t,X(t)+1)=1,\ \ u(t,X(t)+L-1)=\frac{1}{N}.
\ee
Then we see that $L=\log N$ and $A=\kappa L$, where $\kappa={\cal O}(1)$.
The position of the rightmost boundary ($Y$) is the point to the right of which
there are no particles in typical individual realizations.
We will denote it by $x_\text{tip}(t)$.

All in all, writing it for a more general diffusion equation
$\partial_t u=\omega(-\partial_x)u$, the final solution reads
\be
\boxed{
u(t,x)=\kappa\, e^{-\gamma_c(x-X(t))}L\sin\frac{\pi(x-X(t))}{L}
}
\ee
(see Fig.~\ref{fig:cutoff3})
where the size of the front is
\be
L=\frac{\log N}{\gamma_c}
\ee
and its velocity reads
\begin{equation}
\boxed{
V_{\text{BD}}\equiv\frac{dX}{dt}=
V_{\infty}
-\frac{\pi^2\omega^{\prime\prime}(\gamma_c)}
{2\gamma_c L^2}
=\frac{\omega(\gamma_c)}{\gamma_c}
-\frac{\pi^2\gamma_c\omega^{\prime\prime}(\gamma_c)}{2\log^2 N}
}.
\label{eq:velocityBD}
\end{equation}
The subscript BD stands for ``Brunet-Derrida''.
For $\omega(\gamma)=\gamma^2+1$, $\gamma_c=1$, 
$\omega(\gamma_c)=\omega^{\prime\prime}(\gamma_c)=2$
and we recover Eqs.~(\ref{eq:2boundfront}),(\ref{eq:2boundposition}).

\subsection{\label{sec:beyond}
Beyond the deterministic equations: Effect of the fluctuations}

So far, we have actually solved deterministic equations although we were
addressing a model with a discrete number of particles, that therefore
has necessarily fluctuations. Our procedure gave the leading effects.
We shall now incorporate more fluctuation effects,
in a phenomenological way.
(We shall essentially review Ref.~\cite{Brunet:2005bz}).

\subsubsection{Phenomenological model and analytical results}

The two-boundary procedure has led to the following result: 
The front propagates at a velocity $V_\text{BD}$ in Eq.~(\ref{eq:velocityBD})
lower than the velocity predicted by the mean-field equation~(\ref{eq:velocitygeneral0}), 
and its shape is the decreasing exponential $e^{-\gamma_c(x-X(t))}$
down to the position 
\be
x_\text{tip}(t)=V_\text{BD}t
+\frac{\log N}{\gamma_c},
\ee
(up to a global constant independent of $N$)
at which it
is sharply cut off by an absorptive boundary.
This boundary was meant to make the front vanish
over one unit in $x$, hence to implement discreteness
on a deterministic equation.
\begin{figure}
\begin{center}
\epsfig{file=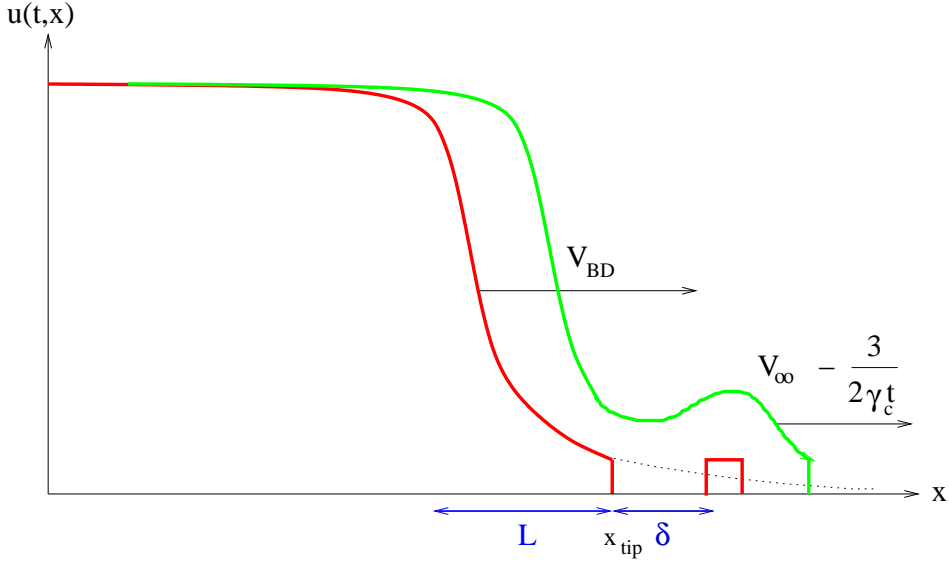,width=0.9\textwidth}
\end{center}
\caption{\label{fig:fluct}
Evolution of the front 
with a forward fluctuation.
At time $t_0$, the primary front extends over a size 
$L$ and is a solution of the branching diffusion equation
with two appropriate boundaries.
An extra particle is stochastically generated at
a distance $\delta$ with respect to the tip of the
primary front.
At a later time, the latter grows deterministically into
a secondary front that is a bit slower, and
that will add up to the primary one.
The overall effect, after relaxation, is a shift to
the right of the distance $R(\delta)$
with respect to the position of the front if a
fluctuation had not occured.
}
\end{figure}

But since the evolution is not deterministic,
it may happen that a few extra 
particles are sent stochastically
ahead of the tip of the front (See Fig.~\ref{fig:fluct}).
Their evolution would pull the front forward.
To model this effect, we assume that the probability 
per unit time
that there be a particle
sent at a distance $\delta$ ahead of the tip simply 
continues the asymptotic
shape of the front, that is to say, the distribution of $\delta$ is
\begin{equation}
p(\delta)=C_1 e^{-\gamma_c\delta},
\label{proba}
\end{equation}
where $C_1$ is a constant.
Heuristic arguments to support this assumption were presented
in Ref.~\cite{Brunet:2005bz}.
Note that while the exponential shape is quite natural since it is the continuation
of the deterministic solution~(\ref{eq:frontgeneral0}) in the linear regime, 
the fact that $C_1$ need to be strictly
constant (and cannot be a slowly varying function of $\delta$)
is a priori more difficult to argue.

Once 
a particle has been produced at position $x_\text{tip}+\delta$, 
say at time $t_0$,
it starts to multiply (see Fig.~\ref{fig:fluct}) and
it eventually develops its own front (after a time 
$t_\text{relax}$
of the order of $L^2$), 
that will add up to
the deterministic primary front made of
the evolution of the bulk of the particles.

Note that the philosophy of our phenomenological approach to
the treatment of the fluctuations is identical to the spirit of the 
statistical
approach in Sec.~\ref{sec:statisticalmethods}
developped for the zero-dimensional model. Whenever the number of particles
is larger than $\bar n$ ($\bar n=1$ here), we apply a deterministic
nonlinear evolution. Fluctuations instead are produced with a probability
which stems from a linear equation.

Let us estimate the shift in the position of the front
induced by these extra forward particles.
Between the times $t_0$ (of the order of 1) and $t=t_0+t_\text{relax}$,
the velocity of the secondary front is given by Eq.~(\ref{eq:velocitygeneral0}).
Hence its position $X^{(2)}(t)$,
after relaxation, will be given by
\begin{equation}
X^{(2)}(t)=X_\text{BD}(t)+\delta+\int_{t_0}^t dt^\prime\, v_{t^{\prime}-t_0}
\sim X_\text{BD}(t)+\delta-\frac{3}{2\gamma_c}\log L^2 
\label{Xt2}
\end{equation}
where $X_\text{BD}(t)=V_\text{BD}t$. Eq.~(\ref{Xt2}) holds
up to a constant independent of $\delta$ and $N$. We have
used Eq.~(\ref{eq:velocitygeneral0}) to express $v_{t^{\prime}-t_0}$.
The observed front will eventually result in the sum of the primary and
secondary fronts, after relaxation of the latter. 
Its position will be $X_\text{BD}(t)$ supplemented by a shift
$R(\delta)$ that may be computed by writing the resulting front shape
in the large-$x$ tail
as the sum of the primary and secondary fronts:
\be
\begin{split}
e^{-\gamma_c(x-X_\text{BD}(t)-R(\delta))}
&=e^{-\gamma_c(x-X_\text{BD}(t))}
+e^{-\gamma_c(x-X(t))}\\
&=e^{-\gamma_c(x-X_\text{BD}(t))}
+C_2 e^{-\gamma_c(x-X_\text{BD}(t)-\delta+\frac{3}{2\gamma_c}\log L^2)},
\end{split}
\label{sumfronts}
\ee
where $C_2$ is an undetermined constant.
From Eq.~(\ref{sumfronts}) 
we get the shift 
\begin{equation}
R(\delta)=\frac{1}{\gamma_c}\log\left(1+C_2\frac{e^{\gamma_c \delta}}{L^3}\right).
\label{R}
\end{equation}
The probability distribution~(\ref{proba}) 
and the front shift~(\ref{R}) 
due to a forward fluctuation
define
an effective theory for the evolution of the position of the front $X(t)$:
\begin{equation}
X(t+dt)=\begin{cases}
X(t)+V_\text{BD} dt & \text{proba.}
\ \ 1-dt \int_0^\infty d\delta p(\delta)\\
X(t)+V_\text{BD} dt+R(\delta)& \text{proba.}
\ \ p(\delta)d\delta dt.
\end{cases}
\label{eq:effectivetheory}
\end{equation} 
From these rules,
we may compute all cumulants of $X(t)$, by writing the evolution
of their generating function, deduced from the effective theory~(\ref{eq:effectivetheory}):
\be
\frac{\partial}{\partial t}\log\left\langle
e^{\lambda X(t)}\right\rangle=
\lambda V_\text{BD}+\int d\delta\, p(\delta)
\left(e^{\lambda R(\delta)}-1\right)
\ee
The left hand-side is a power series in $\lambda$
whose coefficients are
the time derivatives of the cumulants of $X(t)$.
Identifying the powers of $\lambda$ in the left and 
right handside, we get
\begin{equation}
\begin{split}
&V-V_\text{BD}=\int d\delta p(\delta)R(\delta)
=\frac{C_1C_2}{\gamma_c}\frac{3\log L}{\gamma_c L^3}\\
&\frac{[\text{$n$-th cumulant}]}{t}=\int d\delta p(\delta)[R(\delta)]^n
 =\frac{C_1C_2}{\gamma_c}\frac{n!\zeta(n)}{\gamma_c^n L^3}.
\end{split}
\label{eq:cumulants}
\end{equation}
We see that the statistics of the position of the front still depend 
on the product $C_1 C_2$ of
the undetermined constants $C_1$ and $C_2$. We
need a further assumption to fix
its value.

We go back to the expression for the correction to
the mean-field front velocity, given in
Eq.~(\ref{eq:velocityBD}).
From the expressions of $R(\delta)$ (Eq.~(\ref{R})) and of
$p(\delta)$ (Eq.~(\ref{proba})), we see that
the integrand defining $V-V_\text{BD}$ in Eq.~(\ref{eq:cumulants})
is almost a constant function of $\delta$ for 
$\delta<\delta_0=3\log L/\gamma_c$, and is decaying exponentially for
$\delta>\delta_0$. Furthermore, $R(\delta_0)$ is
of order 1, which means that when a fluctuation is sent out at
a distance $\delta\sim\delta_0$ ahead of the tip of the front,
it evolves into a front that matches in position the
deterministic primary front.
We also notice that when a fluctuation has $\delta<\delta_0$,
its evolution is completely linear until it is incorporated
to the primary front, whereas fluctuations with $\delta>\delta_0$
evolve nonlinearly but at the same time have a very 
suppressed probability.
We are thus led to the natural conjecture that the average 
front velocity is given by $V_\text{BD}$ in
Eq.~(\ref{eq:velocityBD}), with the replacement 
\be
L\rightarrow L_\text{eff}=\frac{\log N}{\gamma_c}+\delta_0
=\frac{\log N}{\gamma_c}+3\frac{\log\log N}{\gamma_c},
\ee
namely
\be
V=\frac{\omega(\gamma_c)}{\gamma_c}-\frac{\pi^2\omega^{\prime\prime}(\gamma_c)}
{2\gamma_c\left(
\frac{\log N}{\gamma_c}+\frac{3\log\log N}{\gamma_c}
\right)^2}.
\ee
The large-$N$ expansion of the new expression of the velocity
yields a correction  of the order of
$\log\log N/\log^3 N$
to the Brunet-Derrida result,
more precisely
\be
\boxed{
V=\frac{\omega(\gamma_c)}{\gamma_c}
-\frac{\pi^2\gamma_c\omega^{\prime\prime}(\gamma_c)}{2\log^2 N}
+\pi^2\gamma_c\omega^{\prime\prime}(\gamma_c)
\frac{3\log\log N}{\log^3 N}.
}
\label{eq:vcorr}
\ee
Eqs.~(\ref{eq:cumulants}) and~(\ref{eq:vcorr}) match
for the choice $C_1 C_2=\pi^2\omega^{\prime\prime}(\gamma_c)$. 
From this determination of $C_1 C_2$, 
we also get the full expression of the
cumulants of the position of the front:
\begin{equation}
\boxed{
\frac{[\text{$n$-th cumulant}]}{t}=
\pi^2\gamma_c^2\omega^{\prime\prime}(\gamma_c)
\frac{n!\zeta(n)}{\gamma_c^n\log^3 N}.
}
\label{eq:cumcorr}
\end{equation}
We note that all cumulants are of order 
unity for $t\sim \log^3 N$, which 
is the sign that the distribution of the front position 
is far from being a trivial 
Gaussian. This makes it particularly interesting.
On the other hand, the cumulants 
are proportional to $\kappa=t/\log^3 N$, 
which is the sign that
the position of the front is the result of the sum of $\kappa$ 
independent random variables,
and as such, becomes Gaussian when $\kappa$ is very large.
The properties of the statistics of the front position
were investigated in some more details in Ref.~\cite{Marquet:2006xm}.

Thanks to our discussion in Sec.~\ref{sec:schannel}, we see that
these results should apply to QCD with the relevant substitution
of the kernel $\omega$ and of the parameter $N$ 
according to Tab.~\ref{tab:dictionary}.

\subsubsection{Numerical simulations}

These results rely on a number of 
conjectures that no-one has been able to prove
so far. 
In order to check our results,
let us consider again the model
introduced in Sec.~\ref{sec:reactiondiffusionexample}.
The first step to take before being able to apply our results 
to this particular model
is to extract from the linear part of Eq.~(\ref{mod_mf})
the corresponding function $\omega(\gamma)$, and then to compute $\gamma_c$.
Setting $\Delta x=\Delta t=1$, we get
\begin{equation}
\omega(\gamma)=\log\left(1+\lambda+p_l(e^{-\gamma}-1)+p_r(e^{\gamma}-1)\right),
\label{eq:omegaEGBM}
\end{equation}
and $\gamma_c$ is defined by $\omega(\gamma_c)=\gamma_c\omega^\prime(\gamma_c)$.

\begin{figure}
\begin{center}
\epsfig{file=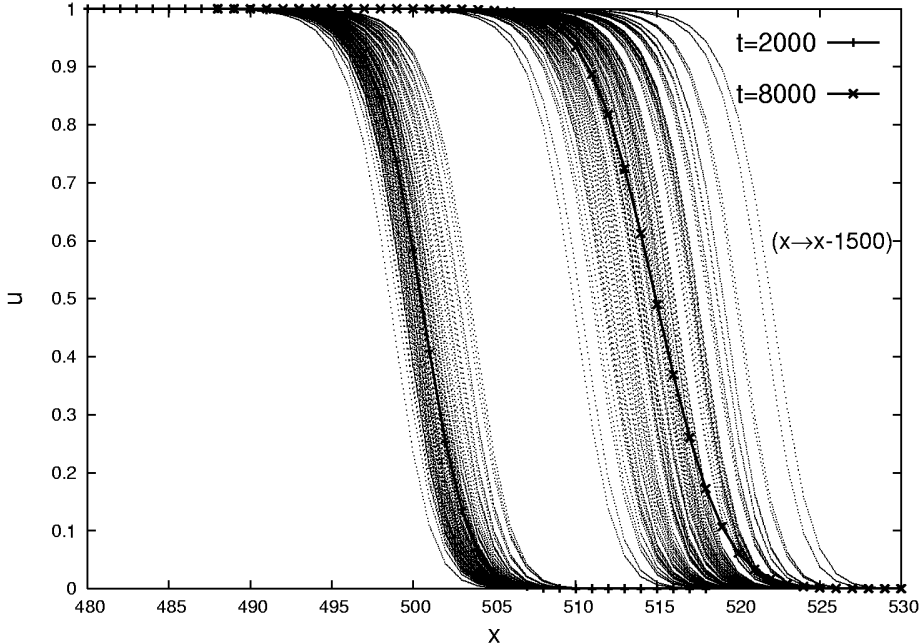,width=0.9\textwidth}
\end{center}
\caption{\label{fig:diffscaling}
1000 realizations of the model introduced in Sec.~\ref{sec:reactiondiffusionexample}
at two different times (dotted lines), and the average of $u$ over the realizations
(full line). One clearly sees that 
$\langle u\rangle$ does not keep its shape upon time evolution,
which shows that the traveling wave property of the FKPP equation
is lost due to the stochasticity.
}
\end{figure}

For the purpose of our numerical study, we set
\begin{equation}
p_l=p_r=0.1\text{\quad and\quad}\lambda=0.2\,.
\end{equation}
Simulated realizations for this set of parameters are shown 
in Fig.~\ref{fig:diffscaling}.

From (\ref{eq:omegaEGBM}), this choice leads to
\begin{equation}
\begin{split}
&\gamma_c=1.352\cdots\ ,\ \ \omega^\prime(\gamma_c)=0.2553\cdots,\\
&\omega^{\prime\prime}(\gamma_c)=0.2267\cdots.
\end{split}
\end{equation}
Predictions for all cumulants 
of the position of the front 
are obtained
by replacing the values of these parameters in 
Eqs.~(\ref{eq:vcorr}),(\ref{eq:cumcorr}).

Technically,
in order to be able to go to very large values of $N$, we replace the 
full stochastic model by its deterministic
mean field approximation $u\rightarrow \langle u\rangle$, where 
the evolution of $\langle u\rangle$
is given by Eq.~(\ref{mod_mf}), in all bins in which the number of
particles is larger than $10^3$ (that is, in the bulk of the front).
Whenever the number of particles is smaller, we use the full 
stochastic evolution~(\ref{stocha}).
We add an appropriate boundary condition 
on the interface between the
bins described by the deterministic equation and the bins described by
the stochastic equation so that the flux of particles is
conserved~\cite{moro-2004-70}.
This setup will be called ``model~I''.
Eventually, we shall use the 
mean field approximation
everywhere except in the rightmost bin (model~II): 
at each time step, a new bin is filled immediately
on the right of the rightmost nonempty site with a number 
of particles given by a Poisson law
of average
$\theta=N\langle u(x,t\!+\!1)|\{u(x,t)\}\rangle.$
We checked numerically that
this last approximation gives indistinguishable results from
those obtained within model~I as far as the statistics of the position of the front
is concerned.

We define the position of the front at time $t$ by
\begin{equation}
X_t=\sum_{x=0}^\infty u(x,t).
\end{equation}
We start at time $t=0$ from the initial condition $u(x,0)=1$ for $x\leq 0$ and
$u(x,0)=0$ for $x>0$. We evolve it up to time $t=\log^2 N$ to get rid
of subasymptotic effects related to the building up
of the asymptotic shape of the 
front, and we measure the mean velocity between times 
$\log^2 N$ and $16\times \log^2 N$. 
For model I (many stochastic bins),  we average the results over $10^4$ such realizations.
For model II (only one stochastic bin), we generate  $10^5$ such realizations
for $N\leq 10^{50}$ and $10^4$ realizations for $N>10^{50}$.
In all our simulations, models~I and~II give
numerically indistinguishable results for the values of $N$ where both
models were simulated, as can be seen on the figures (results for model~I are
represented by a circle and for model~II by a cross).

\begin{figure}
\begin{center}
\epsfig{file=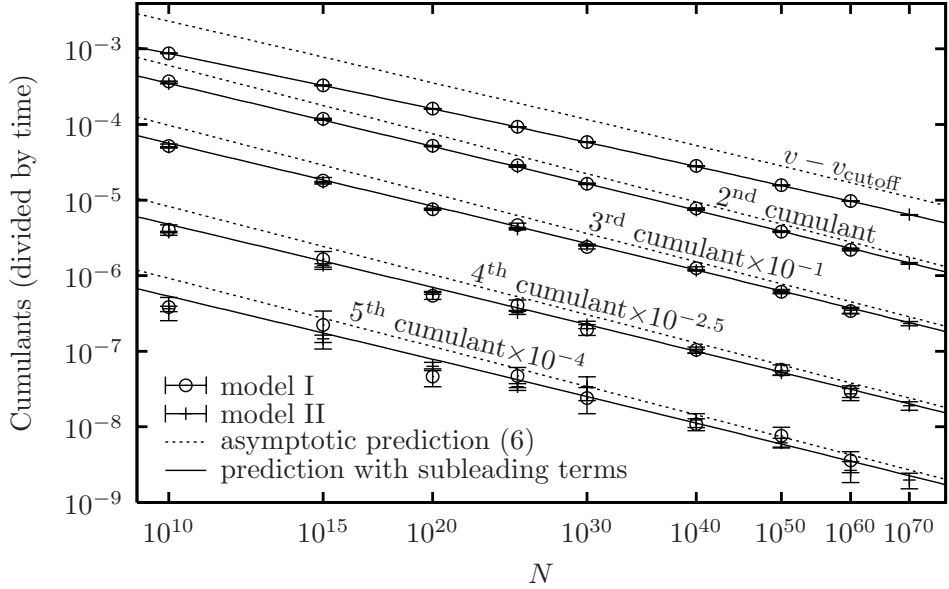,width=0.9\textwidth}
\end{center}
\caption{\label{cumulants}
[From Ref.~\cite{Brunet:2005bz}]
From top to bottom, the correction to the
velocity given by the cutoff theory 
and the cumulants of orders 2 to 5 of the position
of the front in the stochastic model. The numerical data
are compared to our parameter-free 
analytical predictions~(\ref{eq:vcorr}),(\ref{eq:cumcorr}), 
represented by the dashed line.
}
\end{figure}

Our numerical data for the cumulants 
is shown in Fig.~\ref{cumulants} together with
the analytical
predictions obtained from~(\ref{eq:vcorr}),(\ref{eq:cumcorr})
(dotted lines in the figure).
We see that the numerical simulations get very close to the analytical 
predictions at large $N$.
However, higher-order corrections are presumably still important for
the lowest values of $N$ displayed in the figure.

We try to account for these corrections by replacing
the factor $(\log N)/\gamma_c=L$ in the denominator of the expression for the
cumulants in Eqs.~(\ref{eq:vcorr}),(\ref{eq:cumcorr})
by the ansatz 
\begin{equation}
L_\text{eff}
=L+\frac{3 \log(\log N)}{\gamma_c}+c+d \frac{\log(\log N)}{\log N}\ .
\label{ansatzLeff}
\end{equation}
The two first terms in the r.h.s. are suggested by our model.
We have
added two subleading terms which go beyond our theory:
a constant term, and a term that vanishes at large $N$.
The latter are naturally expected to be among the next terms in the
asymptotic expansion for large $N$. We include them
in this numerical analysis
because in the range of $N$ in which we are able to perform our numerical
simulations, they may still bring a significant contribution.

We fit (\ref{ansatzLeff})
to the numerical data obtained in the framework of
model~II, restricting ourselves to values of $N$ larger than $10^{30}$. 
In the fit, each data point is weighted by the statistical 
dispersion of its value in our sample of data.
We obtain a determination of the values of the free parameters
$c=-4.26\pm 0.01$ and $d=5.12\pm 0.27$, with a good quality of the fit 
($\chi^2/d.o.f\sim 1.15$).

Now we see that with this modification, 
the results for the cumulants shown in figure~\ref{cumulants} (full lines), 
are in excellent
agreement with the numerical data over the whole range of~$N$.


\section{\label{sec:application}
Application to the computation of QCD scattering amplitudes}

In this section, we shall study the phenomenological relevance
of the results obtained from the correspondence
with statistical physics. There are two aspects that should be discussed.
First, we go back to the assumptions that were required to
go from QCD to reaction-diffusion, and in particular, the
hypothesis of uniformity of the evolution in impact-parameter space.
Next, we derive new properties of the QCD scattering amplitudes
and discuss their impact on phenomenology.

\subsection{\label{sec:relevance}
Relevance of one-dimensional models: impact-parameter correlations}

So far, we have argued that high-energy scattering in QCD 
at fixed coupling and fixed impact parameter
is in the universality class of the stochastic FKPP equation 
(Sec.~\ref{sec:schannel}), which is
an equation with one evolution variable (time or rapidity in QCD), 
and one spatial dimension ($x$ generically, or $\log k^2\sim \log(1/r^2)$ in QCD).
From the very beginning, we have simply discarded the impact parameter dependence.
It is important to understand that the spatial variable and the impact
parameter play different roles, and thus, the impact parameter may a priori
not be accounted for by a two-dimensional extension of the FKPP equation.

There are general arguments to support the assumption that the
QCD evolution is local enough in impact parameter
for the different impact parameters to decouple through the
rapidity evolution, which we are now going to present.

Let us start with a single dipole at rest, and bring it gradually to a higher
rapidity. As was explained in Sec.~\ref{sec:schannel},
during this process, this dipole may be replaced by
two new dipoles, which themselves may split, and so on, eventually producing
a chain of dipoles.
Figure~\ref{fig:scheme} pictures one realization of such a chain.

According to the splitting rate given in
Eq.~(\ref{splitting}), splittings to smaller-size dipoles
are favored, and thus, one expects that the sizes of the dipoles
get smaller on the average,
and that in turn, the successive splittings become more local.
The dipoles around region ``1'' and those around region ``2'' should
have an independent evolution beyond the stage pictured 
in Fig.~\ref{fig:scheme}: 
further splittings will not mix in impact parameter space, and thus,
the traveling waves around these regions should be uncorrelated.
For a dipole in region 1 of size $r$ to migrate to region 2, it should
first split into a dipole whose size is of the order of the
distance $\Delta b$ between regions 1 and 2, up to
some multiplicative factor of order $1$. 
(We assume in this discussion 
that the dipoles in region 2 relevant to the propagation 
of the local traveling waves, that is, those
which are in the bulk of the wave front, also have sizes of order $r$).
Roughly speaking, the rate of such splittings may be estimated from
the dipole splitting probability~(\ref{splitting}): 
it is of order $\bar\alpha (r^2/(\Delta b)^2)^2$,
while the rate of splittings of the same dipole into a dipole of similar
size in region 1 is of order $\bar\alpha$. Thus the first process is strongly suppressed
as soon as regions 1 and 2 are more distant than a few units of $r$.
Note that for $\Delta b \gtrsim 1/Q_s$, saturation may further reduce
the emission of the first, large, dipole leading to an even stronger
suppression of the estimated rate.

What could also happen is that some larger dipole has, by chance, one of
its endpoints tuned to the vicinity of the coordinate
one is looking at (at a distance which is at most $|\Delta r|\ll 1/Q_s(Y)$),
and easily produces a large number of dipoles there.
In this case, the position of the traveling wave at that impact
parameter would suddenly jump.
If such events were frequent enough, then they would
modify the average wave velocity and thus the one-dimensional
sFKPP picture. We may give a rough estimate of the rate at which
dipoles of size smaller than $\Delta r$ are produced.
%
Assuming local uniformity for the distribution $n$ of the emitting
dipoles, the rate (per unit of $\bar\alpha y$) of such events can be
written
\begin{equation}
  \int_{r_0 > \Delta r} \frac{d^2r_0}{r_0^2}
  \int_{\varepsilon < \Delta r} d^2\varepsilon\:
  n(r_0) \left(\frac{\varepsilon}{r_0}\right)^2
  \frac{1}{2\pi}\frac{r_0^2}{\varepsilon^2 (r_0-\varepsilon)^2},
\end{equation}
where we integrate over large dipoles of size $r_0 > \Delta r$
emitting smaller dipoles (of size $\varepsilon < \Delta r$) with a
probability $d^2\varepsilon\, r_0^2/(2\pi\varepsilon^2 (r_0-\varepsilon)^2)$. The factor
$(\varepsilon/r_0)^2$ accounts for the fact that one endpoint of the
dipole of size $r_0$ has to be in a given region of size $\varepsilon$ in
order to emit the dipoles at the right impact parameter.
To estimate this expression, we first use $n(r_0)=T(r_0)/\alpha_s^2$
and approximate $T$ by
\be
T(r_0) = \theta(r_0 - 1/Q_s)\, + \,(r_0^2Q_s^2)^{\gamma_c}\, \theta(1/Q_s - r_0).
\label{eq:approxt}
\ee
The front is replaced by 1 above the saturation scale
(for $r_0>1/Q_s$) and by an exponentially decaying tail
for $r_0 < 1/Q_s$.
Using $r_0-\varepsilon \approx r_0$ in the emission kernel, the
integration is then easily performed and
one finds a rate whose dominant term is
\begin{equation}
\frac{\pi}{2\alpha_s^2}\frac{((\Delta r)^2Q_s^2)^{\gamma_c}}{1-\gamma_c}.
\end{equation}
For $(\Delta r)^2\ll(\alpha_s^2)^{1/\gamma_c}/Q_s^2$, i.e.
ahead of the bulk of the front, this term is
parametrically less than 1 and is in fact of the order of the probability
to find an object in this region that contributes
to the normal evolution of the front \cite{Brunet:2005bz}.
Hence there is no extra contribution due to the fact that there
are many dipoles around at different impact parameters.

The arguments given here are based on estimates 
of average numbers of dipoles, 
on typical configurations,
and we are not able to account analytically for the
possible fluctuations. As we have seen through this review, the latter 
often play an important role. As a matter of fact, in the physics of
disordered systems, rare events sometimes dominate.
So before studying the phenomenological consequences of 
the statistical picture of high-energy QCD based on a one-dimensional equation,
one should check more precisely locality of the evolution
in impact parameter.

A numerical check was recently achieved in the case of 
a toy model that has an impact-parameter dependence
in Ref.~\cite{Munier:2008cg}.
Let us briefly describe the model.

\subsubsection{A model incorporating an impact-parameter dependence}

In order to arrive at a model that is tractable numerically, we only keep
one transverse dimension
instead of two in 3+1-dimensional QCD. 
However, we cannot consider genuine 2+1-dimensional QCD
because we
do not wish to give up the
logarithmic collinear singularities at $x_2=x_0$ and $x_2=x_1$.
Moreover, QCD with one dimension less has very different properties at high
energies \cite{Ivanov:1998we}.
Starting from Eq.~(\ref{splitting}), 
a splitting rate which complies with our requirements is:
\begin{equation}
\frac{dP}{d(\bar\alpha y)}=\frac{1}{4}\frac{|x_{01}|}{|x_{02}| |x_{12}|}dx_2.
\label{eq6:split0}
\end{equation}
We can further simplify this probability distribution
by keeping only its collinear and infrared asymptotics 
(as in Ref.~\cite{Ciafaloni:1999yw}).
If $|x_{02}|\ll |x_{01}|$ (or the symmetrical case $|x_{12}|\ll |x_{01}|$), 
the probability reduces to $dx_2/|x_{02}|$ ($dx_2/|x_{12}|$ resp.). 
The result of the splitting is a small dipole $(x_0,x_2)$ together with 
one close in size to the parent. So for simplicity we will just add the
small dipole to the system and leave the parent unchanged.
In the infrared region, a dipole of size $|x_{02}|\gg |x_{01}|$ 
is emitted with
a rate given by the large-$|x_{02}|$ limit of the above probability.
The probability laws~(\ref{splitting}),(\ref{eq6:split0}) 
imply that a second dipole of
similar size should be produced while the parent dipole
disappears.
To retain a behavior as close as possible to that in the collinear
limit, we will instead just generate a single large dipole and
maintain the parent. To do this consistently one must include a factor
of two in the infrared splitting rate, so as not to modify the average
rate of production of large dipoles.

In formulating our model precisely, let us focus first on the distribution of the 
sizes of the participating dipoles.
(The simplifying assumptions made above enable one to 
choose the sizes and the impact parameters of the dipoles successively).
We call $r$ the modulus of the emitted dipole, $r_0$ the modulus of its parent
and $Y=\bar\alpha y$.
The splitting rate~(\ref{eq6:split0}) 
reads, in this simplified model
\begin{equation}
\frac{dP_{r_0\to r}}{dY}
 = \theta(r-r_0) \frac{r_0 dr}{r^2} + \theta(r_0-r) \frac{dr}{r},
\end{equation}
and the original parent dipole is kept.
Logarithmic variables are the relevant ones here, so we introduce
\begin{equation}
\rho = \thelog_2(1/r) \qquad \text{or} \qquad r=2^{-\rho}.
\end{equation}
We can thus rewrite the dipole creation rate as
\be
\frac{dP_{\rho_0\to \rho}}{dY}
 = \theta(\rho_0-\rho)\, 2^{\rho-\rho_0}\, \thelog 2\, d\rho
 + \theta(\rho-\rho_0)\, \thelog 2\, d\rho.
\ee
To further simplify the model,
we discretise the dipole sizes in such a way that $\rho$
is now an integer.
This amounts to restricting the dipole sizes to negative
integer powers of $2$.
The probability that a dipole at lattice site $i$ ({\em i.e.} a dipole
of size $2^{-i}$) creates a new
dipole at lattice site $j$ is
\be
\frac{dP_{i\to j}}{dY}
 =\int_{\rho_j}^{\rho_{j+1}} \frac{dP_{\rho_i\to\rho}}{dY}
 = \begin{cases}
\thelog 2 & j\ge i \\
2^{j-i} & j<i
\end{cases}.
\label{splitf}
\ee
The rates $dP_{i\pm}/dY$ for a dipole at lattice site $i$ to split to any
lattice site $j\ge i$ or $j<i$ respectively are then given by
\be
\begin{split}
\frac{dP_{i+}}{dY} & =  \sum_{j=i}^{\imax-1} \frac{dP_{i\to j}}{dY}
 = \thelog 2 (\imax-i),\\
\frac{dP_{i-}}{dY} & =  \sum_{j=0}^{i-1} \frac{dP_{i\to j}}{dY}
 = 1-2^{-i},
\label{lifetimef}
\end{split}
\ee
where we have restricted the lattice to $0\le i <\imax$, for obvious
reasons related to the numerical implementation.

Now we have to address the question of the impact parameter of the emitted dipole.
In QCD, the collinear dipoles are produced near the endpoints of the 
parent dipoles. Let us take a parent of size $r_0$ at impact parameter $b_0$.
We set
the emitted dipole (size $r$) at the impact parameter $b$ such that
\begin{equation}
b=b_0\pm \frac{r_0\pm r\times s}{2}
\label{bf}
\end{equation}
where $s$ has uniform probability between 0 and 1. It is introduced to
obtain a continuous distribution of the impact parameter unaffected by
the discretisation of $r$.
This prescription is quite arbitrary in its details, but the latter
do not influence significantly the physical observables.
Each of the two signs that appear in the above expression
is chosen to be either $+$ or $-$ with equal weights.
We apply the same prescription 
when the emitted dipole is larger than its parent.

{\em Scattering amplitude}

We have explained above (see Sec.~\ref{sec:partonmodel}) 
that in QCD, the scattering amplitude 
of an elementary probe dipole
of size $r_i=2^{-i}$
with a dipole in an evolved Fock state
is proportional to the number of objects which have a size of the same
order of magnitude and which sit in a region of size of order $r_i$ around the
impact point of the probe dipole.
Since in our case, the sizes are discrete, the amplitude is just given,
up to a factor, by the
number of dipoles that are exactly in the same bin of
size as the probe, namely
\begin{multline}
T(i,b_0)=\alpha_s^2
\times\# \{\text{dipoles of size $2^{-i}$} \\
\text{at impact parameter $b$ satisfying }|b-b_0|<r_i/2\}.
\label{Tf}
\end{multline}

{\em Saturation}

We now have to enforce unitarity, that is the condition
\begin{equation}
T(i,b)\leq 1
\label{eq6:unitarity}
\end{equation}
for any $i$ and $b$.
This condition is expected to hold due to gluon saturation in QCD.
However,
saturation is not included in the
original dipole model.
%
%
The simplest choice is to veto splittings that would locally 
drive the amplitude
to values larger than 1.
In practice, for each splitting that gives birth to a new dipole
of size $i$ at impact parameter $b$, we compute
$T(i,b)$ and $T(i,b\pm r_i/2)$, and throw away the
produced dipole whenever
one of these numbers gets larger than one.

Given the definition of the amplitude $T$, this saturation rule implies
that there is a maximum number of objects in each bin of size
and at each impact parameter, which is equal to $N_\text{sat}=1/\alpha_s^2$.

\subsubsection{Numerical results}

We take as an initial condition a number $N_\text{sat}$ 
of dipoles of size 1 ($i=0$),
uniformly distributed in impact parameter between
$-r_0/2$ and $r_0/2$.
The impact parameters $b_j$ that are considered are respectively
$0$, $10^{-6}$, $10^{-4}$, $10^{-2}$ and $10^{-1}$.
The number of events generated is typically $10^4$,
which allows one to measure the mean and variance of the position 
of the traveling waves to a sufficient accuracy.

We have checked that at each impact parameter, we get
traveling waves whose positions grow linearly with rapidity
at a velocity less than the expected mean-field velocity for
this model. $N_\text{sat}$ was varied from 10 to 200.

\begin{figure}
\begin{center}
\epsfig{file=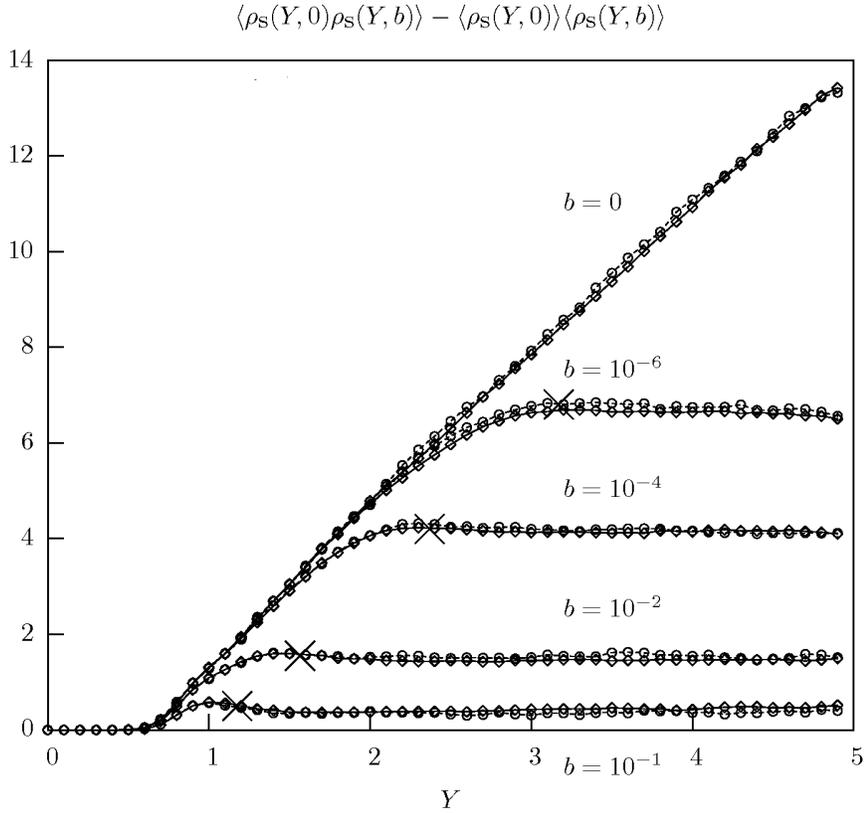,width=0.9\textwidth}
\end{center}
\caption{\label{fig:plot_cor}
Correlations of the positions of the traveling wave fronts between
different impact parameters in the toy model of Sec.~\ref{sec:relevance}. 
The points where the correlations flatten
correspond to the decoupling of the waves in the corresponding regions
of impact parameter.
}
\end{figure}

Fig.~\ref{fig:plot_cor} represents the correlations between the
positions of the wave fronts at different impact parameters in the AIP model, defined
as
\begin{equation}
\langle \rho_\text{s}(Y,b_1)\rho_\text{s}(Y,b_2)\rangle-
\langle \rho_\text{s}(Y,b_1)\rangle\langle\rho_\text{s}(Y,b_2)\rangle.
\end{equation}
We set $N_\text{sat}$ to 25 in that figure, 
but we also repeated the analysis for different values of $N_\text{sat}$
between 10 and 200.

We see very clearly 
the successive decouplings of the different
impact parameters in Fig.~\ref{fig:plot_cor}, from the most distant to the closest one,
as rapidity increases. Indeed, the correlation functions flatten after some
given rapidity depending on the difference in the probed impact parameters, which
means that the evolutions decouple.
This decoupling is expected as soon as the traveling wave front
reaches dipole sizes which are smaller than the distance
between the probed impact parameters, {\em i.e.} at $Y$ such that
$|b_2-b_1|\approx 1/Q_s(Y) = 2^{-\rho_\text{s}(Y)}$.
From the data for $\rho_\text{s}(Y)$, we can estimate quantitatively
the values of the rapidities at which the traveling waves decouple between the
different impact parameters. (It is enough to invert the above formula for the
relevant values of $b_2-b_1$).
These rapidities are denoted by a cross in Fig.~\ref{fig:plot_cor} for the
considered impact parameter differences.
Our numerical results for the correlations are nicely consistent 
with this estimate, since the correlations start 
to saturate to a constant value precisely on the right of each such cross.

We conclude that the different impact parameters indeed decouple,
as was expected from a naive analytical estimate.
What is true for our toy model should go over
to full QCD, since we have included the main
features of QCD.
When looking at the numerical data more carefully however,
it turns out that the model with impact parameter
does not reduce exactly to a supposedly equivalent 
one-dimensional model of the sFKPP type.
This is a point that would deserve more work.
We refer the reader to Ref.~\cite{Munier:2008cg}
for all details of our numerical investigations.

An attempt to build a complete picture of
high-energy QCD that includes the impact parameter
was made in Ref.~\cite{Iancu:2007st}, but it relies on
some more conjectures, that are difficult to prove.
Finding a mathematically sound formulation remains a challenge.

\subsection{Traveling waves, geometric scaling, and consequences of the noise}

As was stated in the Introduction, the initially unplanned opportunity to collect
data in the high-energy regime of deep-inelastic scattering
at HERA triggered a renewed interest in small-$x$ physics among phenomenologists.
The major discoveries in this regime is the (unexpected) 
important fraction of diffractive events,
and a new scaling, {\em geometric scaling}, featured by total 
(and even semi-inclusive) cross sections (see Fig.~\ref{fig:geometricscaling}).

In order to deal theoretically with the small-$x$ regime,
one needs new factorization theorems in order to
single out the elements of the cross sections 
that are computable in perturbation theory. High-energy, 
or $k_\perp$-factorization,
\cite{Catani:1990xk,Catani:1990eg,Collins:1991ty} is the
appropriate tool. A practical way to implement
$k_\perp$-factorization is the color dipole model 
presented in Sec.~\ref{sec:schannel}.

\subsubsection{Dipole models and geometric scaling}

The main observable measured at HERA is the proton structure function $F_2$.
It is proportional to the sum of the virtual photon-proton cross section
for a transversely and longitudinally polarized photon respectively.

A bare photon has no hadronic interactions, since it does not carry any color
charge. However, it may easily fluctuate into a quark-antiquark pair, 
overall color-neutral, thus forming a color dipole.
Subsequently, these dipoles will interact with the target proton.
This picture is represented by the following equations:
\be\label{sigmagamma}
\begin{split}
F_2(x,Q^2)&= \frac{Q^2}{4 \pi^2 \alpha_{\rm em}}
\big(\sigma_T+\sigma_L\big),\\
\sigma_{T,L}(x,Q^2)&= \int dz d^2{r}\, |\Psi_{T,L}(z,{ r},Q^2)|^2\,
\sigma_{\rm dipole}(x,{r}).
\end{split}
\ee
Here, $\sigma_{T,L}$ are the photon-proton cross sections for transversly and
longitudinally polarized virtual photons.
$\Psi_{T,L}$ are light-cone wavefunctions for 
$\gamma^*$, computable
within QED (see, e.g., Ref.~\cite{GolecBiernat:1998js} 
for explicit expressions to lowest order in
$\alpha_{\rm em}$).
Furthermore, $\sigma_{\rm dipole}(x,{r})$ is the cross-section 
for dipole--proton scattering (for a dipole of transverse size ${r}$),
and encodes all the information about hadronic interactions
(including unitarization effects). 
This cross section is related to the
amplitude $A$ discussed so far by an integration over the impact
parameter. (Actually, $A$ was the forward elastic amplitude; the
optical theorem relates it to the total cross section).

In Ref.~\cite{GolecBiernat:1998js,GolecBiernat:1999qd}, the dipole
cross-section was modeled as
\be\label{Golec}
\sigma_{\rm dipole}(x,{r})\,=\,\sigma_0\Big(1 - {\rm e}^{-{r}^2 Q_s^2(x)/4}\Big),
\ee
where $\sigma_0$ is a hadronic cross-section: It stems from
the integration over the impact parameter, when the impact parameter dependence
is supposed to be uniform over a disk of radius $\sim\sqrt{\sigma_0}$.
$Q_s(x)$ plays the role
of the saturation momentum, parametrized as 
$Q_s^2 (x)= (x_0/x)^\lambda\times 1$~GeV$^2$. 
Note that, by construction, this cross section only depends on the combined variable
$r^2 Q_s^2(x)$ instead of $r$ and $x$ separately.
This property is transmitted to the measured photon
cross sections $\sigma_{T,L}(x,Q^2)$, which then depend on $Q^2/Q_s^2(x)$ only
(this scaling is slightly violated by the masses of the quarks).
This is {\em geometric scaling}, predicted to be a feature of
the solutions to the BK equation at large rapidity.

Historically, geometric scaling was discovered first in the data
(see Ref.~\cite{Stasto:2000er}), after Golec-Biernat and W\"usthoff (GBW)
had written down their model: The latter happened to
feature this scaling (up to small violations
induced by the quark masses).
There was no apparent need for finite rapidity scaling violations
in the first HERA data.
However, later analysis revealed
that a significant amount of explicit scaling violations 
in the dipole cross section,
predicted by the BK equation, were actually
required by more accurate data.

A now popular model that describes the HERA data in a way that 
takes a better account of the subasymptotics, beyond the
GBW model, was formulated in Ref.~\cite{Iancu:2003ge}.
The dipole scattering cross section reads
$\sigma_{\rm dipole}(x,{r})=2\pi R^2 {\mathcal N}(y,rQ_s)$, with
\be\label{NFIT}
{\mathcal N}(y,rQ_s)=
\begin{cases}
{\mathcal N}_0\, \left(\frac{{r}^2 Q_s^2}{4}\right)^
{\gamma_c + \frac{\log(2/rQ_s)}{\kappa \lambda Y}}&{\rm for}\quad rQ_s\le 2,
\\
1 - {\rm e}^{-a\log^2(b\, rQ_s)}&{\rm for}\quad rQ_s > 2,
\end{cases}
\ee
where
$Q_s\equiv Q_s(x) = (x_0/x)^{\lambda/2}$ GeV.
The expression for the cross section for $r$ small compared to $2/Q_s$ corresponds
to the solution of the BK equation
(compare to Eq.~(\ref{eq:frontgeneral0}) with the help of Tab.~\ref{tab:dictionary}),
in which we substituted $\omega(\gamma_c)=\chi(\gamma_c)$ 
and $\omega^{\prime\prime}(\gamma_c)=\chi^{\prime\prime}(\gamma_c)$
by the parameters $\lambda$ and $\kappa$ that we subsequently fit to the data.
The expression in the second line also has the correct functional
form for $r\gg 2/Q_s$, as obtained by solving the BK equation \cite{Levin:2000mv}.
This is strictly valid only to leading-order accuracy,
but here it is used merely as a convenient interpolation 
towards the `black disk' limit ${\mathcal N}=1$.
(The details of this interpolation 
are unimportant for the calculation of $\sigma_{\gamma^*p}$.)
The coefficients $a$ and $b$ are determined  uniquely
from the condition that ${\mathcal N}(rQ_s,Y)$ 
and its slope be continuous at $rQ_s=2$.
The overall factor ${\mathcal N}_0$ in the first line of Eq.~(\ref{NFIT})
is ambiguous, reflecting an ambiguity in the definition of $Q_s$. 
This model fits well all HERA data for structure functions, in the range 
$x\leq 10^{-2}$. All details may be found in Ref.~\cite{Iancu:2003ge}.

The model explicitely breaks geometric scaling. However,
effectively, geometric scaling remains a fairly 
good symmetry of the model,
as required by the data.
The small finite-rapidity scaling violations are needed to describe
accurately the high-precision HERA data.

The model may also accomodate less inclusive observables, such as
diffraction \cite{Forshaw:2004xd}.
It has been improved recently by including heavy quarks \cite{Soyez:2007kg}
(The crucial need for taking account of the charm quark 
was emphasized in Ref.~\cite{Thorne:2005kj}).
An impact-parameter dependence was also introduced
\cite{Kowalski:2003hm,Watt:2007nr,Sergey:2008wk} 
that was already missing
in the GBW model.

The range of validity of dipole models has been re-examined recently \cite{Ewerz:2007md}.

\subsubsection{Diffusive scaling}

At still higher energies, according to the discussion 
of Sec.~\ref{sec:reviewtraveling}, one expects
the saturation scale to acquire a dispersion from event to event
that scales with the rapidity like $\sqrt{\bar\alpha y}$ when rapidity increases.
Although this dispersion is not an observable since there is no way
to measure the saturation scale of an individual event,
it manifests itself in the total cross section in the form
of a new scaling, different from geometric scaling.

The physical amplitude for the scattering of a dipole of size $r$ off some
target is given by the average of all realizations
of the evolution at a given $y$:
\begin{equation}
A(y,r)=\langle T(r)\rangle|_y.
\end{equation}
For large enough rapidities and small enough $\alpha_s$,
these realizations are exponentially decaying fronts 
in the variable $\rho=\log(1/r^2)$,
fully characterized by a stochastic saturation
scale, or rather its logarithm $\rho_s=\log Q_s^2(y)$. 
For the purpose of the present discussion,
it may be approximated in the same way as in Eq.~(\ref{eq:approxt}), namely
\begin{equation}
T(\rho)=\theta(\rho_s-\rho)+\theta(\rho-\rho_s)e^{-\gamma_c(\rho-\rho_s)}.
\label{eq:tsimple2}
\end{equation}
The statistics of $\rho_s$ is given by Eqs.~(\ref{eq:vcorr}),(\ref{eq:cumcorr})
(up to the replacements suggested in Tab.~\ref{tab:dictionary} to go from a generic
reaction-diffusion to QCD).
At ultrahigh energies (and very small $\alpha_s$), 
it is essentially a Gaussian centered at
\be
\langle\rho_s\rangle=
\left(\frac{\chi(\gamma_c)}{\gamma_c}-
\frac{\pi^2\gamma_c\chi^{\prime\prime}(\gamma_c)}
{2\left(\log(1/\alpha_s^2)+3\log\log(1/\alpha_s^2)\right)^2}
\right)\bar\alpha y
\ee
and of variance
\be
\sigma^2=\langle \rho_s^2\rangle-\langle\rho_s\rangle^2=
\frac{\pi^4\chi^{\prime\prime}(\gamma_c)}{3\log^3(1/\alpha_s^2)}\bar\alpha y.
\ee
The scattering amplitude may be expressed by the simple formula
\be
A(y,\rho)=\frac{1}{\sigma\sqrt{2\pi}}
\int d\rho_s\, T(\rho)|_y
\exp\left(\frac{(\rho_s-\langle\rho_s\rangle)^2}{2\sigma^2}\right).
\label{eq:adiffusive0}
\ee
The most remarkable feature of this amplitude 
is the scaling form for $A$ that it yields:
\be
A(y,\rho)=A\left(\frac{\rho-\langle\rho_s(y)\rangle}
{\sqrt{\bar\alpha y/\log^3(1/\alpha_s^2)}}
\right).
\label{eq:diffusivescalingA}
\ee
This equation may be obtained by performing
the integration in Eq.~(\ref{eq:adiffusive0}) after the replacement of
$T$ by its approximation~(\ref{eq:tsimple2}).
This scaling obviously violates geometric scaling: If the
latter scaling were satisfied, then $A$ would be a function
of $\rho-\langle \rho_s(y)\rangle$ only.

Mueller and Shoshi had already noted
that geometric scaling had to be violated beyond the BK equation 
in Ref.~\cite{Mueller:2004sea}.
However, the square root in the denominator of the scaling variable
in Eq.~(\ref{eq:diffusivescalingA})
was missing because their approach was relying on mean field
throughout, thus missing the stochastic nature of the evolution.

This new scaling is a firm
prediction of the correspondence with statistical physics.
However, it may not be tested at particle colliders in a simple way.
Let us work out the order of magnitude of the rapidity 
needed for the different effects (saturation, geometric scaling, diffusive
scaling) to show up.
The rapidity that is needed to reach saturation is roughly
\be
y_{\text{BFKL}}\sim
\frac{\log({1}/{\alpha_s^2})}{\bar\alpha\chi(\frac12)}.
\ee
The BK picture is expected to be valid until the asymptotic 
exponential shape of the
front has diffused down to the point where the amplitude becomes
of the order of $\alpha_s^2$. This additional rapidity 
needed to get to the regime of geometric scaling is thus given by
Eq.~(\ref{eq:relaxtime}) once the appropriate replacements have been done
\be
y_\text{BK}\sim
\frac{1}{2\bar\alpha\chi^{\prime\prime}(\gamma_c)}
\left[\frac{\log({1}/{\alpha_s^2})}{\gamma_c}\right]^2,
\ee
and finally, the effect of the fluctuations of the saturation scale
gets important at the rapidity
\be
y_\text{fluct}\sim\frac{3\log^3 (1/\alpha_s^2)}
{\bar\alpha\pi^3\chi^{\prime\prime}(\gamma_c)}.
\ee
The relevant parameters read, in QCD,
\begin{equation}
\gamma_c=0.627549,\ \ \chi(\gamma_c)=3.0645,\ \ 
\chi^{\prime\prime}(\gamma_c)=48.5176.
\end{equation}
For some realistic strong coupling constant, $\alpha_s\sim 0.2$,
we get
\begin{equation}
y_{\text{BFKL}}\sim 6.07879
\ ,\ \
y_\text{BK}\sim 1.41965
\ ,\ \
y_\text{fluct}\sim 0.348244.
\end{equation}
Given that rapidities in the small-$x$ regime at HERA were of the order of 10,
and will be of the order of 15 at the LHC, these figures may give us 
the hope that we may observe these effects.
However, the values of the rapidity that delimitates the different regimes
are largely underestimated given that they rely on the 
leading-order BFKL kernel, which predicts a much too large
growth of the cross section with the rapidity and a too fast diffusion (see the 
large value of $\chi^{\prime\prime}(\gamma_c)$).
One also has to keep in mind that the former estimates should only hold
for very small values of $\alpha_s$.

Furthermore, the effect of the running coupling, which should
be taken into account in any detailed phenomenological
study, is expected
to still reduce the effects of the fluctuations \cite{Dumitru:2007ew}.

Nevertheless,
the effect of diffusive scaling (i.e. of the event-by-event
fluctuations of the saturation scale) on observables
has already been investigated in some detail 
by several groups. Diffractive amplitudes were studied
in Ref.~\cite{Hatta:2006hs}.
The ratio of the gluon distribution
in a nucleus to the same quantity in a proton
was computed in Ref.~\cite{Kozlov:2006qw}.


\section{Conclusion and outlook}

We have reviewed a peculiar way of viewing high-energy scattering in QCD,
based on the physics of the parton model, and its strong similarities
with reaction-diffusion processes (Sec.~\ref{sec:schannel}).
The correspondence is best summarized in the mapping 
of Tab.~\ref{tab:dictionary}.
We have seen that the equations that describe the dynamics of
these processes are in the universality class of the stochastic FKPP
equation, and admit traveling-wave solutions whose
features are likely to be universal, in such a way
that a study of simple reaction-diffusion-like models may lead to
exact asymptotic results also for QCD scattering amplitudes.
Understanding the very mechanism of traveling wave formation
and front propagation was crucial to see how the universality may
come about (see Sec.~\ref{sec:reviewtraveling}).

In zero-dimensional stochastic models, we could perform exact calculations
and get analytical results within different formulations 
(Sec.~\ref{sec:zerodimensional}).
We understood that analyzing the structure of single events
was technically much simpler if one wants
to get leading orders at large $N$ ($=1/\alpha_s^2$),
since in individual realizations, one may factorize the fluctuating part
from the nonlinear effects.
Thanks to this observation,
in one-dimensional models which admit realizations in the form
of stochastic traveling waves,
we could also get precise analytical results on the
form and shape of the traveling waves, which are presumably exact
asymptotically (Sec.~\ref{sec:reviewtraveling}).
Universality enables one to make statements on
the form of the QCD scattering amplitudes at very high energies.
These statements turn into firm phenomenological predictions 
(Sec.~\ref{sec:application}), which however do not seem to be
testable at colliders in the near future.
Nevertheless, getting new analytical results for QCD in some
limit is always an interesting achievement, given the
complexity of the theory.
Furthermore, while our analytical results only apply for
exponentially small $\alpha_s$ ($\log(1/\alpha_s^2)\gg 1$),
the picture itself should be valid in the whole perturbative range,
namely for $\alpha_s^2\ll 1$.


There are still many open questions.
On the statistical physics side, the statistics of the front position
that we have found has not been derived rigorously, but rather
guessed, and rely on many quite ad hoc conjectures. We got confidence
on the validity of our conjectures on the basis of
numerical simulations.
Moreover, although we expect universality up to corrections of order $1/N$
(that is to say $\mathcal{O}(\alpha_s^2)$ in QCD),
we could only get analytical expressions relative to the cumulants
of the position of the front for 
the first terms in an expansion in powers of $1/\log N$, which requires
much larger values of $N$ to be valid.
On a more general footing, the sFKPP equation seems to describe
many physical, chemical or biological problems (in particular
population evolution with selection in evolutionary biology).
We have also found recently an explicit analogy with the theory of spin
glasses \cite{Brunet:2006zn}.
This large universality is maybe the strongest incentive to
try and find more accurate solutions to that kind of equations.

On the QCD side, the correspondence with reaction-diffusion processes
strongly relies on the assumption that there is saturation
of some form of the quark and gluon densities
in the hadronic wave functions.
While this is a reasonable guess that few experts would challenge,
it is clear that we cannot consider that the problem
is solved before the saturation mechanism at work in QCD
has been exhibited.
QCD is formulated as a quantum field theory. To see the similarity
with reaction-diffusion, we basically needed to
translate it into the parton model first.
It would be better to recover the results of 
Sec.~\ref{sec:reviewtraveling}
(and hopefully get more)
directly from field theory \cite{Munier:2008rh}, 
as one could do it in the zero-dimensional model
introduced in Sec.~\ref{sec:zerodimensional}.
This requires to understand the strong field regime of field theory.
This is an exciting challenge for both particle physicists and statistical
physicists.



\section*{Acknowledgements}

I thank Dr Urko Reinosa for his reading of the manuscript and his
helpful comments.
This work was supported in part by the Agence Nationale de la Recherche
(France), contract ANR-06-JCJC-0084-02.




\begin{thebibliography}{100}
\expandafter\ifx\csname url\endcsname\relax
  \def\url#1{\texttt{#1}}\fi
\expandafter\ifx\csname urlprefix\endcsname\relax\def\urlprefix{URL }\fi

\bibitem{Muta:1998vi}
T.~Muta, {Foundations of quantum chromodynamics. Second edition}, World Sci.
  Lect. Notes Phys. 57 (1998) 1--409.

\bibitem{Gribov:1972ri}
V.~N. Gribov, L.~N. Lipatov, {Deep inelastic e p scattering in perturbation
  theory}, Sov. J. Nucl. Phys. 15 (1972) 438--450.

\bibitem{Dokshitzer:1977sg}
Y.~L. Dokshitzer, {Calculation of the Structure Functions for Deep Inelastic
  Scattering and e+ e- Annihilation by Perturbation Theory in Quantum
  Chromodynamics. (In Russian)}, Sov. Phys. JETP 46 (1977) 641--653.

\bibitem{Altarelli:1977zs}
G.~Altarelli, G.~Parisi, {Asymptotic Freedom in Parton Language}, Nucl. Phys.
  B126 (1977) 298.

\bibitem{Lipatov:1976zz}
L.~N. Lipatov, {Reggeization of the Vector Meson and the Vacuum Singularity in
  Nonabelian Gauge Theories}, Sov. J. Nucl. Phys. 23 (1976) 338--345.

\bibitem{Kuraev:1977fs}
E.~A. Kuraev, L.~N. Lipatov, V.~S. Fadin, {The Pomeranchuk Singularity in
  Nonabelian Gauge Theories}, Sov. Phys. JETP 45 (1977) 199--204.

\bibitem{Balitsky:1978ic}
I.~I. Balitsky, L.~N. Lipatov, {The Pomeranchuk Singularity in Quantum
  Chromodynamics}, Sov. J. Nucl. Phys. 28 (1978) 822--829.

\bibitem{Ciafaloni:1998gs}
M.~Ciafaloni, G.~Camici, {Energy scale(s) and next-to-leading BFKL equation},
  Phys. Lett. B430 (1998) 349--354.

\bibitem{Fadin:1998py}
V.~S. Fadin, L.~N. Lipatov, {BFKL pomeron in the next-to-leading
  approximation}, Phys. Lett. B429 (1998) 127--134.

\bibitem{McLerran:2001sr}
L.~D. McLerran, {The color glass condensate and small x physics: 4 lectures},
  Lect. Notes Phys. 583 (2002) 291--334.

\bibitem{Iancu:2003xm}
E.~Iancu, R.~Venugopalan, {The color glass condensate and high energy
  scattering in QCD}.

\bibitem{Gribov:1981ac}
L.~V. Gribov, E.~M. Levin, M.~G. Ryskin, {Singlet Structure Function at Small
  x: Unitarization of Gluon Ladders}, Nucl. Phys. B188 (1981) 555--576.

\bibitem{Gribov:1984tu}
L.~V. Gribov, E.~M. Levin, M.~G. Ryskin, {Semihard Processes in QCD}, Phys.
  Rept. 100 (1983) 1--150.

\bibitem{Mueller:1985wy}
A.~H. Mueller, J.-w. Qiu, {Gluon Recombination and Shadowing at Small Values of
  x}, Nucl. Phys. B268 (1986) 427.

\bibitem{McLerran:1993ni}
L.~D. McLerran, R.~Venugopalan, {Computing quark and gluon distribution
  functions for very large nuclei}, Phys. Rev. D49 (1994) 2233--2241.

\bibitem{McLerran:1993ka}
L.~D. McLerran, R.~Venugopalan, {Gluon distribution functions for very large
  nuclei at small transverse momentum}, Phys. Rev. D49 (1994) 3352--3355.

\bibitem{McLerran:1994vd}
L.~D. McLerran, R.~Venugopalan, {Green's functions in the color field of a
  large nucleus}, Phys. Rev. D50 (1994) 2225--2233.

\bibitem{Balitsky:1995ub}
I.~Balitsky, {Operator expansion for high-energy scattering}, Nucl. Phys. B463
  (1996) 99--160.

\bibitem{JalilianMarian:1997jx}
J.~Jalilian-Marian, A.~Kovner, A.~Leonidov, H.~Weigert, {The BFKL equation from
  the Wilson renormalization group}, Nucl. Phys. B504 (1997) 415--431.

\bibitem{JalilianMarian:1997gr}
J.~Jalilian-Marian, A.~Kovner, A.~Leonidov, H.~Weigert, {The Wilson
  renormalization group for low x physics: Towards the high density regime},
  Phys. Rev. D59 (1998) 014014.

\bibitem{Iancu:2001ad}
E.~Iancu, A.~Leonidov, L.~D. McLerran, {The renormalization group equation for
  the color glass condensate}, Phys. Lett. B510 (2001) 133--144.

\bibitem{Iancu:2000hn}
E.~Iancu, A.~Leonidov, L.~D. McLerran, {Nonlinear gluon evolution in the color
  glass condensate. I}, Nucl. Phys. A692 (2001) 583--645.

\bibitem{Weigert:2000gi}
H.~Weigert, {Unitarity at small Bjorken x}, Nucl. Phys. A703 (2002) 823--860.

\bibitem{Kovchegov:1999yj}
Y.~V. Kovchegov, {Small-x F2 structure function of a nucleus including multiple
  pomeron exchanges}, Phys. Rev. D60 (1999) 034008.

\bibitem{Kovchegov:1999ua}
Y.~V. Kovchegov, {Unitarization of the BFKL pomeron on a nucleus}, Phys. Rev.
  D61 (2000) 074018.

\bibitem{Mueller:1993rr}
A.~H. Mueller, {Soft gluons in the infinite momentum wave function and the BFKL
  pomeron}, Nucl. Phys. B415 (1994) 373--385.

\bibitem{GolecBiernat:1998js}
K.~J. Golec-Biernat, M.~Wusthoff, {Saturation effects in deep inelastic
  scattering at low Q**2 and its implications on diffraction}, Phys. Rev. D59
  (1999) 014017.

\bibitem{GolecBiernat:1999qd}
K.~J. Golec-Biernat, M.~Wusthoff, {Saturation in diffractive deep inelastic
  scattering}, Phys. Rev. D60 (1999) 114023.

\bibitem{Stasto:2000er}
A.~M. Stasto, K.~J. Golec-Biernat, J.~Kwiecinski, {Geometric scaling for the
  total gamma* p cross-section in the low x region}, Phys. Rev. Lett. 86 (2001)
  596--599.

\bibitem{Marquet:2006jb}
C.~Marquet, L.~Schoeffel, {Geometric scaling in diffractive deep inelastic
  scattering}, Phys. Lett. B639 (2006) 471--477.

\bibitem{Levin:1999mw}
E.~Levin, K.~Tuchin, {Solution to the evolution equation for high parton
  density QCD}, Nucl. Phys. B573 (2000) 833--852.

\bibitem{Levin:2000mv}
E.~Levin, K.~Tuchin, {New scaling at high energy DIS}, Nucl. Phys. A691 (2001)
  779--790.

\bibitem{Armesto:2001fa}
N.~Armesto, M.~A. Braun, {Parton densities and dipole cross-sections at small x
  in large nuclei}, Eur. Phys. J. C20 (2001) 517--522.

\bibitem{GolecBiernat:2001if}
K.~J. Golec-Biernat, L.~Motyka, A.~M. Stasto, {Diffusion into infra-red and
  unitarization of the BFKL pomeron}, Phys. Rev. D65 (2002) 074037.

\bibitem{Mueller:2002zm}
A.~H. Mueller, D.~N. Triantafyllopoulos, {The energy dependence of the
  saturation momentum}, Nucl. Phys. B640 (2002) 331--350.

\bibitem{Fisher}
R.~A. Fisher, Ann. Eugenics 7 (1937) 355.

\bibitem{KPP}
A.~Kolmogorov, I.~Petrovsky, N.~Piscounov, Moscou Univ. Bull. Math. A1 (1937)
  1.

\bibitem{Munier:2003vc}
S.~Munier, R.~B. Peschanski, {Geometric scaling as traveling waves}, Phys. Rev.
  Lett. 91 (2003) 232001.

\bibitem{Mueller:2004sea}
A.~H. Mueller, A.~I. Shoshi, {Small-x physics beyond the Kovchegov equation},
  Nucl. Phys. B692 (2004) 175--208.

\bibitem{Iancu:2004es}
E.~Iancu, A.~H. Mueller, S.~Munier, {Universal behavior of QCD amplitudes at
  high energy from general tools of statistical physics}, Phys. Lett. B606
  (2005) 342--350.

\bibitem{Nikolaev:1990ja}
N.~N. Nikolaev, B.~G. Zakharov, {Colour transparency and scaling properties of
  nuclear shadowing in deep inelastic scattering}, Z. Phys. C49 (1991)
  607--618.

\bibitem{Nikolaev:1991et}
N.~Nikolaev, B.~G. Zakharov, {Pomeron structure function and diffraction
  dissociation of virtual photons in perturbative QCD}, Z. Phys. C53 (1992)
  331--346.

\bibitem{Ewerz:2004vf}
C.~Ewerz, O.~Nachtmann, {Towards a Nonperturbative Foundation of the Dipole
  Picture: I. Functional Methods}, Annals Phys. 322 (2007) 1635--1669.

\bibitem{Ewerz:2006vd}
C.~Ewerz, O.~Nachtmann, {Towards a Nonperturbative Foundation of the Dipole
  Picture: II. High Energy Limit}, Annals Phys. 322 (2007) 1670--1726.

\bibitem{Lipatov:1985uk}
L.~N. Lipatov, {The Bare Pomeron in Quantum Chromodynamics}, Sov. Phys. JETP 63
  (1986) 904--912.

\bibitem{Balitsky:1998kc}
I.~Balitsky, {Factorization for high-energy scattering}, Phys. Rev. Lett. 81
  (1998) 2024--2027.

\bibitem{Balitsky:1998ya}
I.~Balitsky, {Factorization and high-energy effective action}, Phys. Rev. D60
  (1999) 014020.

\bibitem{Janik:2004ts}
R.~A. Janik, R.~B. Peschanski, {QCD saturation equations including
  dipole-dipole correlation}, Phys. Rev. D70 (2004) 094005.

\bibitem{Janik:2004ve}
R.~A. Janik, {QCD saturation in the dipole sector with correlations}, Phys.
  Lett. B604 (2004) 192--198.

\bibitem{Levin:2004yd}
E.~Levin, M.~Lublinsky, {Balitsky's hierarchy from Mueller's dipole model and
  more about target correlations}, Phys. Lett. B607 (2005) 131--138.

\bibitem{Rummukainen:2003ns}
K.~Rummukainen, H.~Weigert, {Universal features of JIMWLK and BK evolution at
  small x}, Nucl. Phys. A739 (2004) 183--226.

\bibitem{Chen:1995pa}
Z.~Chen, A.~H. Mueller, {The Dipole picture of high-energy scattering, the BFKL
  equation and many gluon compound states}, Nucl. Phys. B451 (1995) 579--604.

\bibitem{Kovchegov:1997dm}
Y.~V. Kovchegov, A.~H. Mueller, S.~Wallon, {Unitarity corrections and high
  field strengths in high energy hard collisions}, Nucl. Phys. B507 (1997)
  367--378.

\bibitem{Mueller:1996te}
A.~H. Mueller, G.~P. Salam, {Large multiplicity fluctuations and saturation
  effects in onium collisions}, Nucl. Phys. B475 (1996) 293--320.

\bibitem{Avsar:2005iz}
E.~Avsar, G.~Gustafson, L.~Lonnblad, {Energy conservation and saturation in
  small-x evolution}, JHEP 07 (2005) 062.

\bibitem{Avsar:2006jy}
E.~Avsar, G.~Gustafson, L.~Lonnblad, {Small-x dipole evolution beyond the
  large-N(c) limit}, JHEP 01 (2007) 012.

\bibitem{Mueller:2001fv}
A.~H. Mueller, {Parton saturation: An overview}, {\tt arxiv:hep-ph/0111244}.

\bibitem{Mueller:1989st}
A.~H. Mueller, {Small x Behavior and Parton Saturation: A QCD Model}, Nucl.
  Phys. B335 (1990) 115.

\bibitem{Mueller:1999wm}
A.~H. Mueller, {Parton saturation at small x and in large nuclei}, Nucl. Phys.
  B558 (1999) 285--303.

\bibitem{Forshaw:1997dc}
J.~R. Forshaw, D.~A. Ross, {Quantum chromodynamics and the pomeron}, Cambridge
  Lect. Notes Phys. 9 (1997) 1--248.

\bibitem{Kirschner:1994gd}
R.~Kirschner, L.~N. Lipatov, L.~Szymanowski, {Effective action for multi -
  Regge processes in QCD}, Nucl. Phys. B425 (1994) 579--594.

\bibitem{Kirschner:1994xi}
R.~Kirschner, L.~N. Lipatov, L.~Szymanowski, {Symmetry properties of the
  effective action for high- energy scattering in QCD}, Phys. Rev. D51 (1995)
  838--855.

\bibitem{Lipatov:1995pn}
L.~N. Lipatov, {Gauge invariant effective action for high-energy processes in
  QCD}, Nucl. Phys. B452 (1995) 369--400.

\bibitem{Antonov:2004hh}
E.~N. Antonov, L.~N. Lipatov, E.~A. Kuraev, I.~O. Cherednikov, {Feynman rules
  for effective Regge action}, Nucl. Phys. B721 (2005) 111--135.

\bibitem{Blaizot:2005vf}
J.~P. Blaizot, E.~Iancu, K.~Itakura, D.~N. Triantafyllopoulos, {Duality and
  Pomeron effective theory for QCD at high energy and large N(c)}, Phys. Lett.
  B615 (2005) 221--230.

\bibitem{Hatta:2005rn}
Y.~Hatta, E.~Iancu, L.~McLerran, A.~Stasto, D.~N. Triantafyllopoulos,
  {Effective Hamiltonian for QCD evolution at high energy}, Nucl. Phys. A764
  (2006) 423--459.

\bibitem{Bartels:1992ym}
J.~Bartels, {Unitarity corrections to the Lipatov pomeron and the small x
  region in deep inelastic scattering in QCD}, Phys. Lett. B298 (1993)
  204--210.

\bibitem{Bartels:1993ih}
J.~Bartels, {Unitarity corrections to the Lipatov pomeron and the four gluon
  operator in deep inelastic scattering in QCD}, Z. Phys. C60 (1993) 471--488.

\bibitem{Bartels:1999aw}
J.~Bartels, C.~Ewerz, {Unitarity corrections in high-energy QCD}, JHEP 09
  (1999) 026.

\bibitem{Ewerz:2003an}
C.~Ewerz, V.~Schatz, {How pomerons meet in coloured glass}, Nucl. Phys. A736
  (2004) 371--404.

\bibitem{DerridaSpohn}
B.~Derrida, H.~Spohn, {Polymers on disordered trees, spin glasses and traveling
  waves}, J. Stat. Phys. 51 (1988) 817--840.

\bibitem{vansaarloos-2003-386}
W.~{van Saarloos}, Front propagation into unstable states, Physics Reports 386
  (2003) 29.

\bibitem{Enberg:2006aq}
R.~Enberg, {Traveling waves and the renormalization group improved
  Balitsky-Kovchegov equation}, Phys. Rev. D75 (2007) 014012.

\bibitem{Enberg:2005cb}
R.~Enberg, K.~J. Golec-Biernat, S.~Munier, {The high energy asymptotics of
  scattering processes in QCD}, Phys. Rev. D72 (2005) 074021.
\newline\urlprefix\url{http://www.isv.uu.se/\~{}enberg/BK/}

\bibitem{panja-2004-393}
D.~Panja, Effects of fluctuations on propagating fronts, Physics Reports 393
  (2004) 87.

\bibitem{Iancu:2004iy}
E.~Iancu, D.~N. Triantafyllopoulos, {A Langevin equation for high energy
  evolution with pomeron loops}, Nucl. Phys. A756 (2005) 419--467.

\bibitem{Mueller:2005ut}
A.~H. Mueller, A.~I. Shoshi, S.~M.~H. Wong, {Extension of the JIMWLK equation
  in the low gluon density region}, Nucl. Phys. B715 (2005) 440--460.

\bibitem{Iancu:2005nj}
E.~Iancu, D.~N. Triantafyllopoulos, {Non-linear QCD evolution with improved
  triple-pomeron vertices}, Phys. Lett. B610 (2005) 253--261.

\bibitem{Iancu:2005dx}
E.~Iancu, G.~Soyez, D.~N. Triantafyllopoulos, {On the probabilistic
  interpretation of the evolution equations with Pomeron loops in QCD}, Nucl.
  Phys. A768 (2006) 194--221.

\bibitem{Salam:1995zd}
G.~P. Salam, {Multiplicity distribution of color dipoles at small x}, Nucl.
  Phys. B449 (1995) 589--604.

\bibitem{Salam:1995uy}
G.~P. Salam, {Studies of Unitarity at Smallã$x$ Using the Dipole Formulation},
  Nucl. Phys. B461 (1996) 512--538.

\bibitem{Salam:1996nb}
G.~P. Salam, {OEDIPUS: Onium evolution, dipole interaction and perturbative
  unitarisation simulation}, Comput. Phys. Commun. 105 (1997) 62--76.

\bibitem{Doi}
M.~Doi, J. Phys. A 9 (1976) 1479.

\bibitem{Peliti}
L.~Peliti, J. Phys. (Paris) 46 (1985) 1469.

\bibitem{tauber-2007-716}
U.~C. Tauber, Field theory approaches to nonequilibrium dynamics, LECT.NOTES
  PHYS. 716 (2007) 295.

\bibitem{PhysRevE.59.3893}
L.~Pechenik, H.~Levine, Interfacial velocity corrections due to multiplicative
  noise, Phys. Rev. E 59~(4) (1999) 3893--3900.

\bibitem{Shoshi:2005pf}
A.~I. Shoshi, B.-W. Xiao, {Pomeron loops in zero transverse dimensions}, Phys.
  Rev. D73 (2006) 094014.

\bibitem{Shoshi:2006eb}
A.~I. Shoshi, B.-W. Xiao, {Diffractive dissociation including pomeron loops in
  zero transverse dimensions}, Phys. Rev. D75 (2007) 054002.

\bibitem{Levin:2007yv}
E.~Levin, A.~Prygarin, {The BFKL Pomeron Calculus in zero transverse
  dimensions: summation of Pomeron loops and generating functional for the
  multiparticle production processes}, Eur. Phys. J. C53 (2008) 385--399.

\bibitem{Kozlov:2006cu}
M.~Kozlov, E.~Levin, V.~Khachatryan, J.~Miller, {The BFKL pomeron calculus in
  zero transverse dimensions: Diffractive processes and survival probability
  for central diffractive production}, Nucl. Phys. A791 (2007) 382--405.

\bibitem{Kozlov:2007xc}
M.~Kozlov, E.~Levin, A.~Prygarin, {The BFKL Pomeron Calculus in the dipole
  approach}, Nucl. Phys. A792 (2007) 122--151.

\bibitem{Levin:2007wc}
E.~Levin, J.~Miller, A.~Prygarin, {Summing Pomeron loops in the dipole
  approach}, Nucl. Phys. A806 (2008) 245--286.

\bibitem{gardiner}
C.~W. Gardiner, Handbook of Stochastic Methods: for Physics, Chemistry and the
  Natural Sciences (Springer Series in Synergetics), 3rd Edition, Springer,
  2004.

\bibitem{Munier:2006um}
S.~Munier, {Dense-dilute factorization for a class of stochastic processes and
  for high energy QCD}, Phys. Rev. D75 (2007) 034009.

\bibitem{Blaizot:2006wp}
J.~P. Blaizot, E.~Iancu, D.~N. Triantafyllopoulos, {A zero-dimensional model
  for high-energy scattering in QCD}, Nucl. Phys. A784 (2007) 227--258.

\bibitem{Bondarenko:2006rh}
S.~Bondarenko, L.~Motyka, A.~H. Mueller, A.~I. Shoshi, B.~W. Xiao, {On the
  equivalence of Reggeon field theory in zero transverse dimensions and
  reaction-diffusion processes}, Eur. Phys. J. C50 (2007) 593--601.

\bibitem{Bramson}
M.~Bramson, Mem. Am. Math. Soc. 44 (1983) 285.

\bibitem{Albacete:2005ef}
J.~L. Albacete, N.~Armesto, J.~G. Milhano, C.~A. Salgado, U.~A. Wiedemann,
  {Nuclear size and rapidity dependence of the saturation scale from QCD
  evolution and experimental data}, Eur. Phys. J. C43 (2005) 353--360.

\bibitem{Levin:2001et}
E.~Levin, M.~Lublinsky, {Parton densities and saturation scale from non-linear
  evolution in DIS on nuclei}, Nucl. Phys. A696 (2001) 833--850.

\bibitem{Albacete:2003iq}
J.~L. Albacete, N.~Armesto, A.~Kovner, C.~A. Salgado, U.~A. Wiedemann, {Energy
  dependence of the Cronin effect from non-linear QCD evolution}, Phys. Rev.
  Lett. 92 (2004) 082001.

\bibitem{Iancu:2002tr}
E.~Iancu, K.~Itakura, L.~McLerran, {Geometric scaling above the saturation
  scale}, Nucl. Phys. A708 (2002) 327--352.

\bibitem{Triantafyllopoulos:2002nz}
D.~N. Triantafyllopoulos, {The energy dependence of the saturation momentum
  from RG improved BFKL evolution}, Nucl. Phys. B648 (2003) 293--316.

\bibitem{ebert-2000-146}
U.~Ebert, W.~{van Saarloos}, Front propagation into unstable states: Universal
  algebraic convergence towards uniformly translating pulled fronts, PHYSICA D
  146 (2000) 1.

\bibitem{Munier:2004xu}
S.~Munier, R.~B. Peschanski, {Universality and tree structure of high energy
  QCD}, Phys. Rev. D70 (2004) 077503.

\bibitem{brunet-1997-57}
E.~Brunet, B.~Derrida, Shift in the velocity of a front due to a cut-off,
  Physical Review E 57 (1997) 2597.

\bibitem{Kovner:2005jc}
A.~Kovner, M.~Lublinsky, {Remarks on high energy evolution}, JHEP 03 (2005)
  001.

\bibitem{Kovner:2005en}
A.~Kovner, M.~Lublinsky, {From target to projectile and back again: Selfduality
  of high energy evolution}, Phys. Rev. Lett. 94 (2005) 181603.

\bibitem{Kovner:2005uw}
A.~Kovner, M.~Lublinsky, {Dense-dilute duality at work: Dipoles of the target},
  Phys. Rev. D72 (2005) 074023.

\bibitem{Kovner:2005aq}
A.~Kovner, M.~Lublinsky, {More remarks on high energy evolution}, Nucl. Phys.
  A767 (2006) 171--188.

\bibitem{Kovner:2007zu}
A.~Kovner, M.~Lublinsky, U.~Wiedemann, {From bubbles to foam: Dilute to dense
  evolution of hadronic wave function at high energy}, JHEP 06 (2007) 075.

\bibitem{Brunet:2005bz}
E.~Brunet, B.~Derrida, A.~H. Mueller, S.~Munier, {A phenomenological theory
  giving the full statistics of the position of fluctuating pulled fronts},
  Phys. Rev. E73 (2006) 056126.

\bibitem{Marquet:2006xm}
C.~Marquet, G.~Soyez, B.-W. Xiao, {On the probability distribution of the
  stochastic saturation scale in QCD}, Phys. Lett. B639 (2006) 635--641.

\bibitem{moro-2004-70}
E.~Moro, Numerical schemes for continuum models of reaction-diffusion systems
  subject to internal noise, Physical Review E 70 (2004) 045102.

\bibitem{Munier:2008cg}
S.~Munier, G.~P. Salam, G.~Soyez, {Travelling waves and impact-parameter
  correlations}, Phys. Rev. D78 (2008) 054009.

\bibitem{Ivanov:1998we}
D.~Y. Ivanov, et~al., {The BFKL pomeron in 2+1 dimensional {QCD}}, Phys. Rev.
  D58 (1998) 074010.

\bibitem{Ciafaloni:1999yw}
M.~Ciafaloni, D.~Colferai, G.~P. Salam, {Renormalization group improved small-x
  equation}, Phys. Rev. D60 (1999) 114036.

\bibitem{Iancu:2007st}
E.~Iancu, L.~McLerran, {Liouville field theory for gluon saturation in QCD at
  high energy}, Nucl. Phys. A793 (2007) 96--127.

\bibitem{Catani:1990xk}
S.~Catani, M.~Ciafaloni, F.~Hautmann, {GLUON CONTRIBUTIONS TO SMALL x HEAVY
  FLAVOR PRODUCTION}, Phys. Lett. B242 (1990) 97.

\bibitem{Catani:1990eg}
S.~Catani, M.~Ciafaloni, F.~Hautmann, {High-energy factorization and small x
  heavy flavor production}, Nucl. Phys. B366 (1991) 135--188.

\bibitem{Collins:1991ty}
J.~C. Collins, R.~K. Ellis, {Heavy quark production in very high-energy hadron
  collisions}, Nucl. Phys. B360 (1991) 3--30.

\bibitem{Iancu:2003ge}
E.~Iancu, K.~Itakura, S.~Munier, {Saturation and BFKL dynamics in the HERA data
  at small x}, Phys. Lett. B590 (2004) 199--208.

\bibitem{Forshaw:2004xd}
J.~R. Forshaw, R.~Sandapen, G.~Shaw, {Predicting F2(D(3)) from the colour glass
  condensate model}, Phys. Lett. B594 (2004) 283--290.

\bibitem{Soyez:2007kg}
G.~Soyez, {Saturation QCD predictions with heavy quarks at HERA}, Phys. Lett.
  B655 (2007) 32--38.

\bibitem{Thorne:2005kj}
R.~S. Thorne, {Gluon distributions and fits using dipole cross- sections},
  Phys. Rev. D71 (2005) 054024.

\bibitem{Kowalski:2003hm}
H.~Kowalski, D.~Teaney, {An impact parameter dipole saturation model}, Phys.
  Rev. D68 (2003) 114005.

\bibitem{Watt:2007nr}
G.~Watt, H.~Kowalski, {Impact parameter dependent colour glass condensate
  dipole model}, Phys. Rev. D78 (2008) 014016.

\bibitem{Sergey:2008wk}
S.~Bondarenko, {Gluon density and $F_{2}$ functions from BK equation with
  impact parameter dependence}, Phys. Lett. B665 (2008) 72--78.

\bibitem{Ewerz:2007md}
C.~Ewerz, A.~von Manteuffel, O.~Nachtmann, {On the Range of Validity of the
  Dipole Picture}, Phys. Rev. D77 (2008) 074022.

\bibitem{Dumitru:2007ew}
A.~Dumitru, E.~Iancu, L.~Portugal, G.~Soyez, D.~N. Triantafyllopoulos, {Pomeron
  loop and running coupling effects in high energy QCD evolution}, JHEP 08
  (2007) 062.

\bibitem{Hatta:2006hs}
Y.~Hatta, E.~Iancu, C.~Marquet, G.~Soyez, D.~N. Triantafyllopoulos, {Diffusive
  scaling and the high-energy limit of deep inelastic scattering in QCD at
  large N(c)}, Nucl. Phys. A773 (2006) 95--155.

\bibitem{Kozlov:2006qw}
M.~Kozlov, A.~I. Shoshi, B.-W. Xiao, {Total gluon shadowing due to fluctuation
  effects}, Nucl. Phys. A792 (2007) 170--186.

\bibitem{Brunet:2006zn}
E.~Brunet, B.~Derrida, A.~H. Mueller, S.~Munier, {Noisy traveling waves: effect
  of selection on genealogies}, Europhys. Lett. 76 (2006) 1--7.

\bibitem{Munier:2008rh}
S.~Munier, F.~Schwennsen, {Resummation of projectile-target multiple
  scatterings and parton saturation}, Phys. Rev. D78 (2008) 034029.

\end{thebibliography}

\end{document}